\documentclass[twocolumn,twocolappendix,trackchanges]{aastex701}

\usepackage{amsmath}
\usepackage{verbatim}

\usepackage[dvipsnames]{xcolor}
\begin{document}

\title{Long-term optical and near-infrared photometric evolution of SN\,2019vxm, an interacting Type~IIn supernova}

\shorttitle{Photometric evolution of SN\,2019vxm}
\shortauthors{Lelkes et al.}

\author[0009-0007-3760-515X]{Kl\'ara Lelkes}
\affiliation{HUN-REN CSFK, Konkoly Observatory, MTA Centre for Excellence, Konkoly Thege Mikl\'os \'ut 15-17, Budapest, 1121 Hungary}
\affiliation{ELTE E\"otv\"os Lor\'and University, Institute of Physics and Astronomy, P\'azm\'any P\'eter s\'et\'any 1, Budapest, Hungary}
\email{lelkes.klara@csfk.org}

\correspondingauthor{Kl\'ara Lelkes}
\email{lelkes.klara@csfk.org}

\author[0000-0002-8159-1599]{L\'aszl\'o Moln\'ar}
\affiliation{HUN-REN CSFK, Konkoly Observatory, MTA Centre for Excellence, Konkoly Thege Mikl\'os \'ut 15-17, Budapest, 1121 Hungary}
\affiliation{ELTE E\"otv\"os Lor\'and University, Institute of Physics and Astronomy, P\'azm\'any P\'eter s\'et\'any 1, Budapest, Hungary}
\affiliation{MTA–HUN-REN CSFK Lendület “Momentum” Stellar Pulsation Research Group}
\email{molnar.laszlo@csfk.org}

\author[0000-0001-8764-7832]{J\'ozsef Vink\'o} 
\affiliation{HUN-REN CSFK, Konkoly Observatory, MTA Centre for Excellence, Konkoly Thege Mikl\'os \'ut 15-17, Budapest, 1121 Hungary}
\affiliation{ELTE E\"otv\"os Lor\'and University, Institute of Physics and Astronomy, P\'azm\'any P\'eter s\'et\'any 1, Budapest, Hungary}
\affiliation{Department of Experimental Physics, University of Szeged, D\'om t\'er 9, Szeged, 6720 Hungary}
\email{vinko@konkoly.hu}

\author[0000-0002-8585-4544]{Attila Bódi}
\affiliation{Department of Astrophysical Sciences, Princeton University, NJ 08544, USA}
\affiliation{HUN-REN CSFK, Konkoly Observatory, MTA Centre for Excellence, Konkoly Thege Mikl\'os \'ut 15-17, Budapest, 1121 Hungary}
\email{}

\author[0000-0001-6232-9352]{Zsófia Bora}
\affiliation{HUN-REN CSFK, Konkoly Observatory, MTA Centre for Excellence, Konkoly Thege Mikl\'os \'ut 15-17, Budapest, 1121 Hungary}
\affiliation{ELTE E\"otv\"os Lor\'and University, Institute of Physics and Astronomy, P\'azm\'any P\'eter s\'et\'any 1, Budapest, Hungary}
\email{}

\author[0000-0002-6497-8863]{Borbála Cseh}
\affiliation{HUN-REN CSFK, Konkoly Observatory, MTA Centre for Excellence, Konkoly Thege Mikl\'os \'ut 15-17, Budapest, 1121 Hungary}
\affiliation{MTA-ELTE Lend{\"u}let "Momentum" Milky Way Research Group, Hungary}
\email{}

\author[0000-0002-1663-0707]{Csilla Kalup}
\affiliation{HUN-REN CSFK, Konkoly Observatory, MTA Centre for Excellence, Konkoly Thege Mikl\'os \'ut 15-17, Budapest, 1121 Hungary}
\affiliation{ELTE E\"otv\"os Lor\'and University, Institute of Physics and Astronomy, P\'azm\'any P\'eter s\'et\'any 1, Budapest, Hungary}
\affiliation{MTA–HUN-REN CSFK Lendület “Momentum” Stellar Pulsation Research Group}
\email{}

\author[0000-0002-8770-6764]{Réka Könyves-Tóth}
\affiliation{HUN-REN CSFK, Konkoly Observatory, MTA Centre for Excellence, Konkoly Thege Mikl\'os \'ut 15-17, Budapest, 1121 Hungary}
\email{}

\author[0000-0002-1792-546X]{Levente Kriskovics}
\affiliation{HUN-REN CSFK, Konkoly Observatory, MTA Centre for Excellence, Konkoly Thege Mikl\'os \'ut 15-17, Budapest, 1121 Hungary}
\email{}

\author{András Ordasi}
\affiliation{HUN-REN CSFK, Konkoly Observatory, MTA Centre for Excellence, Konkoly Thege Mikl\'os \'ut 15-17, Budapest, 1121 Hungary}
\email{}

\author[0000-0001-5449-2467]{András Pál}
\affiliation{HUN-REN CSFK, Konkoly Observatory, MTA Centre for Excellence, Konkoly Thege Mikl\'os \'ut 15-17, Budapest, 1121 Hungary}
\email{}

\author[0000-0002-3658-2175]{Bálint Seli}
\affiliation{HUN-REN CSFK, Konkoly Observatory, MTA Centre for Excellence, Konkoly Thege Mikl\'os \'ut 15-17, Budapest, 1121 Hungary}
\email{}

\author[0000-0001-9830-3509]{Zsófia Marianna Szabó}
\affiliation{Max-Planck-Institut für Radioastronomie, Auf dem Hügel 69, 53121, Bonn, Germany}
\affiliation{Scottish Universities Physics Alliance (SUPA), School of Physics and Astronomy, University of St Andrews, North Haugh, St Andrews, KY16 9SS, UK}
\affiliation{HUN-REN CSFK, Konkoly Observatory, MTA Centre for Excellence, Konkoly Thege Mikl\'os \'ut 15-17, Budapest, 1121 Hungary}
\email{}

\author[0000-0002-1698-605X]{Róbert Szakáts}
\affiliation{HUN-REN CSFK, Konkoly Observatory, MTA Centre for Excellence, Konkoly Thege Mikl\'os \'ut 15-17, Budapest, 1121 Hungary}
\email{}

\author[0000-0002-6471-8607]{Krisztián Vida}
\affiliation{HUN-REN CSFK, Konkoly Observatory, MTA Centre for Excellence, Konkoly Thege Mikl\'os \'ut 15-17, Budapest, 1121 Hungary}
\email{}

\begin{abstract}
The diversity of Type IIn supernovae is largely driven by the properties of the circumstellar material (CSM) they explode into. We examine the temporal evolution of SN\,2019vxm, an 
interacting supernova that belongs to 
the class of long-lasting Type~IIn events, using multicolor photometry spanning the ultraviolet, optical and near-infrared wavelengths, including over 650 days of optical and 1500 days of IR coverage. The evolution of the spectral energy distribution and bolometric luminosity, as well as the effective temperature and radius of the photosphere, indicates that the supernova was initially surrounded by an optically thick CSM, which was heated and pushed outward by the forward shock of the impacting ejecta. About 80--100 days after the explosion the forward shock and the photosphere decouples, and we observe the receding photosphere of the H-recombination front within the now thinned CSM. Near-IR measurements reveal long-lasting, slowly cooling emission from circumstellar dust around SN\,2019vxm and an IR rebrightening about one year after explosion, which we tentatively identify as a signature of an outer CSM region. We find that due to the moving photosphere and the transition from optically thick to partially thin inner CSM, modeling the explosion and subsequent interaction of the ejecta with the CSM to infer progenitor and CSM masses faces difficulties. Nevertheless, the inferred high %ejecta and CSM 
masses and extremely high mass-loss rates point to a massive progenitor undergoing intense pre-supernova mass loss.
\end{abstract}



\section{Introduction} \label{sec:intro}

Supernovae (SNe), one of the most energetic phenomena in the Universe, were first classified into Type I and II based on the lack or presence of H lines in their spectra by \citet{Minkowski_1941}.
Core-collapse SNe arise from the explosions of massive stars at the end of their lives and include hydrogen-rich Type II as well as hydrogen-poor Type Ib and Ic events.
Among the hydrogen-rich explosions, Type IIn SNe form a rare and highly distinctive subclass, accounting for only $\sim$4-9\% of all core-collapse supernovae \citep{Li-2011,Smith-2011,Cold-Hjorth-2023}. 
The defining characteristic of SNe IIn is the presence of narrow emission components, particularly in the Balmer series \citep{Schlegel_1990}. 
These narrow features typically exhibit full widths at half maximum of $\sim 100 $~km\,s$^{-1}$, indicating the existence of slowly moving circumstellar material (CSM) surrounding the progenitor, into which the SN explodes. 
In addition to their spectroscopic signatures, many SNe~IIn display a slowly evolving blue continuum and are, on average, more luminous than other Type~II SNe, with typical peak absolute magnitudes around $M_B \approx -18.7$ mag \citep{Kiewe-2012}.
In extreme cases, some events reach the regime of the superluminous supernovae (SLSNe), such as SN~2006gy, SN~2008am, ASASSN-14il and SN~2015da \citep{Smith_2007ApJ...666.1116S_SN_2006gy, Chatzopoulos_2011ApJ...729..143C_SN_2008am, Dukiya_2024, Smith_2024MNRAS.530..405S}. 
Overall, SNe~IIn represent a highly diverse class, spanning a wide range of luminosities and light-curve timescales, with several informal subtypes (e.g. \citealt{Taddia2013A&A...555A..10T, Smith_2017hsn..book..403S}). 
This diversity likely reflects a broad range of progenitor systems and pre-explosion mass-loss histories.

A key feature of SNe~IIn is the strong interaction between the rapidly expanding SN ejecta and the dense CSM expelled by the progenitor during the final stages of its evolution. The narrow emission components originate from these interactions. 
In many SNe~IIn, the CSM is sufficiently dense and optically thick to obscure the underlying SN ejecta, so that the observed spectra are dominated by the interaction region and typically lack classical P~Cygni profiles \citep{Smith_2017hsn..book..403S}. 
The collision between the ejecta and the CSM generates strong shocks that efficiently convert kinetic energy into radiation, powering the observed luminosity. 
As a result of sustained interaction, SNe~IIn are among the longest visible optical transients, which can remain observable for years or even decades after explosion \citep{Tartaglia_2020A&A...635A..39T_SN_2015da, Moran_2023A&A...669A..51M}.

The dense and dusty CSM also gives rise to long-lasting infrared (IR) emission observed in many SNe~IIn, often lasting for years to decades after the explosion \citep[see, e.g.,][]{Van-Dyk-2013}. This emission can originate from several processes (e.g., \citealt{Szalai_2013A&A...549A..79S}), including infrared echoes from pre-existing dust heated by the SN radiation, 
as well as newly formed dust in the post-shock region or within the ejecta itself \citep{Gall-2014,Moran_2023A&A...669A..51M}. 

The presence of such dense CSM implies intense mass-loss episodes shortly before explosion \citep{Smith_2014}, which makes SNe~IIn particularly important for studying the poorly understood mass-loss processes in the final stages of massive star evolution.
The early onset of ejecta--CSM interaction in many events further suggests that this material is located close to the progenitor \citep{Smith_2014}. 
The inferred mass-loss rates can be extreme, reaching $\gtrsim 10^{-2}\,\mathrm{M}_\odot \, \mathrm{yr^{-1}}$ or even higher \citep{Smith_2014, Hiramatsu_2024ApJ...964..181H}, and in some cases the total mass of the CSM can exceed $\sim 1\, \mathrm{M}_\odot$. 
Several physical processes have been proposed to explain this extreme mass loss. While steady line-driven winds may contribute to the CSM \citep{Lamers_1999isw..book.....L}, they are generally insufficient to account for the highest mass-loss rates inferred in SNe~IIn. 
Instead, eruptive mass-loss episodes \citep{Smith_2014, Smith_2017hsn..book..403S, Moriya_2023A&A...677A..20M} 
and pulsational pair-instability events in very massive stars are likely required \citep{Woosley_2007Natur.450..390W, Woosley_2017ApJ...836..244W}. 

In particular, luminous blue variables (LBVs) are considered the leading progenitors for strongly interacting SNe~IIn \citep{Trundle-2008,Smith-LBV-2014}. 
These very massive stars (typically $M_{\rm ZAMS} \gtrsim 25$--$40\,\mathrm{M}_\odot$) represent a short-lived evolutionary phase between O-type stars and Wolf-Rayet stars \citep{Humphreys_1994PASP..106.1025H} 
They can undergo giant eruptions ejecting several solar masses of material on  timescales years to decades \citep{Humphreys_1999PASP..111.1124H, Smith-2011}, with mass-loss rates that can reach $\gtrsim  0.1\,\mathrm{M}_\odot\,\mathrm{yr^{-1}}$ \citep{Smith_2014}. Such eruptive behavior is supported by the detection of precursor outbursts in 
approximately half of SNe IIn \citep{Ofek_2013Natur.494...65O, Ofek_2014ApJ...789..104O}. A well-known example is the Great Eruption of $\eta$~Carinae, during which several solar masses of material were expelled over a relatively short timescale \citep{Humphreys_1999PASP..111.1124H}. These eruptions are also linked to so-called SN impostors, which exhibit IIn-like spectra but are generally less luminous and do not result in the complete destruction of the star \citep{Smith_2011MNRAS.415..773S}. 

Red supergiants have also been proposed as potential progenitors of some SNe~IIn, particularly at the lower end of the progenitor mass range  ($\sim8$--$25\,\mathrm{M}_\odot$). 
However, their typical steady mass-loss rates ($\sim10^{-6}$--$10^{-3}\,\mathrm{M}_\odot\,\mathrm{yr^{-1}}$) \citep{Lamers_1999isw..book.....L, Smith_2014} 
are generally insufficient to explain the most extreme CSM densities inferred in strongly interacting events.

Binary interaction likely also plays a crucial role as well. 
A large fraction of massive stars reside in binary or multiple systems \citep{Sana_2012Sci...337..444S}, and processes such as Roche-lobe overflow can significantly enhance mass loss and shape the structure of the circumstellar environment. Spectropolarimetric observational evidence suggests that the CSM in SNe~IIn is often asymmetric \citep{Smith_2024MNRAS.530..405S}, exhibiting disk- or torus-like geometries and clumpy structures \citep{Chandra_2022MNRAS.517.4151C, Bilinski_2024MNRAS.529.1104B}. 
Such asymmetries indicate that the progenitor’s mass-loss history was complex and possibly influenced by binary interaction.

In this paper we analyze and model the temporal evolution of the luminous IIn-type supernova SN\,2019vxm, one of the best characterized SN~IIn explosions to date based on long-term, multicolor photometry. Early evolution of the SN in optical, UV and X-rays, including the shock breakout, was analyzed and modeled by \citet{Lane-2025}. We instead focus on the long-term evolution of the event. 
Analysis of spectroscopic observations of SN 2019vxm will be presented in a separate publication by Smith et al.\ (in prep.).

This paper is organized as follows: In Section \ref{sec:data}, we describe the photometric data sources we use and the reduction and correction steps we applied. In Section \ref{sec:phys_param}, we derive various physical parameters from the observational data, and in Section \ref{sec:models} we discuss the interaction between SN ejecta and the CSM, and we fit the data with various SN ejecta and CSM interaction models.

\section{Observations and data} \label{sec:data}

In this paper we focus on the analysis of the photometric data collected for SN\,20219vxm. We identified multiple sources of photometry: our own photomety collected at Piszkéstető Mountain Station; the photometry published by \citet{Tsvetkov_2024}; the \textit{Gaia} $G$--band and All-Sky Automated Survey for Supernovae (ASAS-SN) survey $g$--band observations \citep{Gaia-2016,Shappee-2014,ASAS-SN-2017}; and near-infrared photometry from the NEO\-WISE mission \citep{Mainzer-2011,Mainzer-2014}. Based on the work of \citet{Lane-2025}, we also incorporated the Sinistro photometry collected by the Las Cumbres Observatory Global Telescope Network \citep[LCOGT,][]{Brown-2013} and the UV-optical photometry of the Neil Gehrels Swift Observatory \citep{Gehrels-2004}. Below we first describe the discovery and initial observations of the SN, then our process of collecting and processing the various data sets, as well as our efforts to combine the Konkoly photometry from Piszkéstető with that of \citet{Tsvetkov_2024} for the joint modeling afterwards.

\subsection{Discovery and initial measurements}
SN\,2019vxm was discovered on 01/12/2019 (JD 2458818.54) by the ASAS--SN survey, cataloged under the designation ASASSN--19acc, at coordinates $\alpha_{2000}$ = 19:58:28.54, $\delta_{2000}$ = +62:08:15.83 \citep{Cacella-2019,Stanek-2019}. Discovery of the transient was also reported by the \textit{Gaia} Photometric Science Alerts system and the Asteroid Terrestrial-impact Last Alert System (ATLAS) survey \citep{Hodgkin-2021,Tonry-2018} about 37 hours later, cataloged as Gaia19fje and ATLAS19bchy, respectively. An early, low-resolution ($R\sim130$) spectrum\footnote{\url{https://www.wis-tns.org/object/2019vxm}}, taken from the Three Hills Observatory at JD 2458819.22, less than a day after discovery, showed the characteristic narrow H emission lines, indicating that the event was a Type IIn SN, at a redshift of $z\approx0.019$ \citep{Leadbeater-2019}. 

The \textit{Fermi} space telescope observed a 7.4\,s long gamma-ray burst cataloged as GRB191117A with its Gamma-Ray Burst Monitor Instrument at JD 2458804.51 \citep{Meegan-2009}. The initial brightening of the SN was observed by the Transiting Exoplanets Survey Satellite \citep[TESS,][]{Ricker-2015} that was collecting continuous photometry of the area,  for nearly 11 days. Based on the correlation between the timing of the GRB and and the onset of the brightening in the TESS light curve (LC), \citet{Lane-2025} were able to constrain the time of first light (the shock breakout through the dense CSM surrounding the supernova) with an accuracy of several hours from the TESS data to MJD $2458804.03\pm0.30$.

\begin{figure}[]
    \centering
    \includegraphics[width=1.0\columnwidth]{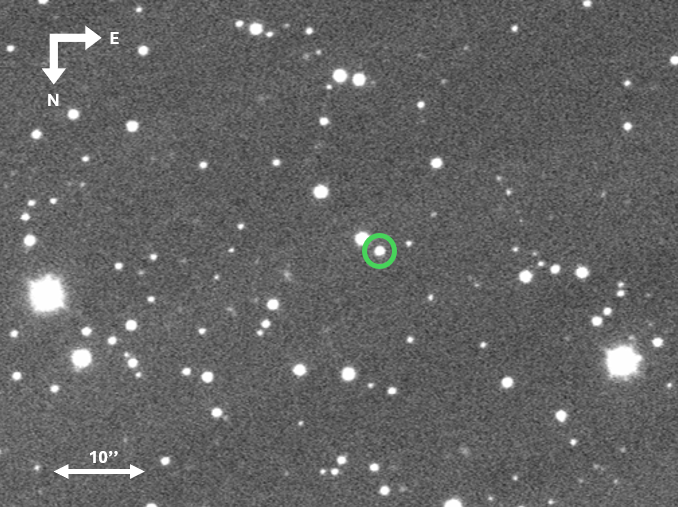}
    \includegraphics[width=1.0\columnwidth]{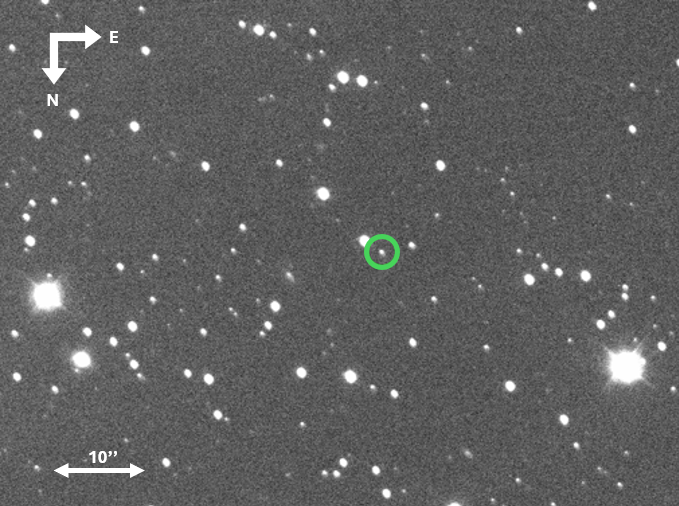}
    \caption{Sloan \textit{r}-band images of SN\,2019vxm obtained with the 0.8\,m Ritchey--Crétien robotic telescope at the Piszkéstető Mountain Station of Konkoly Observatory. The top panel shows the field on 03/12/2019, shortly after discovery and corresponding to the first Konkoly observation used in our analysis. The bottom panel shows the same region on 24/03/2021, representing the last epoch included in this work. The position of SN\,2019vxm is marked with a green circle in both images. Directions and a scale bar corresponding to $10''$ are indicated.}
    \label{fig:piszkes_images}
\end{figure}

The ASAS--SN survey collected early photometry in \textit{g} band, and then continued to follow the SN when visibility allowed. However, the quality of the data, especially at later phases, was poor compared to the other observations. Therefore we decided not to include it in our general analysis, but studied the early light curve in Section~\ref{subsec:Asassn_data}.

\subsection{Konkoly photometry}\label{subsec:Konkoly_data}

Multicolor photometric observations of SN\,2019vxm started two days after the discovery from Piszkéstető Mountain Station of Konkoly Observatory with the 0.8\:m Ritchey-Crétien robotic telescope, in Bessel \textit{BV} and Sloan \textit{griz} filters
(referred to as the \textit{Konkoly dataset} in this paper).
Example images of the supernova field obtained with this telescope are shown in Fig.~\ref{fig:piszkes_images}.
The detector was a Finger Lakes Instrumentation ProLine PL230 camera, equipped with a 2k$\times$2k CCD chip, resulting in a field-of-view of 18.8\arcmin$\times$18.8\arcmin. Further details of the observational setup can be found in \citet{Barna-2023}.

Each observing night consisted of three consecutive exposures per filter, with individual exposure times of 300\,s in the \textit{B} band and 180\,s in all other filters.
The brightness of the supernova was determined via aperture photometry after standard image processing. For the \textit{griz} data we standardized the photometry using the Pan-STARRS PS1 brightnesses of stars within the field-of-view, in the AB magnitude system \citep{PanSTARRS-2016}. 
The \textit{BV} data were standardized in the Vega magnitude system, after converting the PS1 \textit{griz} magnitudes to $B$ and $V$ magnitudes, according to \citet{Tonry_2012}. Galactic interstellar extinction and reddening were estimated to be $A_V = 0.282$\,mag and $E(B-V)=0.091$\,mag, respectively, from the dust map of \citet{Schlafly_2011}. The resulting \textit{A} extinction values in each passband for all datasets are summarized in Table~\ref{tab:ext_coeff}, assuming a coefficient of $R=3.1$. Assumptions on host galaxy extinction will be discussed in Section~\ref{subsec:redshift_distance}.
The uncorrected observed apparent magnitudes are listed in Table~\ref{tab:Konkoly_photometry} in Appendix \ref{sec:Tables} and publicly available on GitHub\footnote{\url{https://github.com/LelkesKlara/SN2019vxm}}.
All observation epochs were converted to rest-frame phases by correcting for cosmological time dilation.
The epoch $t_p$ was chosen to correspond to the time of maximum (or peak) light in the $B$ band, measured by \citet{Tsvetkov_2024} as MJD~58829.64 (12/12/2019).

Observations ran from 03/12/2019 to 04/04/2022, but data after 24/03/2021 were discarded after processing due to low signal-to-noise ratio. The initial few observations were followed by an 80\,d data gap caused by bad weather and poor visibility of the target.

\subsection{Tsvetkov et al.\ (2024) photometry}\label{subsec:Tsvetkov_data}

Multicolor photometry of SN\,2019vxm was also collected by multiple Russian observatories, in \textit{UBVRI} passbands. The standardized photometric data were published by \citet{Tsvetkov_2024}
(referred to as the \textit{Tsvetkov dataset} in this paper).
These data extend the coverage of the supernova considerably: they fill the gaps within the Konkoly data, including the time of peak brightness, and covers the late phase of the LC. It also extends coverage into the $U$ band, which is important for the early, hot phase of the explosion. Therefore we decided to combine the two data sets. However, we found a significant systematic offset between the two \textit{B}--band light curves. 
This likely originates from differences between the effective transmissions between the Konkoly and Tsvetkov B--bands. To remove these differences, we decided to calibrate both observations against synthetic photometry from the \textit{Gaia} mission.

One of the data products the \textit{Gaia} mission collected are the low-resolution (R $\approx$ 25-100) XP spectra from its BP and RP spectro-photometric channels. These spectra make it possible to construct synthetic photometry in any filter that falls within the wavelength range of the combined XP spectra (between 330 and 1050 nm). In \textit{Gaia} Data Release 3, pre-computed synthetic photometric brightness values were released for various filter systems, including Johnson-Krohn-Cousins's \textit{UBVRI}, SDSS \textit{ubvri}, and selected HST passbands for about 220 million stars \citep{GaiaDR3-syth-phot-2023}. We extracted synthetic Johnson \textit{BV} photometry for the comparison stars listed by \citet{Tsvetkov_2024}, and compared their synthetic magnitude values to those from the ground-based data sets. We then determined the corrections as a function of ($B-V$) colors, and applied them to the \textit{B} photometry of SN2019vxm to homogenize the measurements: 
\begin{multline}
B_{Gaia} = B_{\mathrm{Konkoly}} + (0.286 \pm 0.067)(B-V)_{\mathrm{Konkoly}} \\ - (0.212 \pm 0.049) 
\end{multline}
\begin{multline}
B_{Gaia}  = B_{\mathrm{Tsvetkov}} + (0.0317 \pm 0.0251)(B-V)_{\mathrm{Tsvetkov}} \\ - (0.0064 \pm 0.0178).
\end{multline}

\begin{figure*}[]
\includegraphics[width=0.98\textwidth]{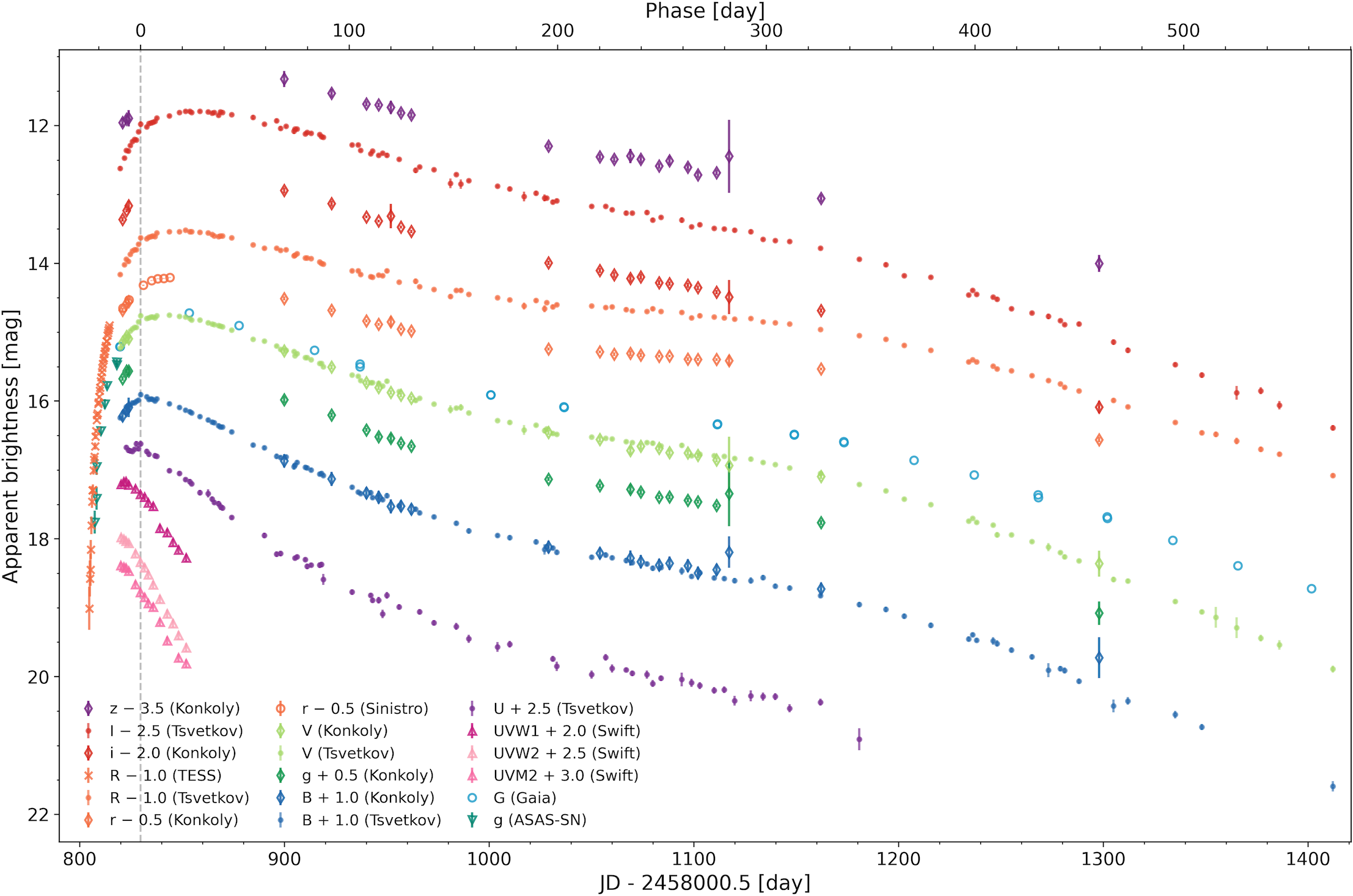}
\caption{The apparent brightness of SN\,2019vxm over time via multi-band photometry. 
Diamond symbols indicate the Konkoly data; dots are from \citet{Tsvetkov_2024}; blue open circles correspond to \textit{Gaia} $G$-band observations; green triangles to ASAS--SN $g$-band measurements; orange crosses to TESS photometry; pink-purple triangles to \textit{Swift} UVOT observations; and orange circles to LCOGT-Sinistro $r$-band data. 
Data in the various passbands have been shifted by constant offsets relative to the Johnson \textit{V}-band LC for clarity. 
The vertical dashed line marks the epoch of peak brightness ($t_p$) in the \textit{B}-band at MJD~58829.64, which is also the starting point of the phase scale.}
\label{fig:LC}
\end{figure*}

\noindent The final light curves are shown in Fig.~\ref{fig:LC}.

\subsection{LCOGT Sinistro photometry}\label{subsec:Sinistro_data}

The supernova was observed by Las Cumbres Observatory Global Telescope Network (LCOGT, \citealt{Brown-2013}), which consists of 12 robotic 1\,m telescopes across the globe. The $r$-band photometry published by \citet{Lane-2025} comes from the McDonald Observatory (Texas, USA) node of the network. The telescopes use the Sinistro cameras specifically designed for the project. We refer to the data with this name, following the nomenclature used by \citet{Lane-2025}.

\begin{deluxetable}{c c | c c | c c}
\tablecaption{Calculated Galactic interstellar extinction values for each passband we use, based on the \citet{Schlafly_2011} dust maps, except for the \textit{Swift} data for which calculations were calibrated to the \citet{sfd1998} maps.}
\tablecolumns{6}
\tablehead{
       \colhead{Band} & \colhead{A (mag)} & \colhead{Band} & \colhead{A (mag)} & \colhead{Band} & \colhead{A (mag)}  } 
\startdata
        UVW2 & 0.580 & UVM2 & 0.723 & UVW1 & 0.508 \\
        Swift-u & 0.427 & Swift-b & 0.353 & Swift-v & 0.266 \\
        $U$ & 0.446 & $B$ & 0.373 & $V$ & 0.282 \\
        $R$ & 0.223 & $I$ & 0.155 & $g$ & 0.341 \\
        $r$ & 0.234 & $i$ & 0.177 & $z$ & 0.132 \\
        W1 & 0.020 & W2 & 0.015 & ~ & ~ 
\label{tab:ext_coeff}
\enddata
\end{deluxetable}

\subsection{Swift UV-optical photometry}\label{subsec:Swift_data}

The \textit{Neil Gehrels Swift Observatory} is a multi-wavelength mission whose primary purpose is to detect gamma-ray bursts in gamma and X-ray wavelengths, and follow up them in the UV and optical, using the UVOT (Ultra-Violet Optical Telescope) instrument on board \citep{Gehrels-2004}. \textit{Swift} observed SN\,2019vxm around and after peak brightness for 32 days. Steps of the initial data processing are described by \citet{Lane-2025}. 

\textit{Swift} observes in six different passbands, marked UVW2, UWM2, UVW1, u, b, and v (not to be confused with SDSS \textit{u} or Strömgren \textit{bv} passbands). Calibrations of these passbands, flux zero points, count rate fo flux conversions, and transformations into other passbands were published by \citet{Poole-2008}. Conversions into Johnson passbands are based on synthetic stellar and GRB spectra and we found the results unsatisfactory for our IIn-type SN data, therefore we decided 
not to convert the measurements and instead work with the original \textit{Swift} passbands. Extinction coefficients were determined by \citet{Yi-2023}, but the extinction correction is based on the older \citet{sfd1998} maps which differ from the \citet{Schlafly_2011} maps by a scaling factor. Therefore we used $E(B-V)_{\rm SFD}=0.104$\,mag to calculate extinctions in the \textit{Swift} passbands. These \textit{A} extinction values are also listed in Table~\ref{tab:ext_coeff}. 

\subsection{NEOWISE infrared photometry}\label{subsec:NEOWISE_data}
The Wide-field Infrared Survey Explorer (WISE) space telescope completed its original, four-channel cryogenic and two-channel post-cryogenic sky surveys in 2010--11 \citep{Wright-2010}. The spacecraft was hibernated and then it was reactivated in 2013 and continued to survey the sky until 2024 with its two short-wavelength channels (W1 and W2, centered on 3.4 and 4.6 $\mu$m, respectively) during the NEOWISE mission \citep{Mainzer-2011,Mainzer-2014}. The main objective of the mission was to detect Near-Earth Objects, but measurements for stationary sources and transients were also extracted from the images. 

We accessed the W1 and W2 single-frame photometry data at the location of SN\,2019vxm through the NASA/IPAC Infrared Science Archive. We limited data points to within 1\arcsec{} of the source coordinates and removed all upper-limit flagged values. The deeper AllWISE catalog, which is based on co-added images, reports the brightness of the assumed host galaxy at ${\rm W1} = 17.16 \pm 0.07$\,mag and ${\rm W2} = 17.31 \pm 0.26$\,mag. This is too faint to be detected in the single NEOWISE images, and thus any pre-explosion data points can also be safely removed from the time series.

\begin{figure}[]
\includegraphics[width=0.98\columnwidth]{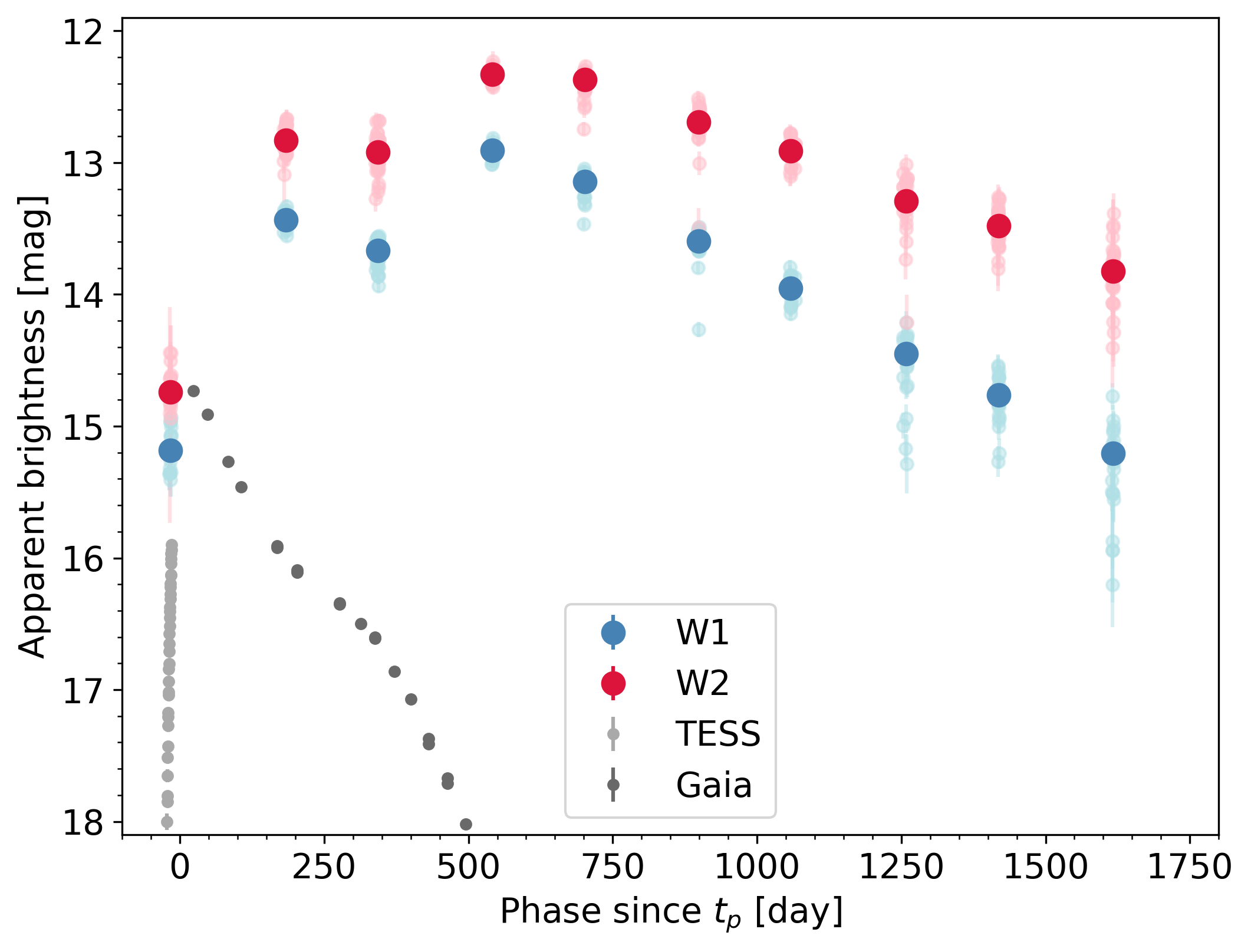}
\caption{
WISE photometry of SN\,2019vxm in the W1 and W2 bands. We combined individual measurements into average brightness values per season. Optical light curves from TESS and \textit{Gaia}, pre-and post-maximum, respectively, are shown for reference. Note the rebrightening after 500~d.
}
\label{fig:WISE_lc}
\end{figure}

The SN was detected once during the brightening phase before maximum light, and then nine further times until the termination of the mission, with several close-by individual measurements separated by $\approx200$ and $\approx 160$ day long gaps, due to the scanning pattern of the survey. We binned these clusters of measurements into ten averaged values. We show these apparent brightness values, along with optical light curves for reference, in Fig.~\ref{fig:WISE_lc}. 

WISE brightness values are published in the Vega magnitude system. We corrected the brightnesses for Galactic extinction based on the \citet{Schlafly_2011} maps (Table~\ref{tab:ext_coeff}), and then converted them to flux densities following the prescription provided in the WISE All-Sky Data Release Explanatory Supplement\footnote{\url{https://irsa.ipac.caltech.edu/data/WISE/docs/release/All-Sky/expsup/sec4_4h.html\#conv2flux}}. The in-band flux density was computed as

\begin{equation}\label{Eg:F_nu_WISE}
    F_\nu = F_{\nu,0} \cdot 10^{-m_{\rm Vega}^{}/2.5},
\end{equation}

\noindent where $F_{\nu,0}$ is the zero-magnitude flux density for the corresponding band. We adopted $F_{W1,0} = 3.09540 \cdot 10^{-21}$ erg~s$^{-1}$~cm$^{-2}$~Hz$^{-1}$ and
$F_{W2,0} = 1.71787 \cdot 10^{-21}$ erg~s$^{-1}$~cm$^{-2}$~Hz$^{-1}$.

\section{Physical parameters}\label{sec:phys_param}

In this section we derive the basic physical parameters of SN\,2019vxm based on its multi-band photometric observations, focusing on its distance, the evolution of its spectral energy distribution (SED), the effective temperature $(T_{\rm eff})$, the bolometric luminosity $(L_{\rm bol})$ and the radius of the photosphere.
By constructing the SED at multiple epochs, we can follow the temporal evolution of the continuum emission, from which $T_{\rm eff}$ and $L_{\rm bol}$ can then be estimated. We studied the evolution of both the hot photosphere, heated by the shock of the SN, and the thermal emission of the circumstellar dust.
These quantities provide direct constraints on the radiative output and the physical conditions of the emitting regions during the strong circumstellar interaction phase and then the late evolution of the SN. Throughout this Section we use phase (time in the rest-frame of the supernova) as the time unit, measured relative to the $B$-band maximum at $t_p = 58829.64$ MJD. Phases measured from the shock-breakout epoch, $t_0 = 58804.03$ MJD, are explicitly labeled as phase since $t_0$. Their difference, $\Delta t = 25.61$ d, corresponds to $t_{\rm rise} = 25.13$\,d in rest frame. The latter is the rest-frame rise time of the SN in the \textit{B} band.

\subsection{Host, redshift and distance} \label{subsec:redshift_distance}

The spectroscopic redshift of SN\,2019vxm is $z = 0.019$, measured from the narrow hydrogen emission lines present in the early spectra \citep{Leadbeater-2019}.
A nearby galaxy, SDSS J195828.83+620824.3, was associated with the supernova in the \textit{Gaia} transient alert as a potential host, although the position of SN was 8.5\arcsec{} South and 2.1\arcsec{} West from the center of the galaxy. Based on Sloan Digital Sky Survey (SDSS) multi-band imaging, a photometric redshift of $z_{\mathrm{ph}} = 0.071 \pm 0.040$ was calculated\footnote{\url{https://skyserver.sdss.org/dr19/VisualTools/explore/summary?ra=299.62015&dec=62.140099}}, but as the large uncertainties and the various warning flags in SDSS indicate, the reliability of this redshift value is questionable. 
Adopting such a high redshift would imply extremely luminous absolute magnitudes for SN\,2019vxm, approaching or exceeding the regime of Type~IIn SNe \citep{Ransome2025ApJ...987...13R}.
To test the plausibility of a higher redshift, we examined the implied peak absolute magnitudes in the red optical bands.

The $r$-band LC is constructed from the Konkoly and Sinistro datasets, which provide partial coverage of the rising phase but do not fully sample the peak, indicating that this value represents a lower limit for the true maximum brightness. 
To better constrain the peak luminosity, we therefore also considered the $R$-band photometry from the Tsvetkov dataset, which offers denser sampling around maximum light.

At the spectroscopic redshift, $z=0.019$, the extinction-corrected peak magnitudes are 
$M_R \approx -20.2$~mag and $M_r \approx -20.0$~mag indicating that SN\,2019vxm is about one magnitude more luminous than the average Type~IIn SN, but still well within the observed population.

In contrast, adopting redshifts corresponding to the SDSS photometric estimate would result in dramatically higher absolute magnitudes. At $z = 0.03$, the inferred magnitudes increase to $M_r \approx -21.0$~mag and $M_R \approx -21.2$~mag. For higher redshifts ($z \approx 0.07 - 0.11$), they would reach extreme values of $M_r \approx-22.8$ to $-23.8$~mag and $M_R \approx - 23.0$ to $-24.0$~mag. Such absolute magnitudes are far above the typical peak magnitude distribution of Type~IIn SNe, which has a mean of $\langle M_r \rangle \approx -19.2$~mag (Malmquist-corrected $\approx -18.7$~mag)  with a scatter of $\sim1$~mag \citep{Ransome2025ApJ...987...13R}.

We conclude that the galaxy is either not associated with the supernova, or its photometric redshift is overestimated by $1.3\sigma$ in SDSS. At this point this question cannot be resolved, and unfortunately the region of the supernova is currently not included in any deep sky survey missions such as the \textit{Euclid} and \textit{Roman} space telescopes, and is not visible from the Vera Rubin Observatory, either. Therefore we adopt $z=0.019$ as the redshift of SN\,2019vxm throughout this work. Assuming the Hubble constant of $H_0 = 73.3$ km s$^{-1}$ Mpc $^{-1}$ \citep{Wong-2020,Riess-2021,Brout-2022}, this corresponds to a luminosity distance of $d_L = 77.7$~Mpc. To account for peculiar velocity uncertainties in the local Universe, we include a velocity dispersion ($\pm 300$\,km s$^{-1}$), resulting in a distance uncertainty of $\pm 2.4$~Mpc.

The possible host galaxy identification also affects the expected amount of host-galaxy reddening. \citet{Tsvetkov_2024} assumed no host-galaxy extinction in their analysis. As we have shown, the supernova is far from the core of the galaxy and the association with it is somewhat uncertain. Because of that,  we do not apply any correction for host-galaxy extinction in this paper. Consequently, our derived absolute magnitudes and luminosities should be regarded as upper limits. An alternative treatment and estimates of possible host-galaxy extinction based on light-curve modeling was presented by \citet{Lane-2025}.

\subsection{Evolution of the UV-optical spectral energy distribution}\label{subsec:SED}

\begin{figure*}[]
\includegraphics[width=0.98\textwidth]{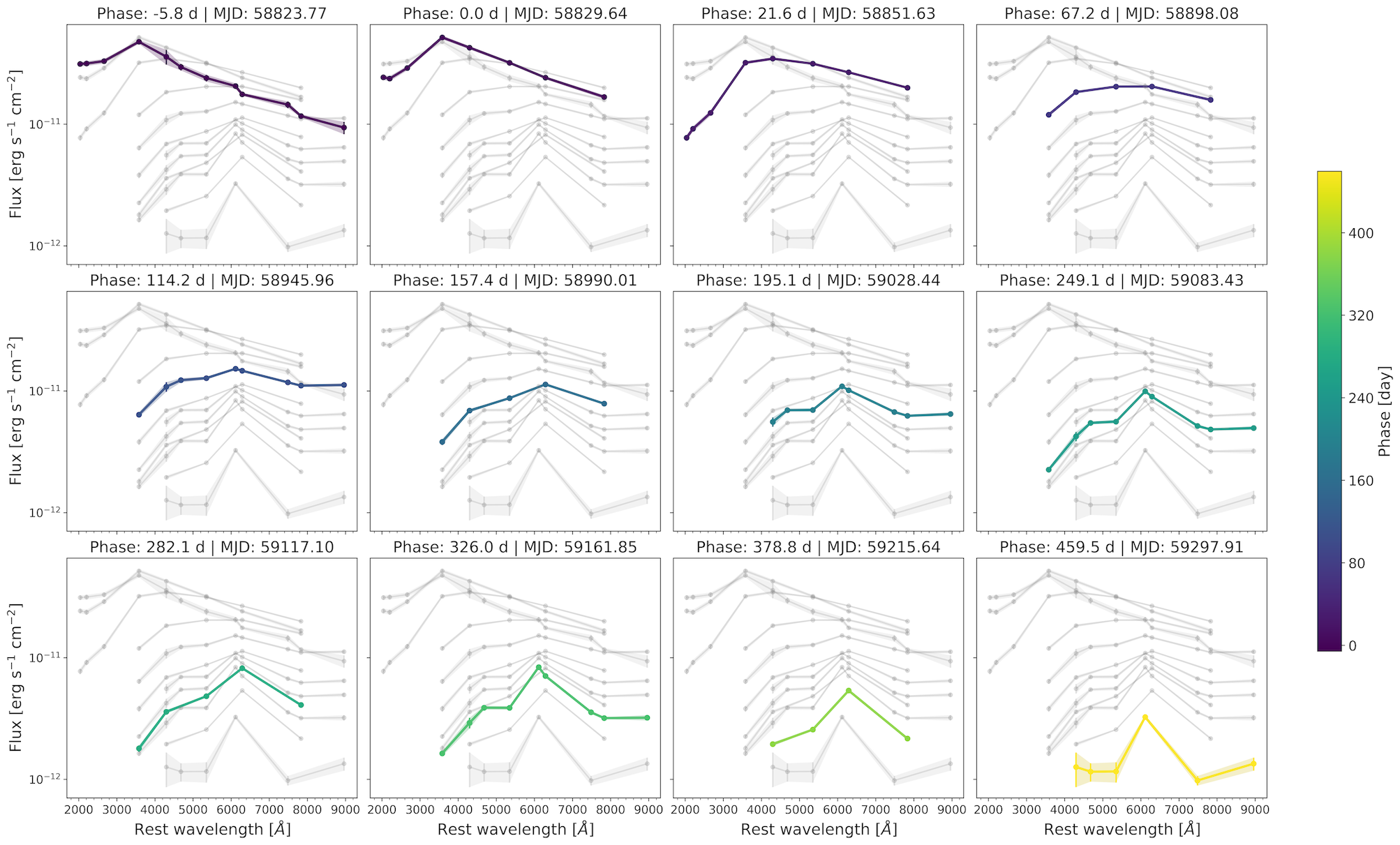}
\caption{UV-optical spectral energy distributions of SN\,2019vxm at selected epochs. All SEDs are shown in grey, and colored curves correspond to the selected epoch indicated above each panel. The SEDs were constructed by combining the Tsvetkov \textit{UBVRI} photometry with the Konkoly \textit{BVgriz} measurements, and $B$ and $V$ fluxes were averaged when both datasets were available. Near maximum light the SEDs are extended with \textit{Swift} UVOT observations.}
\label{fig:SED_12curves}
\end{figure*}

We constructed a series of SEDs of SN\,2019vxm using the available multi-band photometry to follow its evolution over time. For this we selected the Konkoly data set, the observations of \citet{Tsvetkov_2024}, and the \textit{Swift}/UVOT measurements (around maximum). 
Observed magnitudes were corrected for Galactic extinction (see Table~\ref{tab:ext_coeff}) and converted to flux densities following the prescriptions of \citet{Bessell_1998} and \citet{Tonry_2012}. Times for the SEDs are presented in the rest frame of the supernova (see Fig.~\ref{fig:SED_12curves}, with 12 SED realizations).

To minimise temporal evolution effects between data points, only measurements obtained within one day were combined at each SED epoch.
When both Konkoly and Tsvetkov data were available, the $B$ and $V$ band fluxes were averaged. Around the moment of maximum light we extended the optical SEDs into the ultraviolet with contemporaneous \textit{Swift} observations.

Fig.~\ref{fig:SED_12curves} shows the temporal evolution of the SED from shortly before maximum light up to approximately 460 days thereafter at selected phases.
At early phases the continuum is heavily blue-dominated, indicating a high effective temperature of the photosphere. As the supernova evolves, the peak of the SED progressively shifts toward longer wavelengths and the overall continuum flattens, reflecting rapid cooling of the emitting region (as discussed in more detail in Section \ref{subsec:temperature}). Figure~\ref{fig:SED: all Konkoly and Tsvetkov} in Section~\ref{subsec:optical_SED_evolution} of the Appendix shows the full optical SED evolution derived from the Konkoly and Tsvetkov photometry.

An increasing excess is clearly visible in the $r$ and $R$ bands at later phases, consistent with the growing contribution of the strong H$\alpha$ emission typical of Type~IIn supernovae. Although this H$\alpha$ emission is already present in the earliest spectrum obtained one day after discovery, its photometric imprint becomes more and more prominent at later epochs as the continuum cools and fades. 

In the ultraviolet, the flux is significantly suppressed relative to a pure blackbody (BB) continuum. This is most notable in panel three of Fig.~\ref{fig:SED_12curves}, at 21.6~d phase. This suppression is most likely caused by strong line blanketing by metal absorption, as often observed in interacting and superluminous supernovae \citep{Roming-2012,Yan-2018,Ponte-2026}.

Overall, the SED evolution traces the cooling of the emitting region from an initially hot, blue continuum to a progressively redder spectrum increasingly influenced by strong line emission driven by circumstellar interaction.

\subsection{Luminosity} \label{subsec:bol_lum}

With the SEDs at hand, we 
construct the pseudo-bolometric light curve of SN\,2019vxm by integrating the observed SEDs at each epoch.
We estimated the bolometric flux using trapezoidal integration over the observed wavelength range.

Because the photometric coverage does not extend to arbitrarily short and long wavelengths, we applied simple boundary corrections as extrapolations at both ends of the observed SED. 
On the red side, we approximated the unobserved near-infrared contribution by a Rayleigh--Jeans tail anchored to the measured flux at reddest band:
\begin{equation}
F_{\mathrm{IR}} \approx \int_{\lambda_N}^{\infty} F_{\lambda} \,{\mathrm{d}}\lambda 
\simeq \int_{\lambda_N}^{\infty} F_{\lambda,N} \left ( {\lambda_N \over \lambda} \right )^4 \,{\mathrm{d}}\lambda = \frac{F_{\lambda,N}\,\lambda_N}{3},
\end{equation}
where $\lambda_N$ and $F_{\lambda,N}$ is the mean wavelength and the de-reddened flux in the reddest available band, typically the $z$-band. We did not take the WISE observations into account here as they are sparse, and they primarily measure the dust emission beyond the shock-heated photosphere.

On the blue side, a direct BB extrapolation would likely overestimate the flux because the UV continuum of interacting SNe is often strongly suppressed by line blanketing, as mentioned before. Following \citet{Bersten2009ApJ...701..200B}, we estimated the missing flux between a fixed cutoff wavelength at $\lambda_0 = 2000$~\AA\ and the bluest observed band $\lambda_1$ as
\begin{equation}
F_{\mathrm{UV}} \approx \int_{\lambda_0}^{\lambda_1} F_{\lambda}\,{\mathrm{d}}\lambda 
\simeq \frac{F_{\lambda,1}\,(\lambda_1-\lambda_0)}{2}.
\end{equation}
This linear blue extrapolation has been commonly used in the literature and validated with \textit{Swift} UV observations \citep{Lyman2014MNRAS.437.3848L, Martinez2022A&A...660A..40M}, indicating that it provides a reliable estimate of the missing ultraviolet flux for core-collapse supernovae.

To improve wavelength coverage and enable a homogeneous bolometric calculation over the full time span, we estimated missing $z$- and $U$-band fluxes using smoothed light-curve representations. We used the $U$-band measurements from the Tsvetkov dataset to extend the Konkoly SEDs in blue up to $\sim350$~days after maximum, when $U$-band observations were still available. 
Conversely, the $z$-band flux from the Konkoly dataset was used to complement the Tsvetkov SEDs in the red. In both cases, we applied a LOWESS (locally weighted scatterplot smoothing; \citealt{Cleveland01121979}) regression to the measured fluxes to interpolate values contemporaneous with the Konkoly data. 

\begin{figure*}
\includegraphics[width=0.98\textwidth]{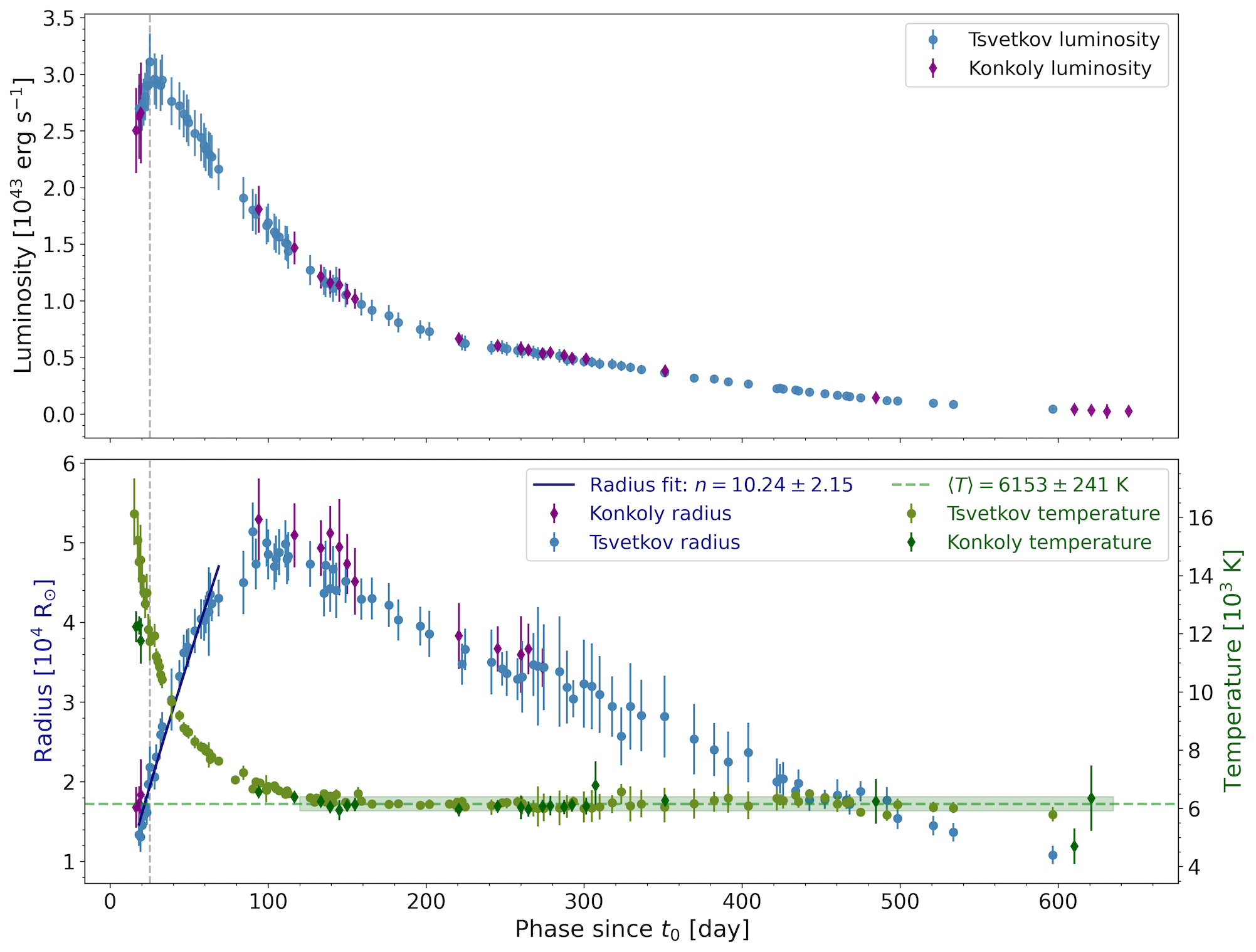}
\caption{\textbf{Top panel:} Pseudo-bolometric light curves of SN\,2019vxm. Red circles correspond to the \textit{BgVriz} Konkoly dataset, where the missing $U$-band flux was interpolated using the Tsvetkov measurements. Blue diamonds represent the \textit{UBVRI} Tsvetkov dataset, complemented with the interpolated $z$-band fluxes from the Konkoly dataset.
\textbf{Bottom panel:} Temporal evolution of the photospheric temperature and radius of SN\,2019vxm derived from the SED fits. The respective units are shown on the right and left axes. Green symbols represent the derived temperatures, with the dashed line marking the mean temperature during the plateau phase, $\langle T \rangle = 6153 \pm 241$ K.
Blue and purple symbols correspond to radii, and the solid blue line shows a power-law fit to the early radius evolution, with a best-fit index of $n = 10.24 \pm 2.15$.
Phase is measured in days relative to the shock breakout epoch of $t_0$, with the vertical dashed line marking $t_p$. The blue line is an \textit{n} parameter fit, which we discuss in Sec~\ref{subsec:opt_depth n_param}.
\label{fig:luminosity_temperature_and_radius}}
\end{figure*}

The bolometric luminosity was calculated assuming isotropic emission as $L_{\mathrm{bol}} = 4\pi d_L^2 F_{\mathrm{bol}}$, where we adopted the luminosity distance derived in Section~\ref{subsec:redshift_distance}. The resulting bolometric light curve is presented in Fig.~\ref{fig:luminosity_temperature_and_radius}. Uncertainties on $L_{\mathrm{bol}}$ include both the bolometric-flux uncertainty and the distance uncertainty. The detailed bolometric luminosity values used in this analysis are reported in  Appendix \ref{sec:Tables} (Table~\ref{tab:LTR}) and publicly available on GitHub\footnote{\url{https://github.com/LelkesKlara/SN2019vxm}}.

The bluest band in our calculation, $\lambda_1$, was typically the \textit{U} band until the late phase of the LC when the supernova became too faint and red and only \textit{B} band data were available (see the last two panels of Fig.~\ref{fig:SED_12curves}). For the few epochs around maximum where \textit{Swift} UV fluxes were measured, we computed the bolometric luminosities with these values included for comparison as well. The differences between the obtained luminosities were between negligible and less than $0.5\sigma$ compared to the values calculated without the \textit{Swift} data, validating the applied UV extrapolation further.

In order to compare SN\,2019vxm with recent population studies of interacting supernovae, we computed the energy radiated over a fixed 200-day interval following the definition of \citet{Ransome2025ApJ...987...13R}. We set $t_p$ as the epoch of maximum brightness, and the integral was evaluated numerically using trapezoidal integration of the merged bolometric light curve. For SN\,2019vxm we obtain a value of $E_{200} = (2.6 \pm 0.04) \times 10^{50}$~erg, corresponding to $\log(E_{200}/{\mathrm{erg}}) \simeq 50.42 \pm 0.01$. \citet{Ransome2025ApJ...987...13R} found that Type~IIn SNe exhibit a broad distribution of radiated energies with a median of $\log(E_{200}/{\mathrm{erg}}) = 50.3_{-0.7}^{+0.5}$ and possible bimodality with peaks near a few $10^{49}$ and a few $10^{50}$~erg. The value derived for SN\,2019vxm therefore places it towards the luminous end of the typical SN~IIn energy distribution.

\subsection{Effective temperature}\label{subsec:temperature}

$T_{\rm eff}$ at the photosphere of SN\,2019vxm was estimated by fitting a Planck function to the optical-UV SEDs at each epoch. 
To constrain the temperature from the continuum slope, we fit the SEDs in relative units by normalizing the fluxes to the $V$ band. 
This removes the need for overall scaling (distance, emitting radius) and makes the fit primarily sensitive to the SED shape, i.e., to temperature. For each epoch we constructed the relative fluxes $F_\lambda/F_V$ with rest-frame wavelengths $\lambda_{\mathrm{rest}}=\lambda_{\mathrm{eff}}/(1+z)$. The relative SEDs were modeled with a normalised BB function, $ f_\lambda^{\mathrm{model}}(T) =
B_\lambda(\lambda_{\mathrm{rest}},T) /B_\lambda(\lambda_V,T) $,
where $B_\lambda$ is the Planck function and $\lambda_V=5450$~\AA\ corresponds to the effective wavelength of the $V$ band. 

Photometric bands significantly affected by strong emission lines where handled with extra 
caution. In particular, the $r$ and $R$ bands include a significant contribution from H$\alpha$ emission. In the (photometric) SEDs this contribution is less apparent near maximum because of the dominating blue continuum, but it is already present in the earliest spectrum obtained one day after discovery \citep{Leadbeater-2019}. The emission becomes increasingly visible at later phases in the photometry as well, as the continuum cools and fades. Accordingly, we excluded $r$- and $R$-band fluxes from the temperature fits. 

For each epoch, we first obtained an initial deterministic solution using a weighted least-squares ($\chi^2$) minimization with the \texttt{curve\_fit} python routine.
The resulting best-fit temperature was then used to initialize a Markov-chain Monte Carlo (MCMC) sampling of the posterior distribution. We sampled the posterior using the \texttt{emcee} ensemble sampler \citep{emcee-2013}, fitting for two free parameters: $T_{\rm eff}$, and an additional scatter term, $\ln f$. We used 32 walkers and 10\,000 steps for each chain, discarding the first 4000 steps as burn-in.

The resulting temperature evolution is shown in Fig.~\ref{fig:luminosity_temperature_and_radius}. The effective temperatures inferred from the BB fits are listed in Table~\ref{tab:LTR} in Appendix \ref{sec:Tables}. The temperature decreases rapidly from $\sim 1.7 \times 10^4$~K around maximum brightness to $\sim 6000$~K within the first $\sim100$~days , and remains approximately constant at $\langle T_{\rm eff} \rangle = 6153 \pm 241$\,K for several hundred days. We attribute this constant temperature phase to the temperature of the hydrogen recombination front, which defines the photosphere at late phases \citep{Faran-2019}.

At the earliest phases we find differences of several thousand Kelvins between temperatures derived from the Konkoly and Tsvetkov data sets. This discrepancy is primarily driven by the contribution of the bluest bands, especially the contribution from the Tsvetkov $U$ band, which strongly constrains the slope of the SED near maximum brightness.

As we discussed in Section~\ref{subsec:bol_lum}, line blanketing suppresses the UV emission, leading to fluxes falling well below what a BB model would predict at the same temperature. This effect was approximated with a modified BB model by \citet{Yan-2018}, who introduced a power-law suppression component above a UV wavelength threshold. We fitted phases when \textit{Swift} photometry was available along with the optical data with this modified BB function:
\begin{equation}\label{eq:modified_BB}
    f_\lambda = f_{\rm \lambda,BB}\,(\lambda/\lambda_0)^\beta,
\end{equation}
where $\beta$ and $\lambda_0$ are free parameters and $\beta = 0$ and $f_\lambda = f_{\rm \lambda,BB}$ when $\lambda > \lambda_0$. We tested this model both with one of the parameters fixed or both fitted simultaneously, with $\lambda_0$ and $\beta$ allowed to vary between 3600--4500 \AA{} and 1--3, respectively. The resulting fit parameters are listed in Table~\ref{tab:Swift_modifiedBBfit_params}. The typical formal relative uncertainties are $\lesssim1\%$ for $T$, $\lesssim6.5\%$ for $\beta$, and negligible for $\lambda_0$. Realistic uncertainties would require us applying more sophisticated fitting methods, such as MCMC, again.

\begin{figure}
    \centering
    \includegraphics[width=0.99\columnwidth]{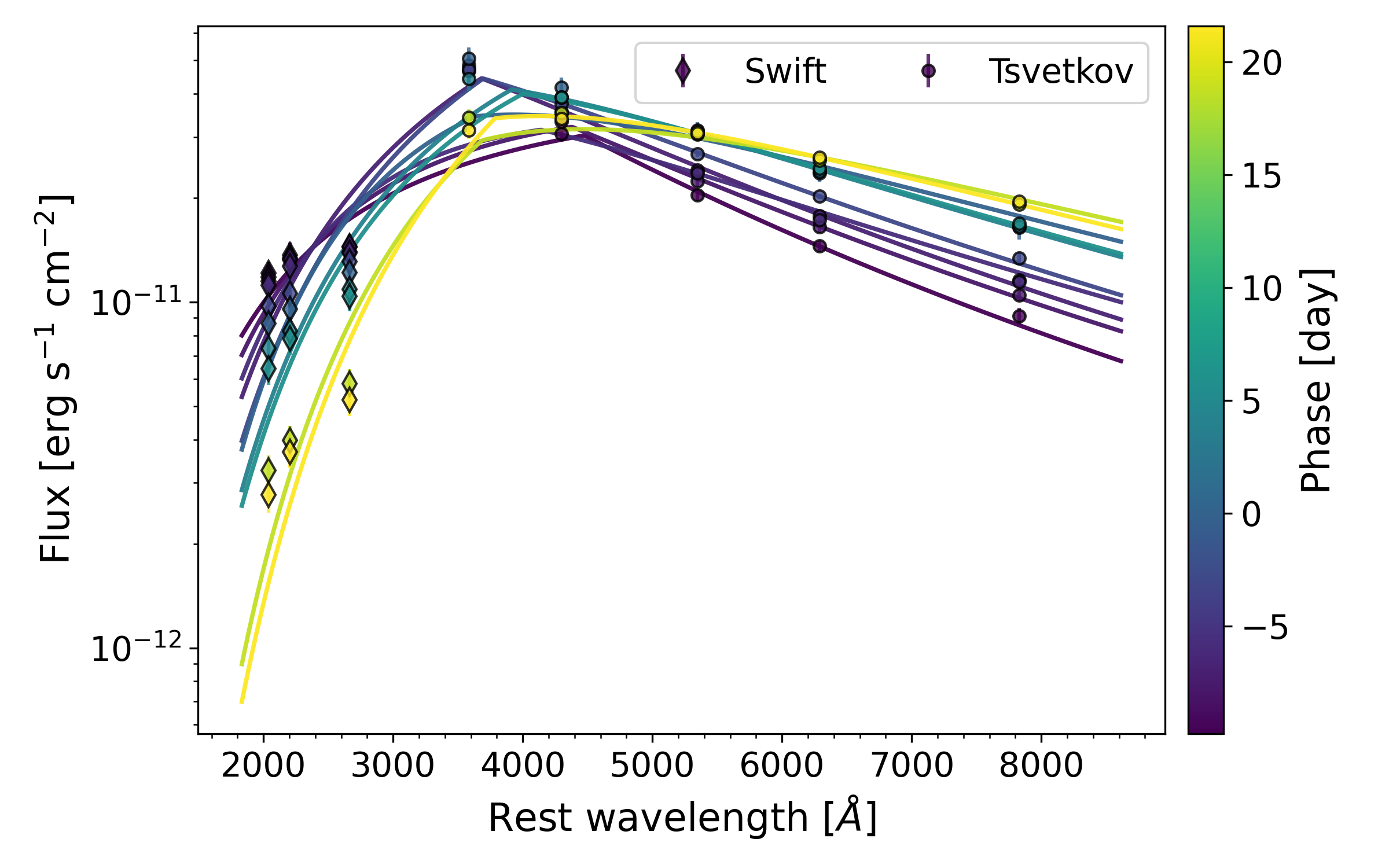}
    \includegraphics[width=0.87\columnwidth]{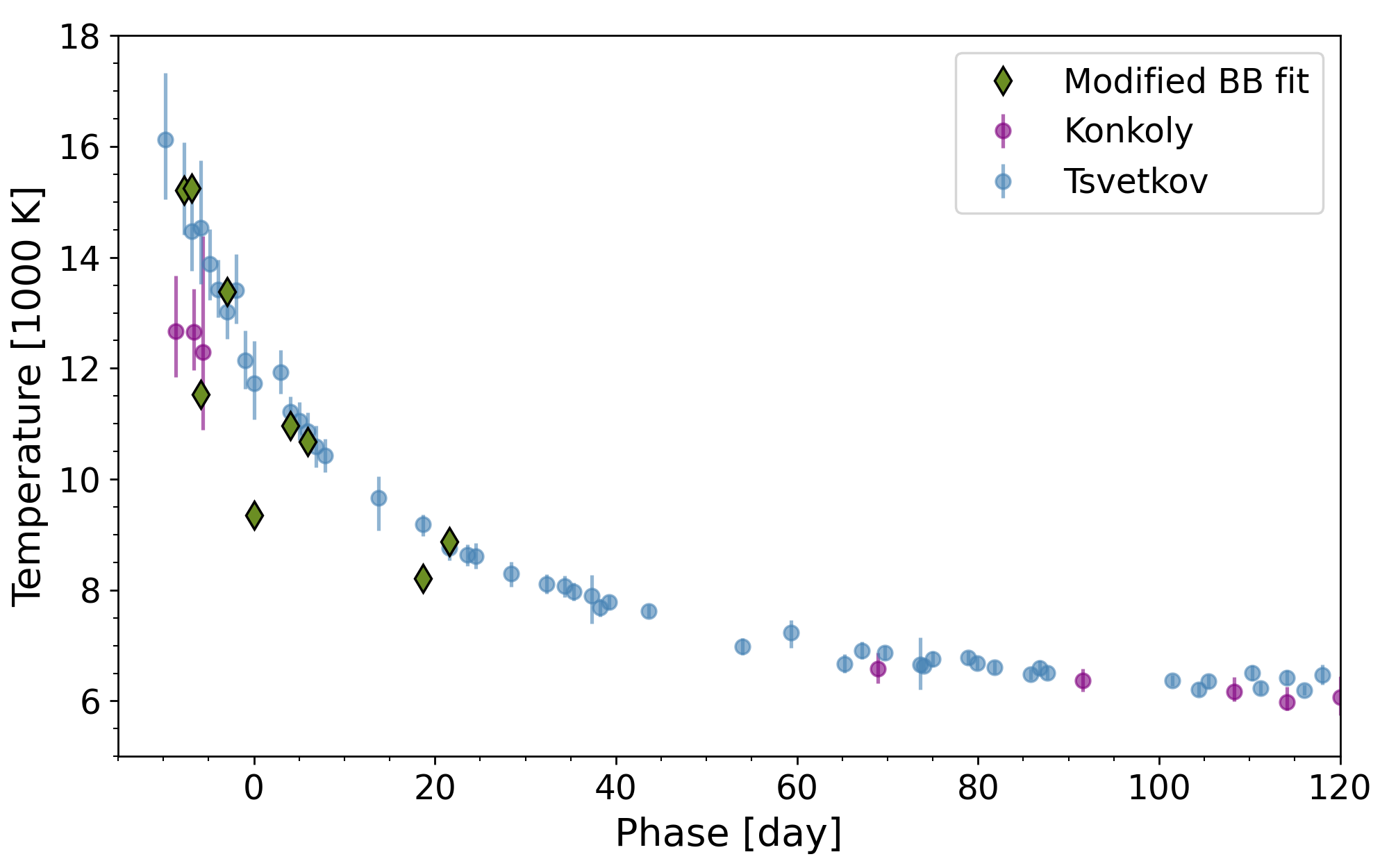}
    \caption{Modified blackbody fits to the early SEDs where the optical data can be extended with UV photometry from \textit{Swift}. \textbf{Top panel:} fits combining the Planck function with a power-law suppression component above a threshold wavelength, following \citet{Yan-2018}. \textbf{Bottom panel:} temperatures derived from the modified BB fits (olive diamonds) compared to the standard BB fit results.}
    \label{fig:Swift_modifiedBBfit_temperatures}
\end{figure}

However, the resulting \textit{T} values did not prove superior to the simple BB fits (without the UV bands). Constant parameters resulted in poor fits, especially at the early stages, and simultaneous fits encountered convergence issues resulting in a large scatter between points. As Fig.~\ref{fig:Swift_modifiedBBfit_temperatures} shows, in the latter case some fits agree well with the BB temperatures, but other instances converged to significantly lower values. Discrepancies between the modified BB fit and the SED points were most noticeable between the \textit{U} and \textit{Swift} UVW1 filters, where the model cannot fully replicate the step-like UV absorption UV passband. We thus did not investigate these fits any further.

\begin{deluxetable}{c c c c c}
\tablecaption{Best-fit parameters of the modified BB model applied to epochs with simultaneous optical and \textit{Swift} UV photometry. The model includes a power-law suppression term below $\lambda_0$ to account for UV line blanketing (see Eq.~\ref{eq:modified_BB}). Lines in italics indicate values that fall significantly below of the simple BB fit values.}
\label{tab:Swift_modifiedBBfit_params}
\tablehead{
\colhead{MJD (d)} & \colhead{$t_p$ phase (d)} & \colhead{T (kK)} & \colhead{$\beta$} & \colhead{$\lambda_0$ (\AA)} }
\startdata
        58819.7 & $-9.8$ & 20.00 & 2.67 & 4473.7 \\
        58821.8 & $-7.7$ & 15.21 & 2.17 & 4380.4 \\
        58822.7 & $-6.9$ & 15.25 & 3.23 & 3682.0 \\
        \textit{58823.6} & \textit{--5.9} & \textit{11.53} & \textit{1.31} & \textit{4141.9} \\
        58826.6 & $-2.9$ & 13.39 & 3.15 & 3695.1 \\
        \textit{58829.6} &  \textit{0.0} &  \textit{9.35} & \textit{1.14} & \textit{3641.9} \\
        58833.7 &  4.0 & 10.96 & 2.45 & 3913.9 \\
        58835.7 &  5.9 & 10.68 & 2.37 & 4000.0 \\
        \textit{58848.7} & \textit{18.7} &  \textit{8.21} & \textit{2.10} & \textit{3686.5} \\
        58851.6 & 21.6 &  8.88 & 3.05 & 3786.3
\enddata
\end{deluxetable}

\subsection{Stefan-Boltzmann radius}\label{subsec:radius}

We estimated the photospheric radius from the Stefan-Boltzmann law: $R = 4\pi \sigma T_{\rm eff}^4/L_{\rm bol}$, using $L_{\rm bol}$ and $T_{\rm eff}$ values derived from the SED fits. This calculation assumes that a spherical photosphere may move within the CSM and it is not fixed to the outside of a non-moving optically dense CSM layer (relative to the velocity of the forward shock). This is a considerable simplification as the structure of the CSM and thus the photosphere is likely to be more complex. However, in lieu of spectroscopic or spectropolarimetric observations of SN\,2019vxm accessible to us, here we treat the photometric observations as an average pseudo-photosphere representation of the supernova. 

In classical SN atmospheres the continuum emission is often described as a diluted BB, where strong electron scattering dominates the opacity over absorption. This means that photons may escape the outer part of the ejecta without thermalizing, and in effect we observed the thermal properties of a layer below the photosphere. In such cases the continuum retains a blackbody-shape, but the Stefan-Boltzmann radius is reduced from the physical radius of the outer edge of the ejecta. However, in strongly interacting Type~IIn SNe the photosphere is expected to be within or close to the dense, optically thick CSM (at least initially), which is directly heated by the forward shock. As such, the thermalization region and the photosphere can be expected to be spatially close.

We show the result in Fig.~\ref{fig:luminosity_temperature_and_radius}. The calculated radius starts from $\approx15000\,R_\odot$, and expands to nearly $50000\,R_\odot$ by about phase 70--100~d. The radius estimates are summarized in Table~\ref{tab:LTR} in Appendix \ref{sec:Tables}. This expansion appears to be almost synchronous with the decrease in $T_{\rm eff}$. From there on, we observe a nearly linear decrease over time, with the Stefan-Boltzmann radius following the decrease in luminosity, as $T_{\rm eff}$ stays mostly constant. By phase 550--600 d, the estimated size shrinks back to and below the starting radius. 

We estimated the velocity of the photosphere by smoothing the radii calculated from the Tsvetkov data with LOWESS regression and then calculating the finite difference derivatives (tabulated values are listed in Section~\ref{sec:Tables}, Table~\ref{tab:LTR} in the Appendix). As Fig.~\ref{fig:velocity} shows, at the time of the first multicolor observations, 17 days after $t_0$, the photosphere was expanding with a velocity of $6700\pm500$~km\,s$^{-1}$. The photosphere then started to slow down significantly after about 35~d. The radius reached its maximum extent between 100--110~d. From there on the photosphere started to shrink, reaching a velocity of $-920\pm200$~km\,s$^{-1}$, but gradually slowing down afterwards. By the end of the multicolor data, between phase 500--600~d, the velocity decreased to $-430\pm70$~km\,s$^{-1}$.

\begin{figure}
    \centering
    \includegraphics[width=0.99\columnwidth]{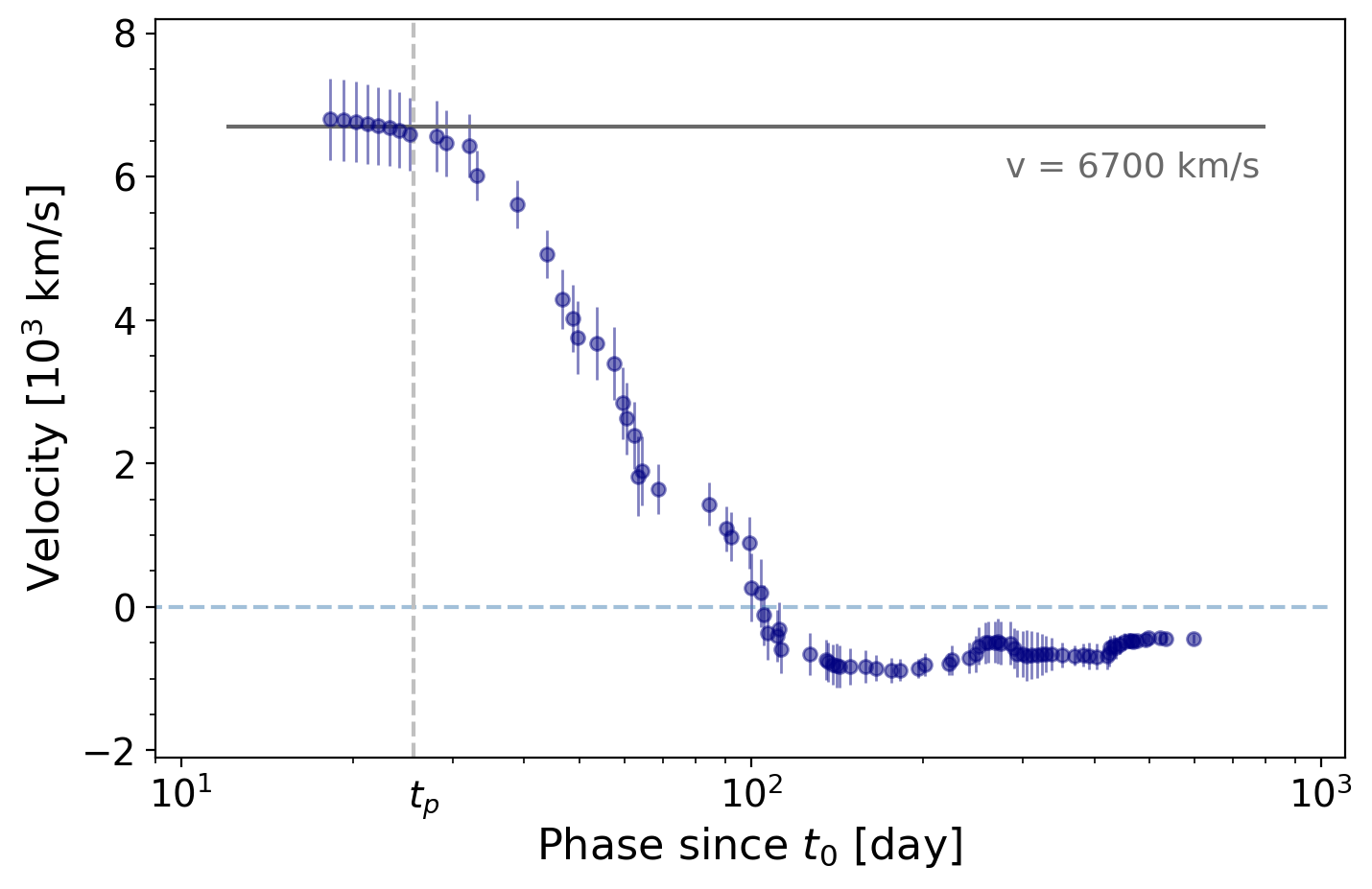}    
    \caption{The velocity of the photosphere of SN\,2019vxm, calculated from the Stefan-Boltzmann radii of the Tsvetkov data set. The vertical dashed line marks $t_p$. The solid gray line shows the average of the velocities before $t_p$, which we discuss further in Section~\ref{subsec:outer_csm}.
    \label{fig:velocity}}
\end{figure}

\subsection{Dust emission}\label{subsec:dust_emission}

To investigate the dust properties in SN\,2019vxm, we analyzed the full set of available WISE $W1$ and $W2$ measurements, spanning phases from approximately $-17$ to $1550$ days. Optical multicolor photometry is available over a more limited interval ($\sim -10$ to $620$ days). Only three WISE epochs fall within the time range covered by this optical data, allowing us to construct optical--IR SEDs for detailed SED modeling without extrapolation. Below we describe the analysis of these observations. The earliest WISE data at $-17$\,d can only be compared to the ASAS--SN and TESS observations which required additional processing: we describe this  Section~\ref{subsec:Asassn_data} and show the results in Figs.~\ref{fig:asassn} and \ref{fig:early_SED} in the Appendix.

\subsubsection{Two-component SEDs}

For each of these three epochs, we constructed SEDs by smoothing the optical photometry using LOWESS regression and interpolating the smoothed light curves to the WISE epochs. We then fitted a two-component BB model using again an MCMC approach with \texttt{emcee}, consisting of a hot component representing the optical photospheric emission and a cool component associated with thermal dust emission. The hot-component temperature was allowed to vary between 5500-10,000~K (initial value 6000~K), while the cool component was constrained to 500-2500~K (initial value 1500~K). In addition to the two temperatures, two scaling factors and an additional noise term were included as free parameters. We used 48 walkers and 16\,000 steps for each chain, discarding the first 10\,000 steps as burn-in. The best-fit values correspond to the maximum of the posterior probability distribution,  while the quoted uncertainties represent the 16th and 84th percentiles of the marginalized posterior distributions.

\begin{figure}
    \centering
    \includegraphics[width=0.99\columnwidth]{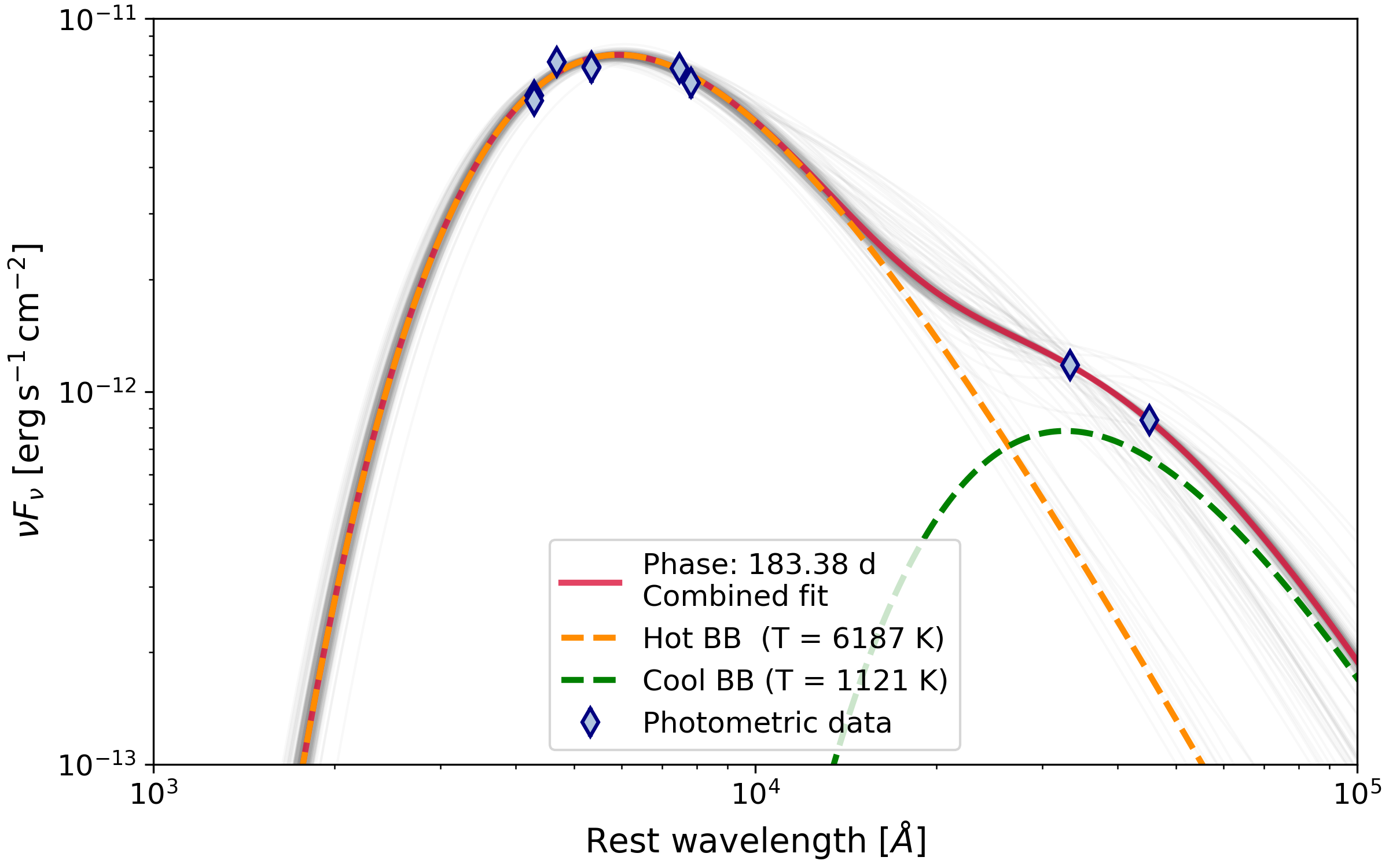}
    \includegraphics[width=0.99\columnwidth]{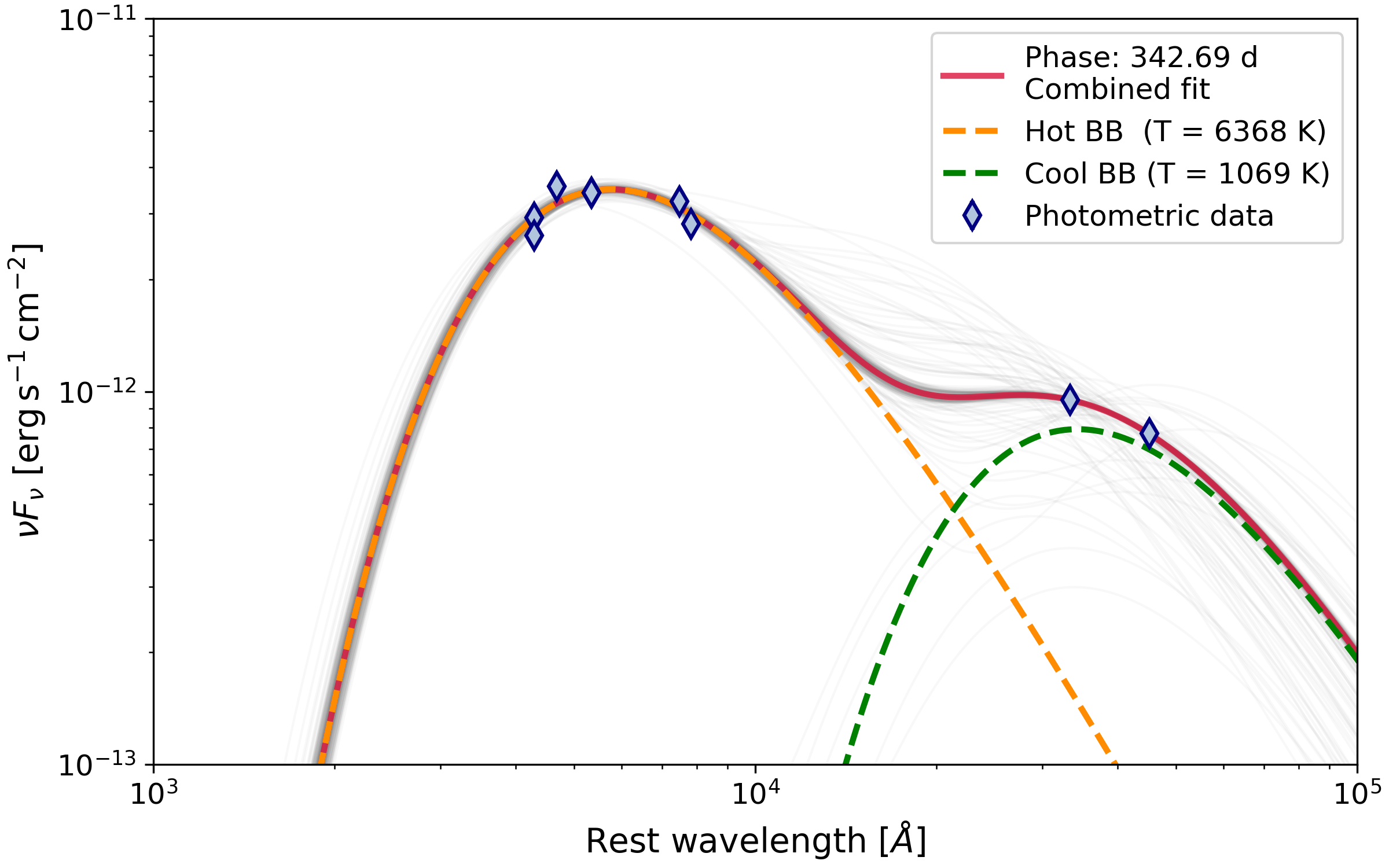}
    \includegraphics[width=0.99\columnwidth]{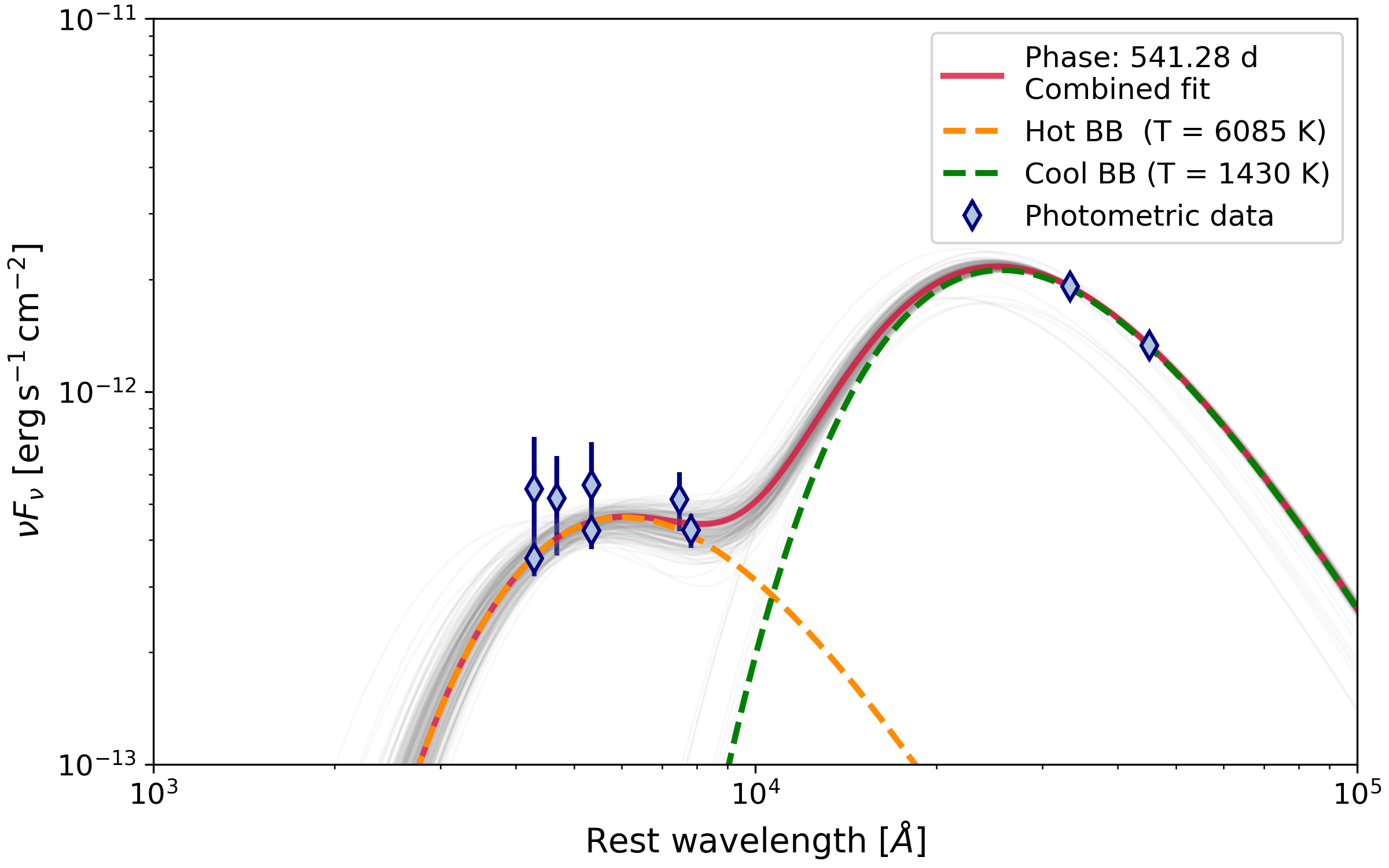}
    \caption{Two-component BB fits to the optical--IR spectral energy distributions of SN\,2019vxm at the three contemporaneous Konkoly--Tsvetkov--WISE epochs. Blue diamonds indicate the observed photometric data points. The $r$ and $R$ bands were excluded as they are affected by strong H$\alpha$ emission. The faint gray curves show 200 fits randomly drawn from the posterior distribution. The dashed orange and green curves represents the hot optical and cool IR components, respectively, while the solid red curve shows the sum of the two components.}
    \label{fig:double_BB_fit}
\end{figure}

\begin{figure}
    \centering
    \includegraphics[width=0.99\columnwidth]{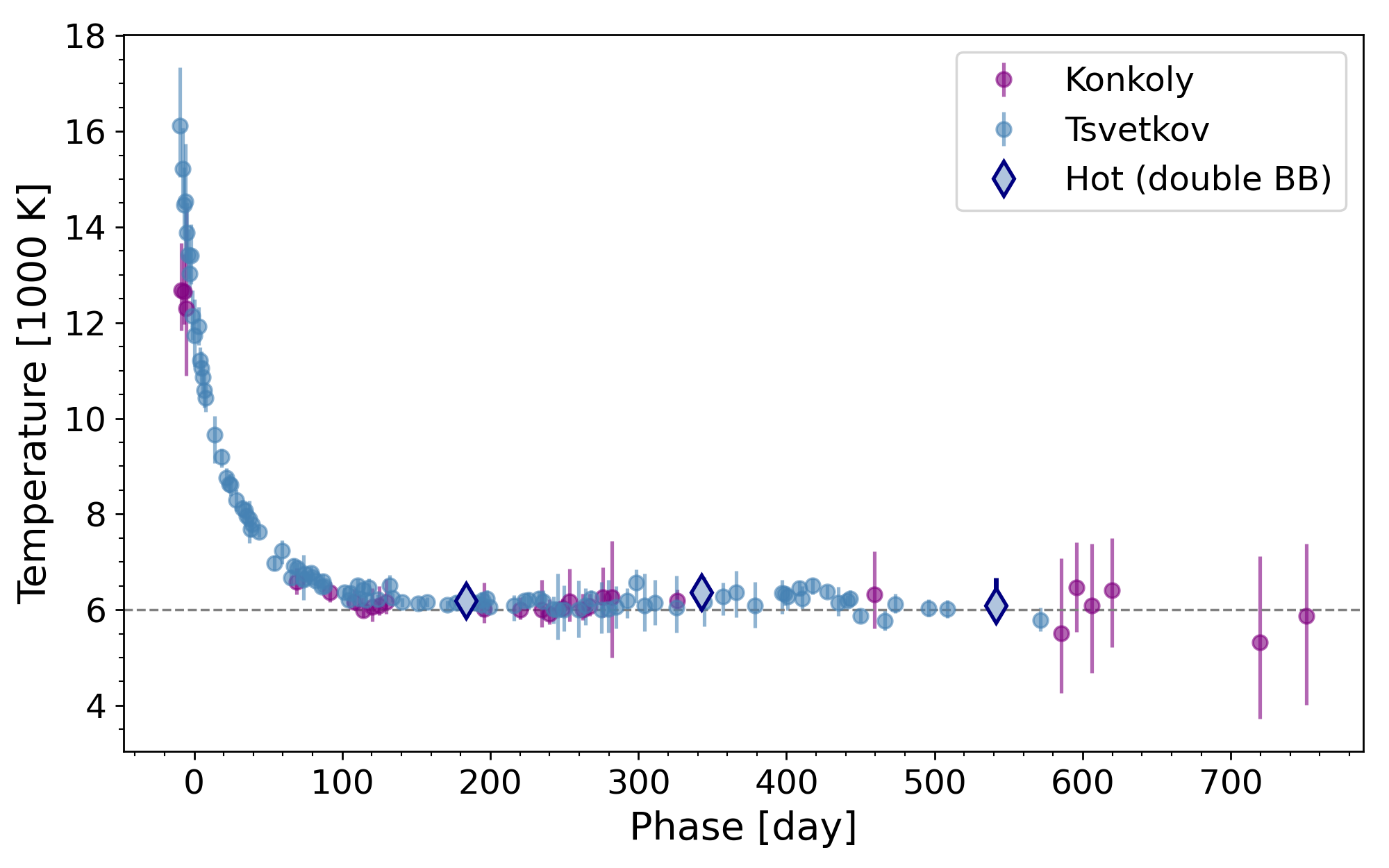}
    \includegraphics[width=0.99\columnwidth]{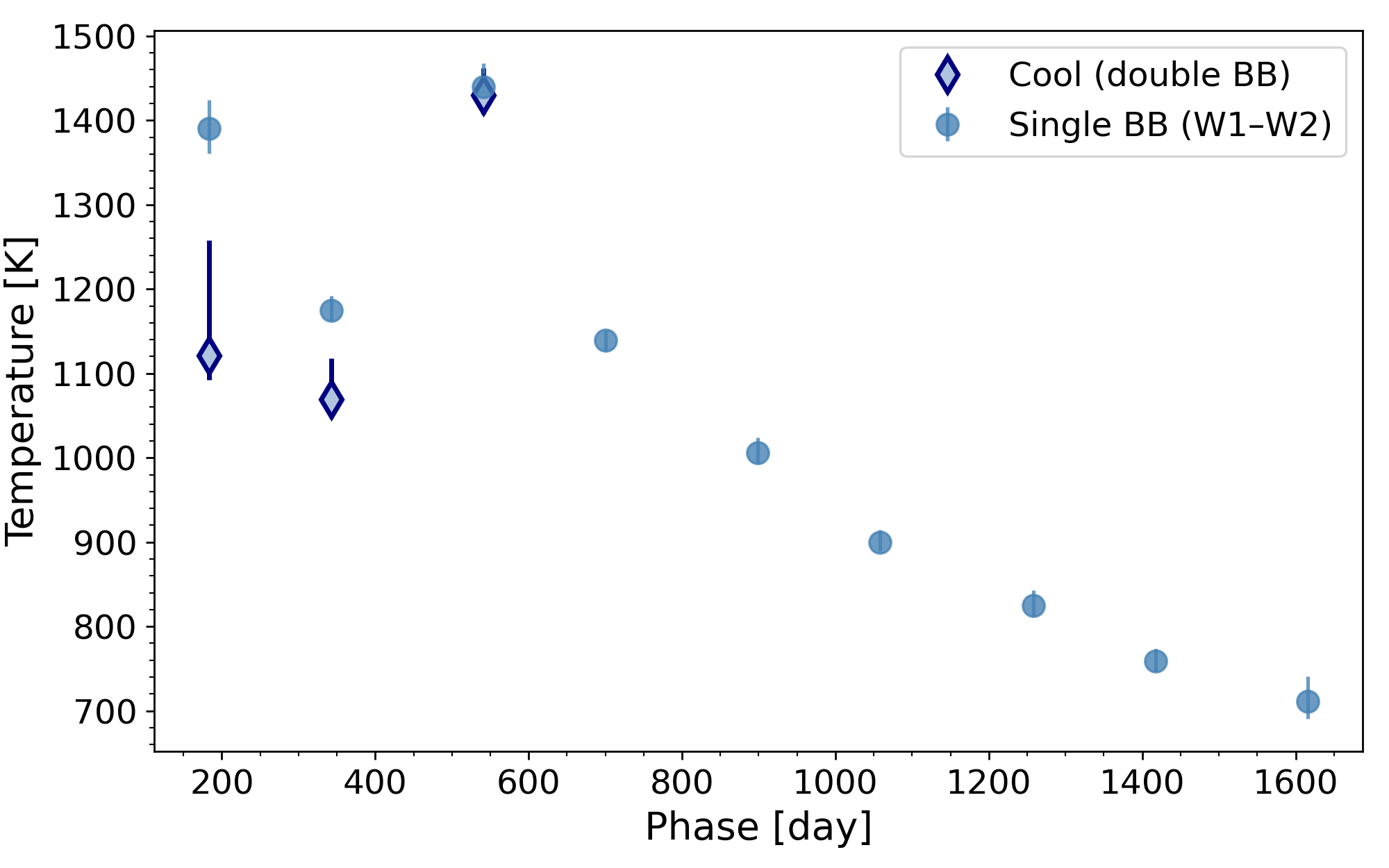}
    \caption{Temperature evolution of the hot and cool BB components inferred from the SEDs of SN\,2019vxm.
    \textbf{Top panel:} purple and blue circles show the temperatures from Fig.~\ref{fig:luminosity_temperature_and_radius}, based on the UV-optical SEDs, while blue diamonds indicate the hot-component temperatures obtained from the two-component optical--IR BB fits. The dashed horizontal line marks 6000~K. \textbf{Bottom panel:} blue circles show the temperatures obtained from single-blackbody fits to the \textit{WISE} $W1$ and $W2$ fluxes, while blue diamonds indicate the cool-component temperatures from the two-component optical--IR fits.}
    \label{fig:temperature_double_BB_simple_BB_WISE}
\end{figure}

The fits are shown in Fig.~\ref{fig:double_BB_fit}, and the obtained temperatures are listed in Table~\ref{tab:double_bb_temperatures}. The values for the hot component agree well with the previous temperature fits based on the UV-optical data alone, as shown in the upper panel of Fig.~\ref{fig:temperature_double_BB_simple_BB_WISE}, with the photosphere aligning with the H recombination zone. We observe an inversion of the flux contributions with the dust emission exceeding the optical by the time of the third contemporaneous WISE measurement, at phase 541.3\,d after $t_p$.  

\subsubsection{IR-only SEDs and temperatures}

To estimate the dust temperature evolution over the full time span ($\sim 1600$ days), we also MCMC-fitted a single BB model to the IR data at every WISE visit within the same 500-2500~K temperature bounds. The fits were performed with 32 walkers and 12\,000 steps, discarding the first 6000 steps as burn-in. The obtained temperature values are listed in Table~\ref{tab:only_WISE_bb_temperature}, and a few example plots of the fits are included in the Appendix in Fig.~\ref{fig:simple_BB_fit_WISE} in Section~\ref{subsec:WISE_cold_db_BBfit}. Although WISE only covered two passbands, the relative changes in the 3.4 and 4.6 $\mu$m fluxes provide good diagnostics on the approximate dust temperature, within reasonable limits. The IR fits to the three occasions contemporaneous with optical fluxes agree with the cool component derived from the double BB modeling within the uncertainties, as we show in the lower panel of Fig.~\ref{fig:temperature_double_BB_simple_BB_WISE}. 

The dust is initially hot at the early phase of the explosion, and the temperature starts to drop from $\sim$1700--2000\,K to $\approx$1200\,K over the course of one year (until 343\,d). We then observe both a rebrightening in the fluxes, as shown in Fig.~\ref{fig:WISE_lc}, and an increase in temperature up to 1440\,K at the next data point, one and a half year (541\,d) after shock breakout. From there the emission declines gradually to about 700-800~K at 1500--1600 days (see Fig.~\ref{fig:temperature_double_BB_simple_BB_WISE}). Similar rebrightening events were observed in other IIn supernovae, as well. SN\,2015da brightened and heated up in IR around 4--500\,d, while SN\,2014ab and SN\,2017hcc both showed brightening between 3--400 d, based on WISE and \textit{Spitzer} space telescope observations \citep{Moran_2023A&A...669A..51M,Moriya-2020A&A...641A.148M,Tartaglia_2020A&A...635A..39T_SN_2015da}. This is not entirely universal behavior, however: another Type IIn event, KISS15b, displayed a slow rise for 700 days, with no clear signs of a dip or rebrightening \citep{Kokubo-2019ApJ...872..135K}.

\begin{deluxetable}{c|ccc|ccc}
\tablecaption{Double BB fit: hot and cold BB temperatures and their 16th and 84th percentile uncertainty edges, in Kelvins.}
    \label{tab:double_bb_temperatures}
\tablehead{ \colhead{Phase} &
        \colhead{$T^{\mathrm{hot}}_{16}$} & \colhead{$T^{\mathrm{hot}}_{\mathrm{best}}$} & \colhead{$T^{\mathrm{hot}}_{84}$} &
        \colhead{$T^{\mathrm{cold}}_{16}$} & \colhead{$T^{\mathrm{cold}}_{\mathrm{best}}$} & \colhead{$T^{\mathrm{cold}}_{84}$}   }
\startdata
        183.4 & 6079 & 6188 & 6359 & 1092 & 1121 & 1258 \\
        342.7 & 6248 & 6368 & 6530 & 1051 & 1069 & 1118 \\
        541.3 & 5827 & 6085 & 6674 & 1408 & 1430 & 1462 \\
\enddata
\end{deluxetable}

\begin{deluxetable*}{c|cccccccc}
\tablecaption{Simple BB fit results and derived luminosity and radius values for the WISE data only. For temperatures luminosities we list the 16th and 84th percentile uncertainty edges. }
    \label{tab:only_WISE_bb_temperature}
\tablehead{ \colhead{Phase} &  \colhead{$T_{16}$} &  \colhead{$T_{\mathrm{best}}$} &  \colhead{$T_{84}$} & \colhead{$L_{\rm bol,16}$} & \colhead{$L_{\rm bol,best}$ } & \colhead{$L_{\rm bol,84}$ }  &
\colhead{$R$} & \colhead{$\sigma R$} \\ 
 \colhead{(day)} & \colhead{(K)} & \colhead{(K)} &  \colhead{(K)} & \colhead{($10^{42}$ erg\,s$^{-1}$ )} & \colhead{($10^{42}$ erg\,s$^{-1}$ )} & \colhead{($10^{42}$ erg\,s$^{-1}$ )}  &
\colhead{($10^{15}$ cm )} & \colhead{($10^{15}$ cm )} }
\startdata
 183.4 & 1361 & 1390 & 1424 & 1.112 & 1.293 & 1.485 & 22.05 & 1.23 \\
 342.7 & 1162 & 1175 & 1192 & 0.860 & 0.952 & 1.051 & 26.48 & 1.14 \\
 541.3 & 1415 & 1440 & 1468 & 1.911 & 2.167 & 2.435 & 26.60 & 1.28 \\
 700.7 & 1126 & 1139 & 1151 & 1.384 & 1.527 & 1.667 & 35.68 & 1.46 \\
 898.4 &  993 & 1006 & 1024 & 0.896 & 1.014 & 1.142 & 37.27 & 1.94 \\
1057.9 &  889 &  900 &  915 & 0.682 & 0.772 & 0.870 & 40.65 & 2.17 \\
1258.3 &  812 &  825 &  843 & 0.451 & 0.531 & 0.617 & 40.11 & 2.76 \\
1417.9 &  747 &  759 &  774 & 0.377 & 0.446 & 0.518 & 43.43 & 3.07 \\
1616.1 &  691 &  711 &  741 & 0.234 & 0.331 & 0.436 & 42.62 & 5.78 
\enddata
\end{deluxetable*}

\subsubsection{Dust emission radius}

We calculated the average radii from where the near-IR emission originates with the Stefan-Boltzmann law using the WISE fluxes and BB fits we presented in the previous section. 
Uncertainties were estimated using Monte Carlo sampling of the temperature ranges and standard error propagation. We find that this colder emission first appears at $3\cdot10^5\, R_\odot$, or $R = (6.8 \pm 1.8)\cdot10^{15}$\,cm and then gradually expands to $6\cdot10^5\, R_\odot$, or $R = (4.3 \pm 0.6)\cdot 10^{16}$\,cm. A jump from 3.8 to $5.1\cdot 10^5\, R_\odot$ happens at 541\,d, but afterwards the expansion of the IR radius stabilizes around $6\cdot10^5\, R_\odot$, or $R = (4.1 \pm 0.2)\cdot 10^{16}$\,cm from 1058\,d, with some indications of a possible slow expansion. These values are in the same order of magnitude as the dust emission radii found in several other Type~IIn SNe, between $\sim 0.6 - 6\cdot 10^{16}$ cm \citep{Sarangi-2018,Moriya-2020A&A...641A.148M,Taddia-2020,Tartaglia_2020A&A...635A..39T_SN_2015da,Moran_2023A&A...669A..51M}. 

\begin{figure}
    \centering
    \includegraphics[width=0.99\columnwidth]{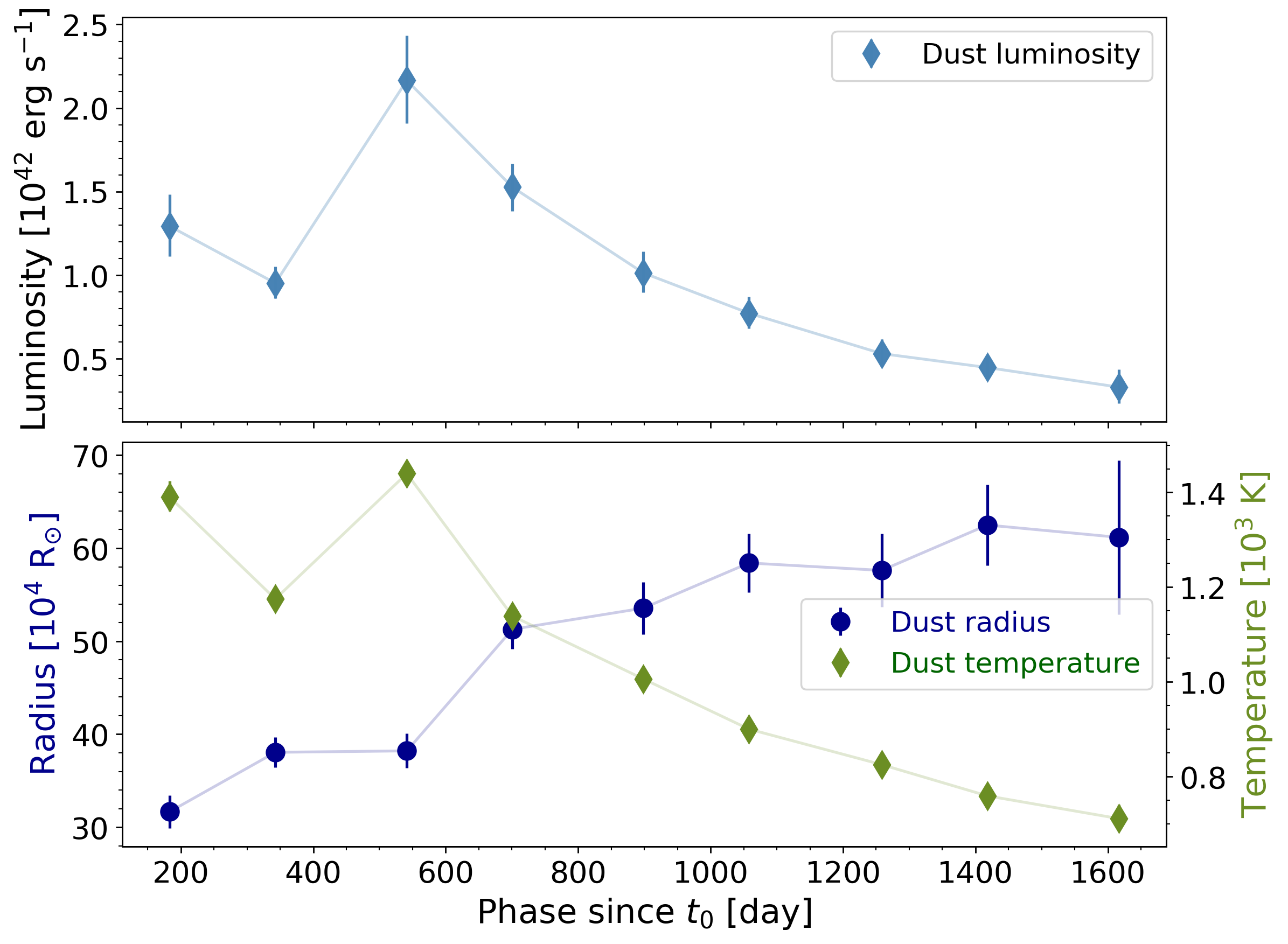}
    \caption{Temporal evolution of the dust properties as a function of phase since $t_0$. The top panel shows the dust luminosity (blue diamonds), while the bottom panel displays the temperature (green diamonds) and radius (blue circles).}
    \label{fig:LRT_WISE_dust}
\end{figure}

The bolometric luminosity of the IR component was then calculated from the derived temperatures and radii using the Stefan–Boltzmann law, with uncertainties obtained through standard error propagation. The measured IR emission reaches a peak luminosity of $L_{\mathrm{max}} = \left( 2.17^{+0.27}_{-0.26} \right) \cdot 10^{42}$ erg\,s$^{-1}$ after the rebrightening, although this is a lower limit of the true peak brightness given the sparseness of the WISE data, and then drops to $L_{\rm end} = \left( 3.31^{+0.11}_{-0.97} \right) \cdot10^{41}$ erg\,s$^{-1}$ by the end of the WISE coverage. We display the luminosity, temperature and radius evolution of the dust in Fig.~\ref{fig:LRT_WISE_dust}, and list the numerical values in Table~\ref{tab:only_WISE_bb_temperature}.

\section{Models} \label{sec:models}

In Type~IIn supernovae the explosion does not expand into a near-vacuum, but, instead, it runs into a previously ejected, hydrogen-rich CSM. The large velocity contrast between the fast SN ejecta ($v_{\rm ej}\sim10^{3} - 10^{4}$~km~s$^{-1}$) and the much slower CSM ($v_{\rm w}\lesssim 10^{2}$~km~s$^{-1}$) naturally produces a double-shock structure: a forward shock (FS) propagates outward into the CSM, heating and accelerating it, while a reverse shock (RS) propagates back into the ejecta \citep{Chevalier_1982ApJ...258..790C, Chevalier_1994ApJ...420..268C, Chevalier_2003}.

The interaction region is often described in terms of four zones, starting from outside: (i) unshocked CSM, (ii) CSM shocked by the forward shock, (iii) ejecta shocked by the reverse shock, and (iv) unshocked freely expanding ejecta \citep{Chevalier_1994ApJ...420..268C, 2024ApJ...972..140K}.

Both the FS and the RS heats up and accelerates the medium in which they are propagating. 
If the CSM density is sufficiently high, both shocks become radiative. In this case a contact discontinuity forms between the two shocks, and a cold dense shell (CDS) develops there \citep{Chevalier_2003}. The CDS can absorb some or all of the high-energy radiation from the RS and thermalize it, producing a significant excess emission in the UV-optical regime. 

In strongly interacting events, CSM-powered emission can remain dominant for months to years. Classical photospheric P~Cygni features may be weak or absent in the spectra, if the shocked regions and/or the CDS remain optically thick during the interacting phase. 

To quantify the physical parameters of the interaction and the shock structure in SN\,2019vxm, we consider two idealized frameworks: an optically thin CSM model (Model A) and an optically thick CSM model (Model B). These models differ in their treatment of the photosphere and radiative transfer. In Model A, the photosphere follows the radius of the expanding shock and the observed light traces the luminosity from the RS reprocessed by the CDS. In Model B, the photosphere resides within the outer part of CSM, and it is static or moves very slowly. The emergent luminosity is governed by the diffusion of the shock power through the optically thick part of the CSM \citep{Chatzopoulos_2012, Chatzopoulos_2013}. 

In both cases, the interaction between the homologously expanding SN ejecta and the CSM produces the double-shock structure described above: a forward shock, a reverse shock. The density structure of both component is approximated by power-law profiles 

\begin{equation} 
\label{eq:rho_ej rho_CSM}
    \rho_{\rm ej} ~=~ g^n t^{n-3} r^{-n}  ~~; ~~ \rho_{\rm CSM} = q\,r^{-s},
\end{equation}
following the self-similar solution of \citet{Chevalier_1982ApJ...258..790C, Chevalier_1994ApJ...420..268C}. Here $n$ and $s$ correspond to the power-law indices of the ejecta and CSM density distributions, respectively.
When $s=0$, this solution describes a constant-density environment, whereas $s=2$ corresponds to a stellar wind having constant velocity. In this case the CSM density scales as $r^{-2}$ and the normalization constant is $q = \dot{M} / 4 \pi v_w$, where $\dot{M}$ is the mass-loss rate and $v_w$ is the wind velocity. The wind parameter $\dot{M}/v_w$ is an important quantity characterizing the shock propagation and radiation. 

The power-law index of the outer density profile of the ejecta is usually assumed 
to be in between $ n = 7$ and $12$ \citep[e.g.][]{Smith_2007ApJ...666.1116S_SN_2006gy, Ofek_2014ApJ...781...42O}. The self-similar solution is valid only for $n > 5$ and $s < 3$.  For $s>3$ the shock accelerates, while for $n < 5$ the CDS is decoupling from the FS \citep{Chevalier_1982ApJ...258..790C}.

The temporal evolution of the radius of the shock 
can be described by the following power-law \citep{Ofek_2014ApJ...781...42O}:
\begin{equation}
\label{R_s}
R_s(t) = R_0 \cdot \left ( \frac{t}{t_0} \right )^{\frac{n-3}{n-s}},     
\end{equation}
where $t_0$ is a reference moment (usually the start of the interaction) and $R_0$ is the shock radius at $t_0$. It can be shown that $R_0$ depends on the ejecta and CSM density parameters as $R_0 \propto ( g^n / q )^{1/(n-s)}$ \citep{Chevalier_2003, Ofek_2014ApJ...781...42O}.

\subsection{Optical depth and the n parameter}\label{subsec:opt_depth n_param}

The (inner) ejecta density exponent $n$ plays a central role in analytic models of supernova-CSM interaction, as it governs both the evolution of the shock radius and the temporal decline rate of the interaction-powered luminosity. Its value is usually set between 7--12, with higher values representing a more extended progenitor \citep{Chevalier_1982ApJ...258..790C}.
Lower values ($n \sim 7$) are generally associated with more compact progenitors or radiative envelopes (e.g., LBV or Wolf-Rayet stars), while higher values ($n \sim 10$--12) correspond to extended, convective envelopes typical of red supergiants. 
Based on model fits to several SNe IIn, \citet{Ransome2025ApJ...987...13R} found a median value of $n=9.44\pm1.80$. 

In the optically thin, self-similar interaction limit for a steady wind ($s=2$), the reverse-shock luminosity follows a power law in time,
\begin{equation}
L(t) \propto t^{-3/(n-2)},
\label{eq:L_n_param}
\end{equation}
\noindent implying that in log-log space the bolometric luminosity should follow a straight line, whose slope directly yields $n$. A persistent power-law LC implies a well-defined ejecta density exponent. This phase lasts until the amount of shocked, swept-up CSM matches the mass of the ejecta, at which point the reverse shock ceases, and the \citet{Chevalier_1982ApJ...258..790C} formalism does not apply anymore. The luminosity begins to decrease faster and the SN reaches the so-called `snow-plow' phase (when assuming a wind-like CSM, \citealt{Svirski-2012}).

We estimated the value of $n$ using two independent approaches.
First, during the early expansion phase we assumed that the photospheric radius is tracing the shock radius. The radius evolution was fitted with a power-law of the form
\begin{equation}\label{eq:R_n_param}
R(t) = a_R \, t^{m_R},
\end{equation}
which becomes linear in log–log space. From the fitted slope $m_R$, the density exponent can be expressed as
\begin{equation}\label{eq:n_param from radius}
n = \frac{3 - 2m_R}{1 - m_R}.
\end{equation}
Using this method, we obtain $n \approx 10.2$ for the early phase (see Fig.~\ref{fig:luminosity_temperature_and_radius}).

Second, we derived $n$ directly from the luminosity evolution using sliding-window fits in log–log space. As shown in Fig.~\ref{fig:n_param}, a linear fit was performed to 30 consecutive data points, shifting the window by one data point at each step. The slope of each fit was then converted to $n$ using Eq.~\ref{eq:L_n_param}. This method reveals that $n$ evolves almost continuously with phase rather than settling to a constant value, and only short segments can be approximated by linear trends. The fit even crosses the critical threshold of $n=5$, which marks the validity limit of the self-similar solution. Moreover, the inferred values depend systematically on the adopted reference time: using $t_0$ or $t_p$ shifts the quasi-plateau region of $n$ toward $\sim 4.5$ or $\sim 5.0$, respectively.

The changes in $n$ are further illustrated in the lower panel of Fig.~\ref{fig:n_param}, where the bolometric light curve is plotted in log-log space, where fitted three segments with different $n$ parameters. This {plot again shows that $n$ is not constant, but evolves strongly with time. Overall, the inferred values range from $n \sim 10$-12 at early phases to $n \lesssim 3$ during the late decline.

This behavior demonstrates that the bolometric light curve of SN~2019vxm cannot be described by a single power-law interaction model over any extended phase. Instead, the evolution reflects changing physical conditions in the shock interaction. It also suggests that an optically thin CSM interaction scenario, at least in its analytic form, cannot provide a self-consistent description of SN\,2019vxm over the entirety of the light curve. 

The steep decline from $\sim350$\,d at $n \approx 2.5$ is indicative of a transition in the shock interaction. In the semi-analytical 1D framework this could signal the onset of the `snow-plow' phase, when in a wind-like CSM the cumulative mass of the shocked CSM reaches the ejecta mass, at which point the reverse shock ceases and the forward shock slows down significantly. Alternatively, in a more shell-like CSM, it could signal that the forward shock reached the edge of the dense CSM and entered a region where the density falls off rapidly \citep{Svirski-2012,Ofek_2014ApJ...781...42O}. 
However, if the CSM structure is more complex, as it was the case for, e.g., SN 2010jl \citep{Moriya-2014,Ofek_2014ApJ...781...42O}, both the modeling and the subsequent interpretation may not be straightforward. We discuss this transition in relation with the IR emission further in Section~\ref{sec:results}.

\begin{figure}[]
\includegraphics[width=0.98\columnwidth]{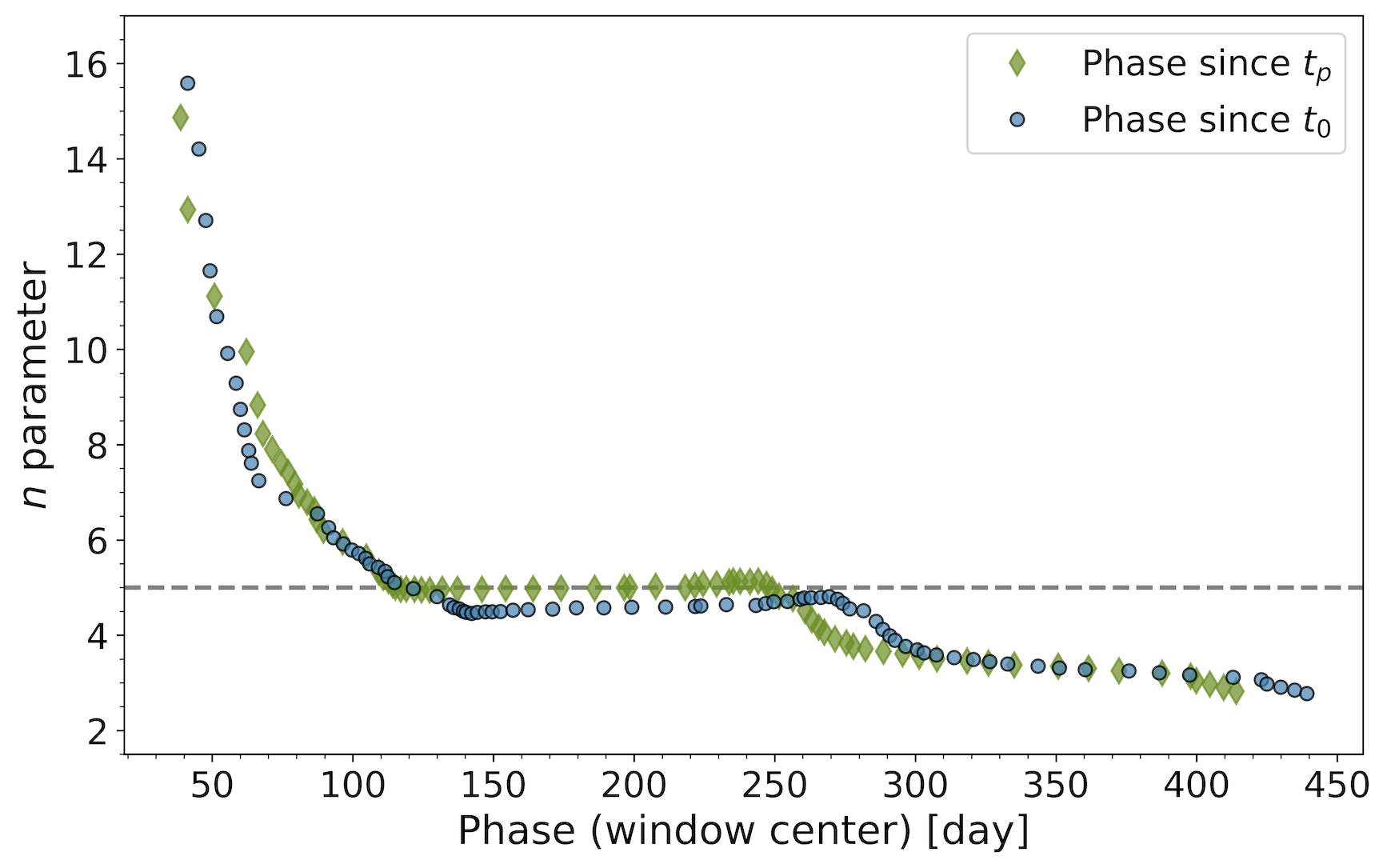}
\includegraphics[width=0.98\columnwidth]{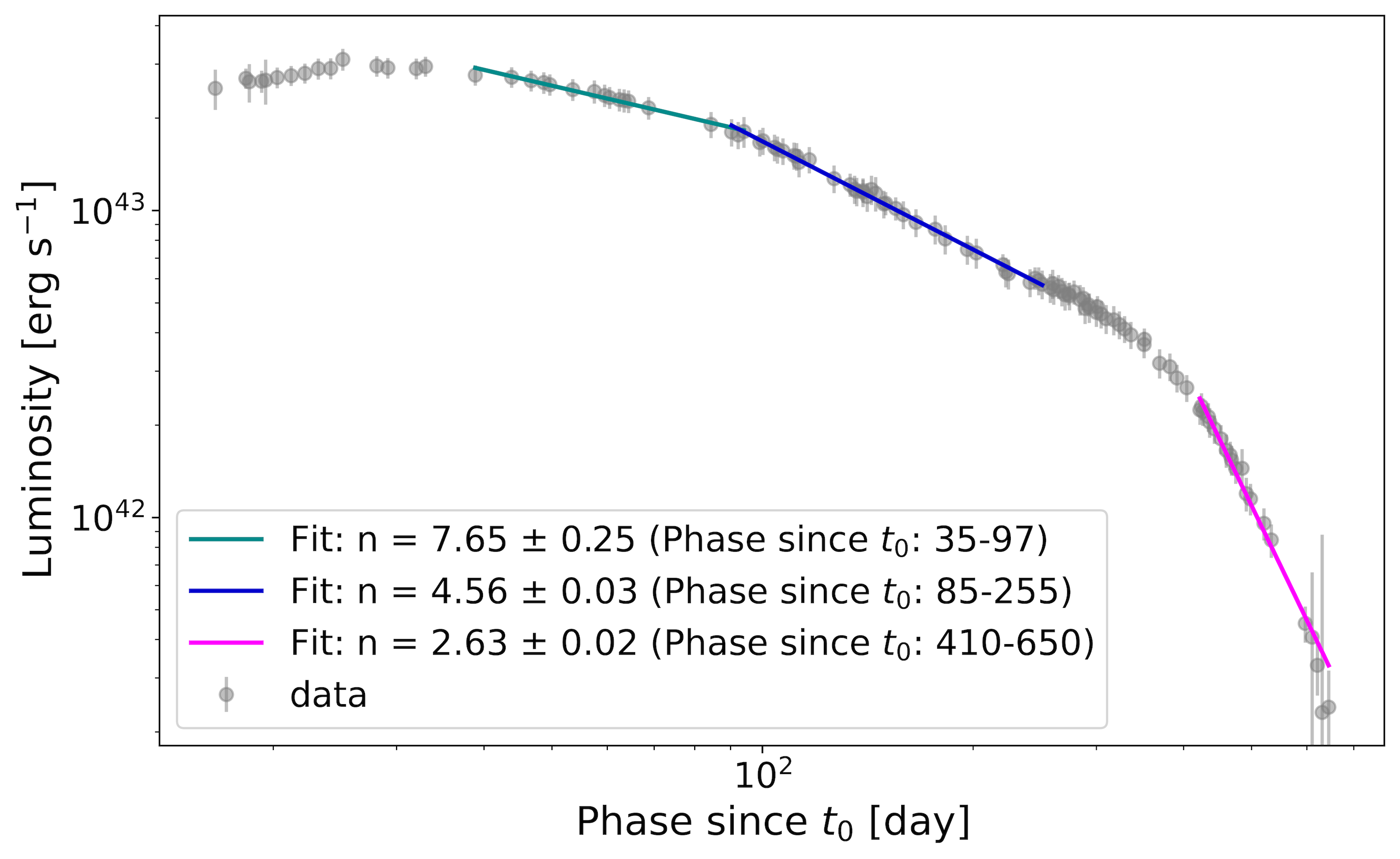}
\caption{Temporal evolution of the power-law index $n$ describing the luminosity decline ($L \propto t^{-n}$) in SN\,2019vxm. \textbf{Top panel:} instantaneous $n$ values, based on a sliding-window linear fit to the logarithmic bolometric light curve. Blue diamonds show the results when the phase is measured relative to $t_p$, while orange circles correspond to phases measured relative to $t_0$. The dashed horizontal line marks $n = 5$, which represents the lower-limit value in the self-similar interaction model of \citet{Chevalier_1982ApJ...258..790C}. \textbf{Bottom panel:} The evolution of $L_{\rm bol}$ for SN\,2019vxm in log–log space as a function of phase since $t_0$. Colored lines illustrate power-law fits to different sections of the LC, demonstrating the temporal evolution of the decline rate. \label{fig:n_param}}
\end{figure}

\subsection{Model A: Optically thin CSM} \label{subsec:opt_thin_CSM_model}

In this model we adopt the approximation that the emitted radiation is mainly produced by the reverse shock. This assumption is justified if we assume that although the temperature behind the FS is very high, it propagates through a low-density CSM. Consequently, the luminosity arising from the deceleration of the FS remains relatively weak. In this scenario, the RS produces a much higher luminosity, which therefore dominates the observed emission.

Within the optically thin framework the luminosity from the RS can be expressed as 
\begin{equation}
L_{\rm rev} = 2 \pi R_s^2 \rho_{\rm ej} v_s^3 = 1.6\times10^{40} \frac{(n-3)(n-4)}{(n-2)^2} \frac{\dot{M_6}}{v_{w1}} v_{s4}^3, 
\label{Eq:L_rev}
\end{equation}
where $\dot{M_6}$ is the mass-loss rate in $10^{-6}$~M$_\odot$yr$^{-1}$ units, $v_{w1}$ is the wind velocity in 10 km~s$^{-1}$ and $v_{s4}$ is the shock velocity in $10^4$ km~s$^{-1}$ units  
\citep{Nymark_2006A&A...449..171N}. The time-dependent shock velocity is given by 
\begin{equation}
    v_{s4} = 0.1157 \cdot \frac{(n-3)}{(n-2)} \cdot R_{s13} \cdot t_d^{-1/(n-2)},
\label{Eq:vs}
\end{equation}
where $R_{s13}$ is the shock radius at 1 day after the start of interaction in $10^{13}$ cm units. Practically, $R_{s13}$ is close to the progenitor radius if there is no gap between the CSM and the progenitor at the moment of the explosion.

Most of this radiation, however, is absorbed by the CDS, at least during the early phases of interaction. Thus, the observed luminosity (neglecting the absorption by the optically thin CSM above the CDS) is 
\begin{equation}
L_{\rm obs} = L_{\rm rev} \cdot \exp[-N_{\rm CDS} \sigma_\lambda], 
\label{Eq:L_rev_obs}
\end{equation}
where 
\begin{equation}
N_{\rm CDS} = 1.8\times10^{22} \cdot ((n-4)/\mu_A) (\dot{M_6}/v_{w1} (v_{s4} t_d)^{-1})
\end{equation}
is the (time-dependent) column density of the CDS, $t_d$ is the time since the beginning of the interaction in days and $\sigma_\lambda \approx 2.2\times10^{-25} \cdot \lambda^{8/3}$ is the absorption cross-section as a function of wavelength. Since most of the RS luminosity is produced by X-rays, we parametrize the bolometric cross-section as $\sigma \approx 2.2\times10^{-25} \cdot \sigma_0$, where $\sigma_0$ is an adjustable parameter when fitting the data.
Since the CDS is diluting when the shock expands, $N_{\rm CDS}$ is decreasing with time, and the (continuously decreasing) luminosity from the RS becomes more and more observable. 

We fit Equation~\ref{Eq:L_rev_obs} to the bolometric light curve of SN\,2019vxm 
via adjusting $R_{s13}$, $\dot{M}$ and $\sigma_0$. We set $\dot{M} = 0.1$ M$_\odot$~yr$^{-1}$ and $v_w = 10$ km~s$^{-1}$, $\mu_A = 1.29$ (solar composition). After several experiments, we find that $n = 7$ produces the best-fitting light curves during the early phases.  

\begin{figure}
    \centering
    \includegraphics[width=1.0\linewidth]{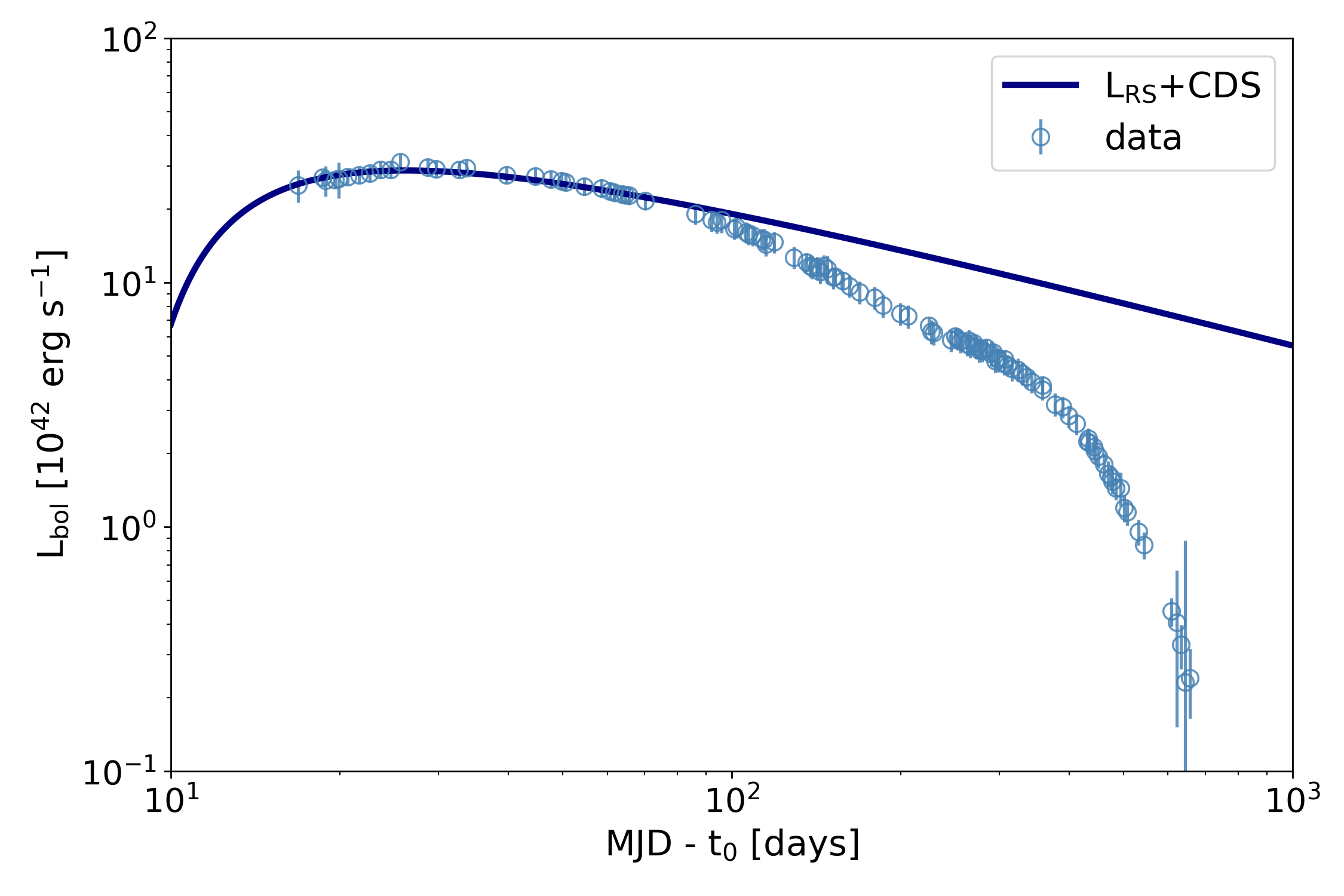}
    \caption{The best-fit light curve from Model A (blue curve) with the bolometric luminosities for SN\,2019vxm (circles). Note that the model LC was shifted horizontally by 7 days after explosion to get the best match between the model and the data.}
    \label{fig:fit_modelA}
\end{figure}

The best-fit LC is plotted together with the data in Figure~\ref{fig:fit_modelA}. The parameters of this model are $R_{s13} = 8.98$, $\dot{M} = 0.083$ M$_\odot$~yr$^{-1}$ and $\sigma_0 = 0.0089$. Note that due to the strong parameter correlation the formal uncertainties of these parameters are very high. 

The best-fit parameters give $v_s \simeq 8314$ km~s$^{-1}$ and $L_{\rm rev} \simeq 3.65 \times 10^{44}$ erg~s$^{-1}$ as the shock velocity and luminosity at $t_d = 1$, which, according to Equation \ref{Eq:L_rev}, predicts an ejecta density of $\rho_{\rm ej} \simeq 6.27 
\times 10^{-12}$ g~cm$^{-3}$ at $r=R_s$. Integrating the assumed $n=7$ density profile from $r=R_s$ back to $r=0.1 R_s$, this corresponds to $M_{\rm ej} \gtrsim 70$ M$_\odot$ as a lower limit for the ejecta mass. This suggests that the progenitor of SN~2019vxm was a very massive star having a huge extended envelope ($R \approx R_s \sim 1300$ R$_\odot$).

Figure~\ref{fig:fit_modelA} shows that Model A can explain the early part of the Lc, roughly up to $t \sim 100$ days. After that the observed bolometric light curve drops below the RS luminosity. The reason for that is not obvious: perhaps the RS gets too deep into the ejecta, or the CSM becomes optically thick due to the accumulation of the swept-up mass. Alternatively, our initial assumption of an optically thin CSM could be false: we explore the optically thick CSM model (Model B) in Section~\ref{subsec:opt_thick_model}.

\subsection{Radioactive decay}\label{subsec:rad_decay}

While the luminosity of Type~IIn SNe is generally dominated by ejecta--CSM interaction, radioactive heating from the dominant decay chain $^{56}$Ni $\rightarrow$ $^{56}$Co $\rightarrow$ $^{56}$Fe may still contribute to the late-time LC. During the explosion a significant amount of radioactive $^{56}$Ni is synthesized in the inner layers of the ejecta. The radioactive energy input is described by the full decay chain, which gives the bolometric luminosity (e.g, Eq.~(2) by \citealt{Bora_2022PASP..134e4201B}) as:

\begin{equation}\label{eq:L_Ni_Co}
    L(t) = \frac{M_{\mathrm{Ni}}}{\mathrm{M}_\odot} \left [ (\epsilon_{\mathrm{Ni}}^{} - \epsilon_{\mathrm{Co}}^{}) e^{-t/\tau_{\mathrm{Ni}}^{}} + \epsilon_{\mathrm{Co}}^{} e^{-t/\tau_{\mathrm{Co}}^{}} \right],
\end{equation}

\noindent where $\epsilon_{\mathrm{Ni}} = 7.9 \cdot 10^{43}$~erg~s$^{-1}$ and $\epsilon_{\mathrm{Co}} = 1.45 \cdot 10^{43}$~erg~s$^{-1}$ are the specific heating rates, and $\tau_{\mathrm{Ni}} = 8.8$ and $\tau_{\mathrm{Co}} = 111.3$~days are the corresponding e-folding timescales.

Since $^{56}$Ni has a short half-life ($t_{1/2, \, \rm Ni}=6.1$~days), most of this isotope has already decayed within the first few weeks after the explosion. The subsequent decay of $^{56}$Co to stable $^{56}$Fe has a longer half-life of $t_{1/2\, \rm Co}=77.2$~days, and therefore dominates the radioactive energy input at later epochs.
In interacting SNe, ejecta--CSM interaction may dominate the luminosity, making the contribution of radioactive decay difficult to isolate. A comparison with the expected $^{56}$Co decay slope provides a diagnostic of radioactive heating. In the Co-dominated regime, the bolometric luminosity can be approximated as:

\begin{equation}\label{eq:L_Co}
    L(t) = a_{\mathrm{Co}} \cdot e^{-\frac{t}{\tau_{\mathrm{Co}}}},
\end{equation}

\noindent where $a_{\mathrm{Co}} = \epsilon_{\mathrm{Co}} M_{\mathrm{Ni}}$.

To examine whether the observed luminosity decline is consistent with radioactive heating, we estimated the local decay timescale of the LC using sliding-window fits in $\ln L$--$t$ space, as shown in Fig.~\ref{fig:ln-lin slope}.
In each step, a linear fit was performed to 10 consecutive data points, from which the logarithmic decline rate $s = d(\ln L)/ dt$ was obtained, the characteristic timescale was computed as $\tau = -1/s$. This analysis shows that the inferred decay timescale approaches the characteristic $^{56}$Co value ($\tau_{\rm Co} = 111.3$ d) in two distinct intervals. In terms of the median phases assigned to the sliding windows, these occur at approximately $\sim20$--120~days and $\sim410$--480~days.

Motivated by this behavior, radioactive decay models were fitted to the bolometric light curve (see Fig.~\ref{fig:Co_decay_fit}) at two different phases. For the earlier phase, in between 35 -- 200 days, we used the full $^{56}$Ni+$^{56}$Co decay model (Eq.~\ref{eq:L_Ni_Co}), while for the late phase, in between 380 -- 625 days, a pure $^{56}$Co exponential decay (Eq.~\ref{eq:L_Co}) was fit. In both cases, the only free parameter was the synthesized initial nickel mass.

From the fit to the early phase, we obtain $M_{\mathrm{Ni}} \sim 2.8 \, \mathrm{M}_\odot$. For the late phase, the Co-only fit yields an even higher value of $M_{\mathrm{Ni}} \sim 7.0 \, \mathrm{M}_\odot$. These values are far above the typical $^{56}$Ni masses inferred for core-collapse supernovae ($M_{\rm Ni} \lesssim 0.1$--$0.2 \, \mathrm{M}_\odot$, \citealt{Nomoto_2006NuPhA.777..424N}), thus, they are physically implausible. For comparison, models assuming a realistic nickel mass ($\sim 0.2\,\mathrm{M}_\odot$) lie well below the observed luminosity, as indicated by the dashed lines in Fig.~\ref{fig:Co_decay_fit}. Moreover, the $^{56}$Ni+$^{56}$Co and pure $^{56}$Co models converge after $\sim$35 days, indicating that the contribution of nickel decay is negligible at the epochs considered.

We therefore conclude that that radioactive heating alone cannot explain the observed luminosity and that the LC is primarily powered by ejecta--CSM interaction. The apparent agreement of the decline rate with the $^{56}$Co slope in two separate intervals is likely coincidental.

\begin{figure}[]
    \includegraphics[width=0.98\columnwidth]{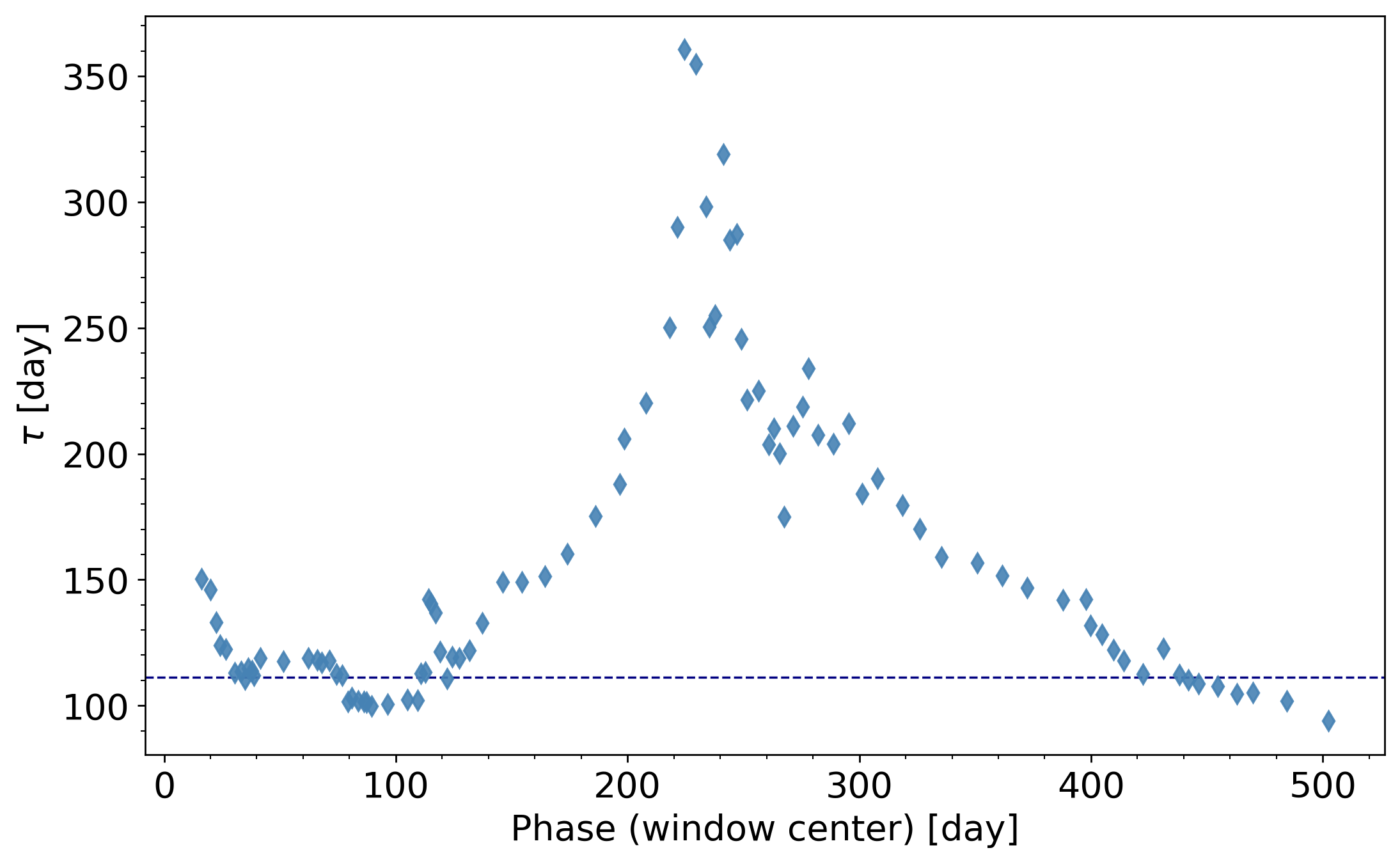}
    \caption{Time evolution of the local decay timescale inferred from the decline of $L_{\rm bol}$, computed using a sliding-window fit. Blue diamonds show the results when the phase is measured relative to $t_0$. Each point is calculated from a window of 10 consecutive data points, and its position on the x-axis corresponds to the median phase of the window. The dashed navy-blue line indicates the characteristic decay timescale of $^{56}$Co ($\tau_{\rm Co}=111.3$\,d).
    \label{fig:ln-lin slope}}
\end{figure}

\begin{figure}[]
\includegraphics[width=0.98\columnwidth]{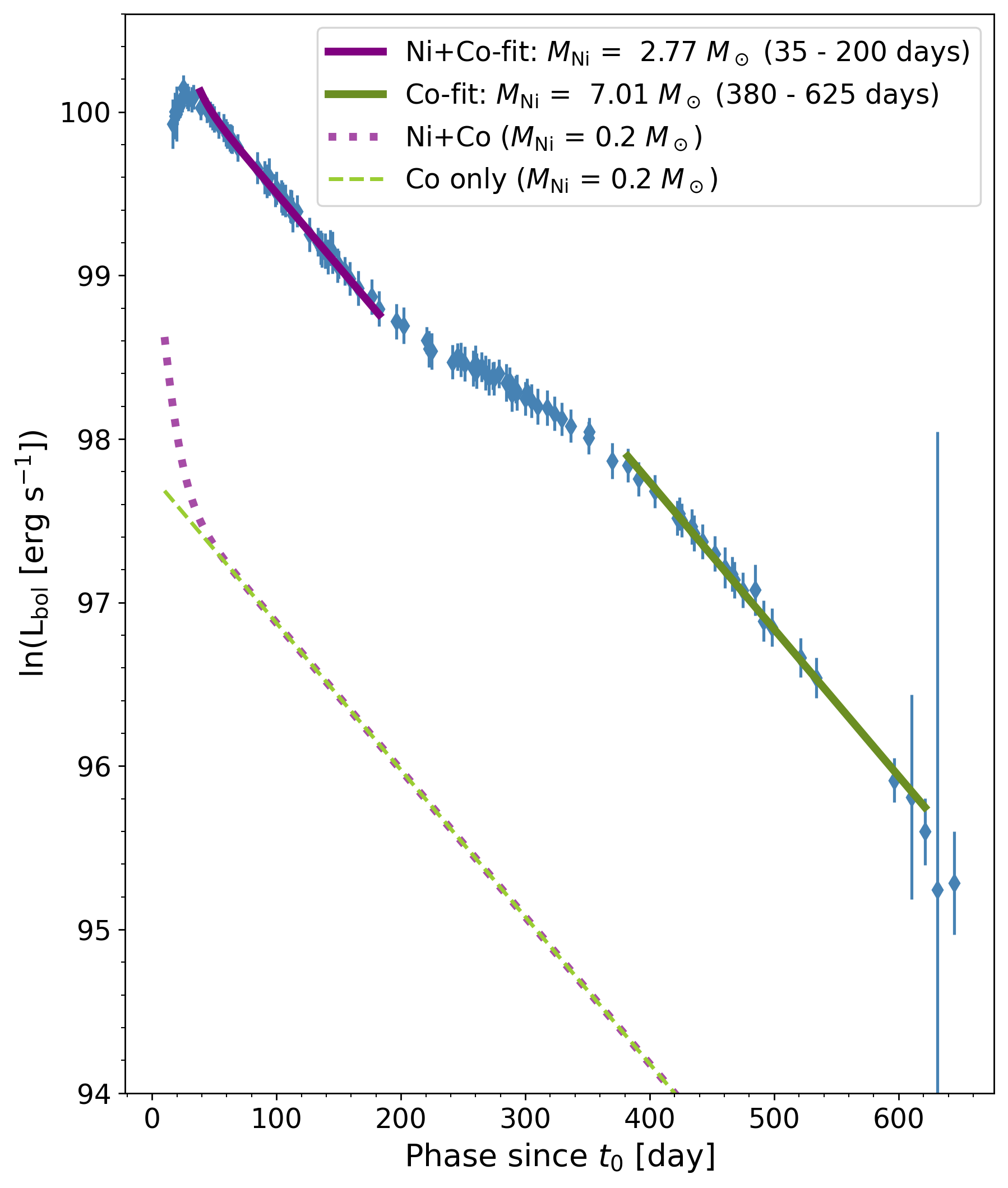}
\caption{Logarithmic bolometric luminosity evolution compared to the expected slope of $^{56}$Co radioactive decay. The solid purple line shows the fit $^{56}$Ni+$^{56}$Co model for the early phase, while the solid green line represents the best-fit pure $^{56}$Co model. The inferred nickel masses in both phases are unrealistically high. 
Dashed purple and green lines indicate the corresponding models assuming a nickel mass of $M_{\rm Ni} = 0.2 \, \mathrm{M}_\odot$ for comparison.
\label{fig:Co_decay_fit}}
\end{figure}

\subsection{Model B: optically thick CSM}\label{subsec:opt_thick_model}

In addition to the optically thin CSM scenario, we also explored the possibility of an optically thick CSM. If the progenitor was a luminous blue variable (LBV) star, we expect the presence of one or more dense CSM shells, as such stars are known to eject substantial amounts of material through intense stellar winds or episodic mass-loss events. We applied a dense CSM model based on the original analytical formulation by \citet{Arnett_1980} and  \citet{Arnett_1982}, which was later extended by \citet{Chatzopoulos_2012} to describe the light curves of superluminous supernovae (SLSNe) as well as interacting SNe.

The model assumes homologous expansion of the SN ejecta, with the outermost layer having an initial velocity of $v_{\text{SN}}$ at radius $R_{\text{max}}$. The expansion is characterized by the timescale $t_h = R_p / v_{\text{max}}$, where $R_p$ is the progenitor radius at explosion. The mass of the ejecta is $M_{\text{SN}}$ and its density profile is assumed to be a broken power-law: a constant density inner core up to a radius of $x_0 = r_0 / R_{\text{max}}$ and a power-law envelope with $\rho_{\text{SN}} \propto r^{-n}$ (see Equation~\ref{eq:rho_ej rho_CSM}).

Opacity is dominated by Thomson scattering, which is assumed to be constant ($\kappa = \kappa_{\text{Th}}$, both spatially and temporally \citep{Moriya_2011MNRAS.415..199M}). We set $\kappa = 0.34$ cm$^2$~g$^{-1}$ as both the ejecta and the CSM is assumed to be H-rich, close to solar composition. 

The luminosity is generated by the FS and the RS heating the shocked CSM and the ejecta, as well as the usual radioactive decay of $^{56}$Ni and $^{56}$Co. Since core-collapse SNe produce only moderate amount of radioactive $^{56}$Ni ($M_{\rm Ni} \lesssim 0.2$ M$_\odot$), the contribution of the Ni-Co decay is expected to be minor in the case of Type IIn SNe, as we have already shown in Fig.~\ref{fig:Co_decay_fit}.

\begin{figure*}
    \centering
    \includegraphics[width=0.98\linewidth]{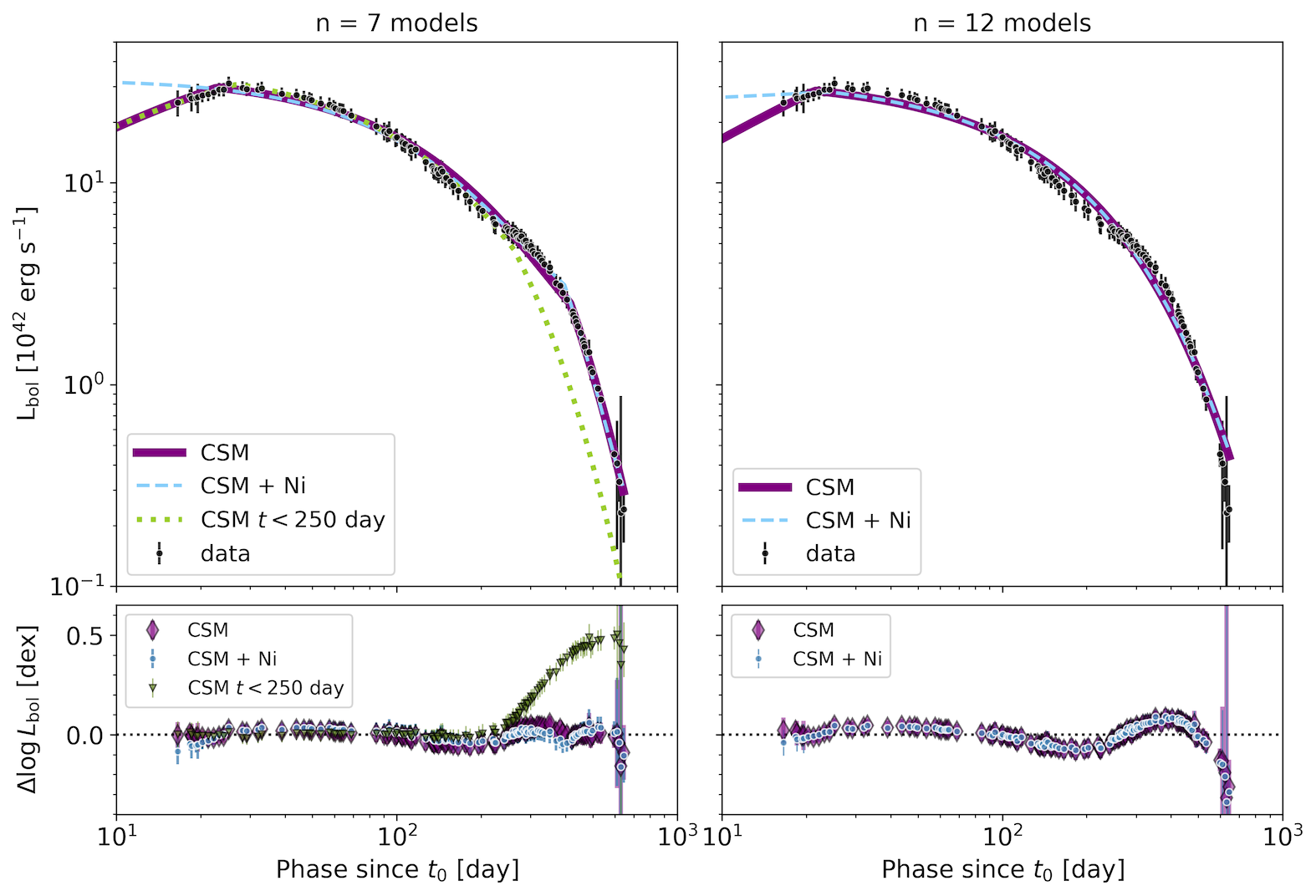}
    \caption{\textbf{Top:} The best-fitting models of the optically thick CSM (Model B) to the bolometric luminosity of SN~2019vxm. 
    \textbf{Bottom:} The logaritmic luminosity difference, ($\log L_{\rm data} - \log L_{\rm model}$). See Table~\ref{tab:opt_thick_model_params} for the model parameters.}
    \label{fig:fit_minim}
\end{figure*}

The model assumes that both the FS and the RS luminosities are thermalized with 100 \% efficiency, providing a time-dependent heating of the ejecta + CSM. This energy is then diffused out to the photosphere. 
Since this model assumes a dense, optically thick CSM, the photosphere remains fixed within the CSM, effectively obscuring the SN itself, as long as the FS is within the optically thick part of the CSM. Although simplified, the fixed photosphere approximation captures the essential features of interaction-powered light curves \citep{Chatzopoulos_2012}. 

The parameters of this model are the following: the progenitor radius $R_p$ at the moment of explosion, the ejecta mass $M_{SN}$, the maximum expansion velocity of the SN $v_{\rm SN}$ at the start of the interaction, the total mass of the CSM $M_{\rm CSM}$, the mass-loss rate $\dot{M}$, the wind velocity $v_w$, the optical opacity $\kappa$, and the density exponents $n$ and $s$. In order to decrease the number of free parameters, we set $\kappa = 0.34$ cm$^2$~g$^{-1}$, $v_w = 10$ km~s$^{-1}$ and $s = 2$, and kept them fixed during the fitting process. Note that the assumed $v_w$ affects only the value of the inferred $\dot{M}$, because the LC depends on only their ratio, $\dot{M}/v_w$. Thus, choosing $v_w = 100$ km~s$^{-1}$ instead of 10 km~s$^{-1}$ would result in higher $\dot{M}$ rates by a factor of 10. 
The $n$ parameter was also fixed during the fitting, but we explored several models by using $n=7$ (radiative envelope)  and $n=12$ (convective envelope). 

Constant-density CSM models having $s=0$ were also considered in a similar way. They resulted in similar best-fit light curves, but their reduced $\chi^2$ values were found to be a factor of $\sim 2 - 3$ times higher than those of the wind-like CSM models ($s=2$). Thus, in the following we focus on the results obtained from fitting the $s=2$ models.

\begin{deluxetable*}{lccccc}
\tablecaption{Parameters of the best-fit models for Model B. Uncertainties are given in parentheses. }
\label{tab:opt_thick_model_params}
\tablehead{ \colhead{Parameter} & \colhead{$t<250$d} & \colhead{$t<1000$d} & \colhead{$t<1000$d} & \colhead{$t<1000$d} & \colhead{$t<1000$d}  }
\startdata
n         & 7 & 7 & 7 & 12 & 12 \\ 
$M_{\rm Ni}$ (M$_\odot$) & 0.0 &  0.0 & 0.2 & 0.0 & 0.2 \\
$R_p$ ($10^{13}$ cm) & 30.6 (0.7) & 39.8 (2.2) & 17.7 (2.5) & 30.4 (1.0) & 18.9 (0.8) \\
$M_{\rm SN}$ (M$_\odot$) & 46.1 (3.4) & 82.0 (1.3) & 79.3 (2.1) & 26.0 (3.8) & 99.5 (2.3) \\
$M_{\rm CSM}$ (M$_\odot$) & 5.74 (0.07) & 6.24 (0.09) & 2.37 (0.18) & 6.70 (0.14) & 1.95 (0.10) \\
$\dot{M}$ (M$_\odot$/yr) & 0.227 (0.003) & 0.284 (0.005) & 0.239 (0.009) & 0.396 (0.005) & 1.048 (0.030) \\
$v_{\rm SN}$ ($10^3$ km/s) & 19.67 (0.31) & 17.64 (0.04) & 21.15 (0.29) & 26.39 (0.42) & 35.95 (0.41) \\
\hline
$E_{\rm SN}$ (foe) & 2.14 & 3.06 & 4.25 & 1.40 & 9.90 \\
$R_{\rm CSM}$ ($10^{15}$ cm) & 1.10 & 1.09 & 0.49 & 0.84 & 0.25 \\
$t_{\rm FS}$ (days) & 27.4 & 23.5 & 7.0 & 21.8 & 1.32 \\
$t_{\rm RS}$ (days) & 267.4 & 422.8 & 405.0 & 23.2 & 24.7
\enddata
\end{deluxetable*}

Following \citet{Chatzopoulos_2013}, the fitting was computed with the Price-algorithm, a Monte-Carlo $\chi^2$-minimization method. The number of random vectors was set as $N = 200$ and the code was allowed to run until the change of the $\chi^2$ became $\Delta \chi^2 \leq 1$. The first models for both $n=7$ and $n=12$ were computed assuming $M_{\rm Ni} = 0$, thus, only shock heating, then a radioactive heating from $M_{\rm Ni} = 0.2$ M$_\odot$ was added. In addition, a model with $n=7$ was also fit to only the first 250 days of the data. The parameters of the best-fit models are collected in Table~\ref{tab:opt_thick_model_params}. The lower panel lists some additional parameters that can be calculated from the equations given by \citet{Chatzopoulos_2012}: $E_{SN}$ is the kinetic energy of the ejecta, $R_{\rm CSM}$ is the outmost radius of the CSM, while $t_{\rm FS}$ and $t_{\rm RS}$ are the lifetime of the FS and the RS, respectively.

Figure~\ref{fig:fit_minim} shows that Model B can adequately describe the  entire observed bolometric light curve, if the model is fitted to the entire dataset ($t < 1000$ days). If the fit is restricted to the early part only (for $t < 250$ days), the model underpredicts the observed luminosities at $t > 300$ days. When radioactive heating is added, the fitted light curves differs from the CSM-only light curves only at the earliest data ($t < 20$ days). 
The lower panels show the residuals (defined as $\log L_{\rm model} - \log L_{\rm data}$ in dex), indicating that the fits reproduce the overall evolution well, although small systematic deviations are present. 

When inspecting the model parameters in Table~\ref{tab:opt_thick_model_params} more closely, essential differences can be found between these best-fit models. When comparing the $n=12$ models with the $n=7$ ones, it is seen that in the former ones both the FS and the RS terminates quickly (at $t \leq 25$ days), and their model light curves are practically due to the diffusive cooling of the shock heated ejecta+CSM. 
On the other hand, in the $n=7$ models the RS propagates on a much longer timescale, terminating only at $t \geq 250$ days up to $t \sim 400$ days. These models describe the observed LC somewhat better, giving a reasonable fit even on the late-phase declining part. Within this context the drop of the decline rate for $t > 400$ days can be explained by the termination of the RS. 

Thus, although all these models are consistent with the data, the one that best describes the observations is the $n=7$ model assuming shock-heating only. This model assumes a very massive SN ejecta ($M_{\rm SN} \gtrsim 80$ M$_\odot$) surrounded by a massive CSM ($M_{\rm CSM} \gtrsim 6$ M$_\odot$). 
It must be kept in mind, however, that the parameters of this model are highly correlated \citep{Chatzopoulos_2013}, thus, it is possible that such extreme values are not real, and may be caused by the parameter correlations. It can be suspected from e.g., the comparison of the ejecta and CSM masses of the two $n=12$ models: they are very different from each other, even though the models produced essentially the same light curves. 

We also explored a more restricted region of the parameter space by limiting the ejecta masses to physically more plausible values (below $50 \, \mathrm{M}_\odot$) and fitting only the early evolution ($t \lesssim 90$--$160$~d). In these tests, both $n=7$ and $n=12$ ejecta profiles and both $s=0$ and $s=2$ CSM density profiles were considered. 
The resulting models generally favor progenitor radii of $R_0 \sim (5$--$15) \cdot 10^{13}$~cm and a maximum SN expansion velocity of $v_0 \sim 2\cdot 10^4$~km~s$^{-1}$. 
The inferred ejecta mass values were reduced to $\sim15$--$45\,\mathrm{M}_\odot$, depending on the adopted density structure, while the CSM masses were typically $\sim1$--$2\,\mathrm{M}_\odot$ for constant-density ($s=0$) CSM models and $\sim4$--$6\,\mathrm{M}_\odot$ for wind-like ($s=2$) CSM models. 
However, these  more physically more constrained solutions require extremely high mass-loss rates, on the order of several $\mathrm{M}_\odot\,\mathrm{yr^{-1}}$. These models highlight that the observed evolution likely reflects a complex, time-dependent ejecta--CSM interaction that cannot be captured uniquely by simple semi-analytic prescriptions.

\section{Results and discussion} \label{sec:results}

The detailed analysis of the temporal evolution of SN\,2019vxm revealed a complex interplay between the CSM surrounding the progenitor and the ejecta and shocks it induced in the CSM. Based on the early evolution of the supernova, \citet{Lane-2025} concluded that the progenitor was surrounded by an asymmetric and/or inhomogeneous CSM, composed of denser and thinner clumps and shells. This is corroborated by their estimate of an intermediate CSM density profile exponent of $s\sim 1.5$. Below we summarize our findings and discuss their implications.

\subsection{Initial evolution of the explosion and the CSM}

The initial phase of the SN appears to be consistent with either an optically thin model or with a thick CSM model obscuring the explosion, but the findings of \citet{Lane-2025} prefer the latter scenario. However, we find that the photosphere does not stay at a fixed radius, as assumed by the optically thick models (Model B), but is expanding. This may indicate that the photosphere generated by the forward shock can still propagate within the CSM, or the CSM itself may start expanding. Alternatively, as the forward shock travels through the structured CSM, the average radius may appear moving outwards.

After 80 days (after $t_p$) the photosphere reaches its maximum radius at $R_{\rm hot, max}=(3.4\pm0.2)\cdot10^{15}$\,cm, or $(48\pm 3)\cdot 10^3$\,R$_\odot$, and starts to shrink. By day 120, temperature of the photosphere cools to $6153\pm241$\,K, consistent within uncertainties with the minimum temperature of $6310^{+147}_{-144}$  K derived by \citet{Lane-2025}, signaling that it now follows the receding hydrogen ionization front. At this stage the photosphere and the forward shock have separated, the latter propagating into regions beyond the close-by, dense CSM. 

The rise time of SN~2019vxm is closer to the slow-rising Type~IIn subpopulation of \citet{Nyholm_2020}, characterized by a mean rise time of $\sim50$~days, than to their fast-rising subpopulation with a mean rise time of $\sim20$~days. We measure a rise time of $\approx47$~days in the $R$ band relative to $t_0$, while the $r$-band value is a lower limit of $\sim39$~days due to incomplete peak coverage. This latter value agrees well with the median $r$-band rise time of $39^{+23}_{-16}$~days found by \citet{Ransome2025ApJ...987...13R} from a larger sample of Type~IIn events.

The decline is unusually slow, the time required for the LC to fade by one magnitude is $\approx202$~days in $R$-band, and $\approx221$~days in $r$-band. Such long decline times are rare among interacting SNe; in the sample of \citet{Ransome2025ApJ...987...13R}, only three out of 142 objects exhibit fall times longer than 200 days.

The extinction-corrected peak luminosity of SN\,2019vxm also place it among the more luminous members of the Type~IIn population, with $M_R \approx -20.2$~mag and $M_r \approx -19.7$~mag, approximately one magnitude brighter than the average peak brightness reported by \citet{Ransome2025ApJ...987...13R}.

The double-BB fits suggest a changing balance between the optical and infrared components. By the epoch of the third contemporaneous \textit{WISE} measurement, at phase 541.3~day, the contribution of the cold component exceeds of the hot photosphere, indicating that dust emission has became increasingly important in shaping the observed SED.

\begin{figure*}
    \centering
    \includegraphics[width=1.0\textwidth]{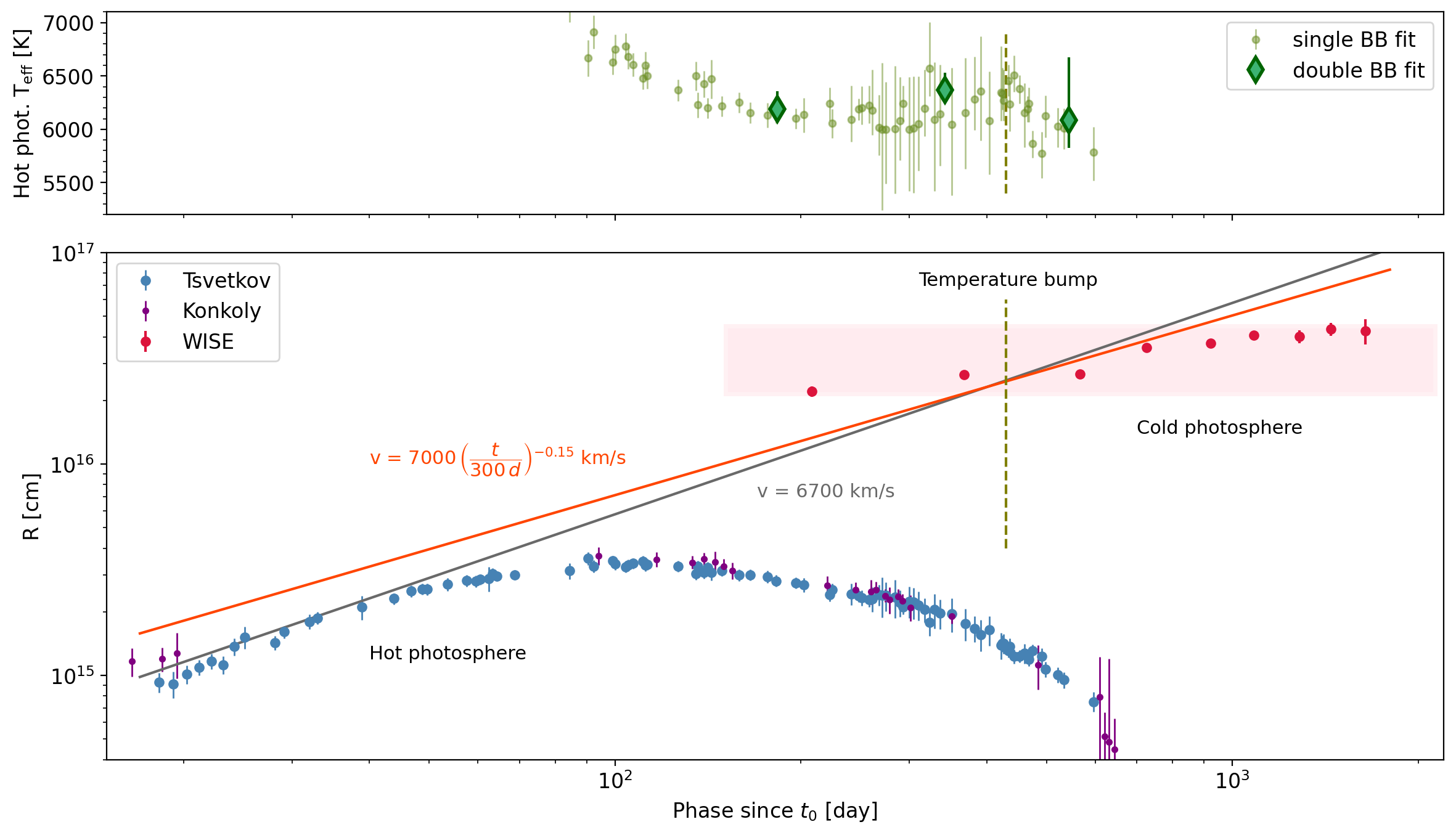}
    \caption{Evolution of the photospheres. \textbf{Top:} temperature of the hot photosphere, same as in Fig.~\ref{fig:temperature_double_BB_simple_BB_WISE}, highlighting a late increase related to the contribution of the outer CSM, with the vertical dashed line marking the approximate time of the peak. \textbf{Bottom:} radii of the hot and cold photospheres (blue-purple and red points, respectively). Grey and red lines indicate constant, $v_{\rm sh}=6700$ km\,s$^{-1}$ shock velocity and a shock velocity slowly decreasing from an initial 7000 km\,s$^{-1}$ velocity. The faint red rectangle marks the estimated extent of the outer CSM layer.}
    \label{fig:radius-cold-hot}
\end{figure*}

\subsection{The IR emission and the outer CSM}
\label{subsec:outer_csm}
The supernova shows an initial burst of IR radiation, and then a second brightening, followed by long-term IR emission from a slowly cooling source. These could originate, for example, from an outer CSM and an expanding light echo, or from newly formed dust closer to the explosion (or both). Here we propose that the initial IR emission, the first two points at 183.4 and 342.7 d, are caused by a light echo. The radiation of the explosion after shock breakout from the CSM reaches the estimated dust radius with a time delay of less than 10 days. In contrast, newly formed dust in the CDS is unlikely as an explanation, as it would need a very high velocity, in excess of $10^4$ km\,s$^{-1}$, to reach the inferred distance by the time of the first post-maximum WISE data point.  

Between 342.7 d and 541.3 d the IR radiation increases along with the temperature of the cold component, but this radiation comes from largely the same distance as the initial echo, between $\sim2.2-4.3\cdot10^{16}$\,cm ($\sim3.2-6.2\cdot10^{4}$\,R$_\odot$, $\sim1500-2900$ AU). The brightening coincides with the final break in the bolometric luminosity of the hot photosphere, between 350--400 d. We hypothesize that these effects are caused by the forward shock reaching this outer CSM region. In Fig.~\ref{fig:radius-cold-hot} we show how the radius of the hot and cold photospheres relate to each other. If we extrapolate the initial expansion velocity of the hot photosphere at $v_{\rm sh} = 6700$ km\,s$^{-1}$, it could reach the distance of the outer CSM in about 400 d (grey line in Fig.~\ref{fig:radius-cold-hot}). The orange line shows a shock that is gradually slowing down from a slightly higher initial velocity, at a rate of $v_{\rm sh}(t)=7000\,(t/(300\,{\rm d}))^{-0.15}$~km\,s$^{-1}$, reaching the outer CSM region at the same time as the constant-velocity assumption. These velocities are rather high but still physically plausible compared to the typical range of $v_{\rm sh}\sim3000-6000$ km\,s$^{-1}$ inferred from observations or used in models of Type IIn SNe \citep{Chevalier_1994ApJ...420..268C,Fransson-2014,Taddia-2020}. 

If the front shock travels through this outer CSM it should heat it up, which we would expect to influence the optical light curves too. A closer inspection of the bolometric light curve and the inferred $T_{\rm eff}$ values of the hot photosphere shows a small bump at around 300 d (e.g., on Fig.~\ref{fig:n_param}). The $T_{\rm eff}$ curve in Fig.~\ref{fig:luminosity_temperature_and_radius} shows an increase of $\sim 200$\,K between 350--450 d. We consider this increase as a potential optical contribution from the heated outer CSM, which temporarily increases the steepness of the SED towards shorter wavelengths. Unfortunately, this section of the LC is only covered by the Tsvetkov \textit{BVI} data in optical, which is not detailed enough for a two-component BB fit. As the outer CSM cools, its peak emission shifts into the IR, outshining the fading light echo. 

The appeal of this hypothesis is that it connects the IR rebrightening to two separate but plausible physical mechanisms at approximately at the same radius and thus in the same CSM layer. The initial IR emission is too far for newly condensed or shock-heated dust, but can be attributed to a light echo. Conversely, the rebrightening in this layer after $\sim350$ d is too close for a light echo but cannot be attributed to the CDS behind the font shock either as the IR radiating layer appears to be largely stationary. Instead, its slow expansion can be attributed to the original CSM expansion velocity, which is usually in the $\sim 100$--200 km\,s$^{-1}$ range for IIn~SNe \citep{Smith-2010,Fransson-2014,Taddia-2020}. However, this hypothesis will have to be validated with spectroscopic data that can directly measure the velocities of various layers.

\subsection{Modeling results}

The radioactive-decay models (Sect.~\ref{subsec:rad_decay}) cannot explain the observed luminosity. The inferred nickel masses are unrealistically high: $M_{\rm Ni} \approx 2.8\,\mathrm{M}_\odot$ for the early phase and $M_{\rm Ni} \approx 7.0\,\mathrm{M}_\odot$ for the late phase. Models with a realistic nickel mass ($\mathrm{M}_{\rm Ni}=0.2\,\mathrm{M}_\odot$) remain well below the observed luminosity, confirming that radioactive heating is negligible compared to CSM interaction.

In the optically thin analytic framework based on a single ejecta density exponent $n$, the inferred $n$ varies strongly with time, ranging from $n \sim 10$--$12$ at early phases to values below the formal self-similar limit of $n=5$ during the late decline. 
This evolution indicates that the LC cannot be described by a single power-law interaction model. The transition to $n<5$ after $\sim 350$ days may mark a change in the shock-interaction regime, such as the weakening or termination of the reverse shock, or the forward shock reaching the outer edge of the dense CSM in this framework. 

The optically thin CSM model (Model A, Sect.~\ref{subsec:opt_thin_CSM_model}), in which the observed luminosity is produced by the reverse shock, partly attenuated by the CDS, provides a good description of the early-time bolometric light curve up to $t \sim 100$ days, with the best agreement obtained for an ejecta density exponent of $n \approx 7$. The corresponding parameters are $R_{s}({\rm 1d}) = 8.98 \cdot 10^{13}$ cm, $\dot{M} = 0.083$ M$_\odot$~yr$^{-1}$ and $\sigma_0 = 0.0089$, although these values are highly uncertain due to strong parameter correlations. The inferred mass-loss rate is high, suggesting an intense pre-supernova mass-loss episode. Integration of the $n\approx 7$ density profile suggests $M_{\rm ej} \gtrsim 70$ M$_\odot$. At later epochs, the observed luminosity falls below the predicted reverse-shock luminosity, indicating that the optically thin approximation breaks down. This behaviour may reflect changes in the shock interaction, such as the RS penetrating deeper into the ejecta or an increase in optical depth due to the accumulation of swept-up material.

The optically thick CSM model (Model B, Sect.~\ref{subsec:opt_thick_model}) provides the best overall description of the bolometric light curve, especially when fitted to the full dataset up to $t<1000$ days. The best-fitting optically thick models favor a wind-like CSM density profile ($s=2$), broadly consistent with the CSM density exponent inferred by \citet{Lane-2025} ($s = 1.40^{+0.08}_{-0.08}$), whereas constant density CSM models ($s=0$), although also being consistent with the data, provide somewhat poorer fitting results around the LC peak. The model fits also prefer an $n=7$ ejecta density profile over $n=12$, consistent with the value of $n = 7.46^{+0.47}_{-0.29}$ inferred by \citet{Lane-2025}. According to the best-fit parameters of this model, the RS remains active for a significantly longer time in the $n=7$ models, up to $\sim 400$ days. The FS, however, terminates rapidly in both cases, within $\lesssim 30$ days. In this framework, the change in the decline rate after $\sim 400$ days can be interpreted as the termination of the reverse shock.

The preferred model implies a very massive ejecta and a dense CSM, with $M_{\rm SN} \gtrsim 80 \, \mathrm{M}_\odot$ and $M_{\rm CSM} \gtrsim 6 \, \mathrm{M}_\odot$, although the best-fit parameters sweep a broad interval, $M_{\mathrm{SN}}$ spanning approximately $26$--$100 \, \mathrm{M}_\odot$, and $M_{\mathrm{CSM}}$ ranging from about $2$ to $7 \, \mathrm{M}_\odot$. 
However, these values should be treated with caution, as the model parameters are strongly correlated. 
Independent modeling by \citet{Lane-2025} also favors a massive progenitor system, although with somewhat lower masses: $M_{\rm ej} = 38.8^{+6.6}_{-6.0} \, \mathrm{M}_\odot$, $M_{\rm CSM} = 1.48^{+0.14}_{-0.13} \mathrm{M}_\odot$ and a total mass of $M_{\rm tot} = 40.3^{+6.7}_{-6.2} \mathrm{M}_\odot$. 
The inferred mass-loss rates from our models are of order $\dot{M} \sim 0.2$--$1 \, \mathrm{M}_\odot \, \mathrm{yr^{-1}}$ (or even higher in the more restricted cases), significantly exceeding the values expected from steady winds of red supergiants and instead pointing toward eruptive pre-supernova mass loss. Similar values have been inferred for other strongly interacting SNe~IIn, such as SN~2014ab, KISS15s and SN~2015da \citep{Bilinski_2020MNRAS.498.3835B,Kokubo-2019ApJ...872..135K, Tartaglia_2020A&A...635A..39T_SN_2015da}

Overall, both the optically thin and thick CSM interaction models give a phenomenologically accurate solution.

\section{Conclusions} \label{sec:conclusions}

%1. Konkoly photomerty ($griz$ bands)

In this paper we presented Konkoly photometry of SN~2019vxm, obtained with the 0.8~m RC80 telescope at Piszkéstető Mountain Station, in Bessell \textit{BV} and Sloan \textit{griz} filters. 
The observations began at two days after discovery and span nearly 650 days, providing the first published multicolor light curves of the event in Sloan \textit{griz} passbands. We combined our observations with data from various other sources, including that of \citet{Tsvetkov_2024,Lane-2025}, and the WISE space telescope. With extensive multicolor LC at hand, we inferred the bolometric luminosity, photospheric temperature and radius evolution of the SN, and applied model fits to it. Whereas the work of \citet{Lane-2025} focused on the early-phase evolution, we investigated the long-term behavior of SN\,2019vxm.

%2. Model A

The optically thin CSM (Model A) reproduces the bolometric light curve well during the first $\sim100$ days. However, this agreement is likely only phenomenological. 
The public spectra \citep{Leadbeater-2019} and the spectra obtained during the first $\sim100$~days by the Las Cumbres Observatory (LCO) telescopes show signatures of only the shocked CSM including strong, narrow Balmer-lines in emission superimposed on a hot continuum, with no clear P~Cygni profiles from the SN ejecta (Smith et al, in prep.). 
This suggests that the CSM remained optically thick during this phase; the photosphere may have receded inwards through the CSM, but the underlying ejecta did not become directly visible. 
Altough Model A can formally fit the early luminosity evolution, its physical assumptions are not fully consistent with all the observations. Further spectroscopic analysis would provide many new essential information on the ejecta and CSM properties.

%3. Model B

The optically thick CSM model (Model B) also provides a good description of the entire bolometric light curve. Both the $n=7$ and $n=12$ ejecta density profiles yield acceptable fits, with the $n=7$ model being only marginally preferable over the other. Although \citet{Lane-2025} argued that their preferred $n \approx 7.5$ solution potentially favors a compact LBV/Wolf-Rayet-like progenitor, the strong parameter degeneracies in our fits, together with the time-dependent $n$ values inferred from the light-curve evolution, prevent us from drawing firm conclusions about the nature of the progenitor.
We find that the contribution of radioactive decay is negligible compared to the energy input from ejecta--CSM interaction. 
However, the assumption of a fixed photospheric radius in Model B is not consistent with the radius evolution inferred from fitting blackbody fluxes to the photometry. 
Moreover, the model parameters are strongly degenerate, limiting the physical interpretation of the best-fit values. A more accurate treatment would require a dedicated radiative hydrodynamic model with a time-dependent photosphere, which is beyond the scope of this work. But we emphasize, again, that although semi-analytical 1D models can provide phenomenologically accurate fits, their physical validity must be carefully evaluated before accepting the results.

%4. WISE

The long-lasting \textit{WISE} infrared emission likely indicates the presence of an extended, dusty CSM surrounding the progenitor. The IR LC shows an initial excess, followed by a second brightening and a slow decline over several years. 
We interpret this behavior as evidence for at least two distinct mass-loss episodes prior to explosion. The late-time IR evolution suggests that the progenitor could have experienced substantial changes in its mass-loss history during the final centuries before core collape. This behavior is consistent with the picture emerging for other long-lasting Type~IIn SNe \citep{Fox-2010, Hillenkamp_2026arXiv260119891H}. Late shock heating and brightening of an outer CSM ring, albeit on a much larger scale, was also observed around SN\,1987A \citep[see, e.g.,][]{Arendt-2016,Cendes-2018,Kangas-2023}.  However, the scenario we outlined here still requires testing and confirmation via spectroscopic velocities of various components, to further constrain the long-term evolution of the CSM interactions.

Assuming that the temperature of the outer CSM follows a simple exponential decrease, we estimate that by now it cooled to about 550\,K, which would put its blackbody peak wavelength to about 5\,$\mu$m. Therefore, ground-, and space-based IR facilities like the UK Infrared Telescope Facility (UKIRT), the \textit{James Webb} Space Telescope, or the SPHEREx space mission would be, in principle, suitable for long-term follow-up. While such a follow-up is beyond the scope of this work, we highlight here the importance of long-term observations of interacting SNe, which allows us to map the extended CSM structure and the mass-loss histories of the progenitors, as shown by \citet{Szalai_2021ApJ...919...17S} using \textit{Spitzer} Space Telescope data.

\section*{Acknowledgments} \label{sec:Acknowledgments}

The authors would like to thank Zachary Lane, Ashley Villar, Conor Ransome and Ryan Ridden-Harper for fruitful discussions. 

The operation of the RC80 robotic telescope at Konkoly Observatory has been supported by the project ``Transient Astrophysical Objects'' (GINOP 2.3.2-15-2016-00033)) from the Government of Hungary based on funding from the European Union.   
This research was supported by the following grants: NKFIH-OTKA K-142534 and K-138962 grants from the Hungarian Research, Development and Innovation Office (NKFIH); `SeismoLab' KKP-137523 and KKP-143986 \'Elvonal grants from NKFIH; LP2025-14/2025 and LP2021-9 Lendület/Momentum grants of the Hungarian Academy of Sciences. K.\,L.\ thanks the financial support provided by the undergraduate research assistant program of Konkoly Observatory. 

This publication makes use of data products from the Near-Earth Object Wide-field Infrared Survey Explorer (NEOWISE), which is a joint project of the Jet Propulsion Laboratory/California Institute of Technology and the University of California, Los Angeles. NEOWISE is funded by the National Aeronautics and Space Administration. 
This research has made use of the NASA/IPAC Infrared Science Archive, which is funded by the National Aeronautics and Space Administration and operated by the California Institute of Technology. 

This research is based on observations made with the Neil Gehrels Swift Observatory, obtained from the MAST data archive at the Space Telescope Science Institute, which is operated by the Association of Universities for Research in Astronomy, Inc., under NASA contract NAS 5–26555.  
This research has made use of the Astrophysics Data System, funded by NASA under Cooperative Agreement 80NSSC21M00561.

This paper used the LLM model ChatGPT--5.5 by OpenAI for supervised assistance with code debugging, literature search, and language editing.

\appendix
\section{Additional analysis} \label{sec:Appendix}

This section presents supplementary analyses related to the  ASAS--SN and early-time WISE observations of SN\,2019vxm, representative examples of the IR single-blackbody fits, the full optical SED evolution constructed from the Konkoly and Tsvetkov photometry.

\subsection{Early ASAS--SN light curve and the early SED}\label{subsec:Asassn_data}

Most of the rising phase was only covered by the ASAS--SN telescopes, in the $g$ band. We downloaded the various photometric outputs from the Sky Patrol service \citep{ASAS-SN-2017,ASAS-SN-2023}. We then tried to match both the aperture photometry and the  differential-image photometry (with the reference flux added back) versions of the LC with the Konkoly $g$--band data, but found significant offsets in the ASAS--SN data. The source of these discrepancies is unclear. Since the $g$ band in itself is not sufficient to estimate the bolometric flux, especially during the rising phase when we expect the temperature of photosphere to increase rapidly, we decided not to investigate the origin of this discrepancy further. 

We also examined the differential image flux with the no reference flux added option. These data sets also included discrepancies, especially in the data from the \texttt{bs} and \texttt{bt} cameras, where measured flux levels alternated between two different average values. We combined data from these two cameras by manually separating `high' and `low' flux points, plus data from cameras \texttt{bC} and \texttt{bD}, and shifting each subset to a flux level where the average of the points before shock breakout is zero. We then scaled this LC by a factor of 1.38 to the match Konkoly \textit{g} fluxes. The final flux curve is shown in Fig.~\ref{fig:asassn}, and available in Table~\ref{tab:asassn-data}. 

Multicolor observations only began shortly before maximum light, with the earliest points being at $-8.6$ and $–5.9$\,d (relative to $t_p$) from the Konkoly and Tsvetkov data sets, respectively. The WISE mission observed SN\,2019vxm at $-17.06$\,d. At such an early phase we expect the IR flux to come from the Rayleigh--Jeans tail of the hot photosphere instead of a separate IR source, like dust, unless the observation is early enough that shock heating has not fully diffused through the CSM yet. Although no standard multicolor photometry is available at this point, we attempted to reconstruct the SED based on the early observations from ASAS--SN and TESS.  

\begin{figure}
\centering
\includegraphics[width=1.0\columnwidth]{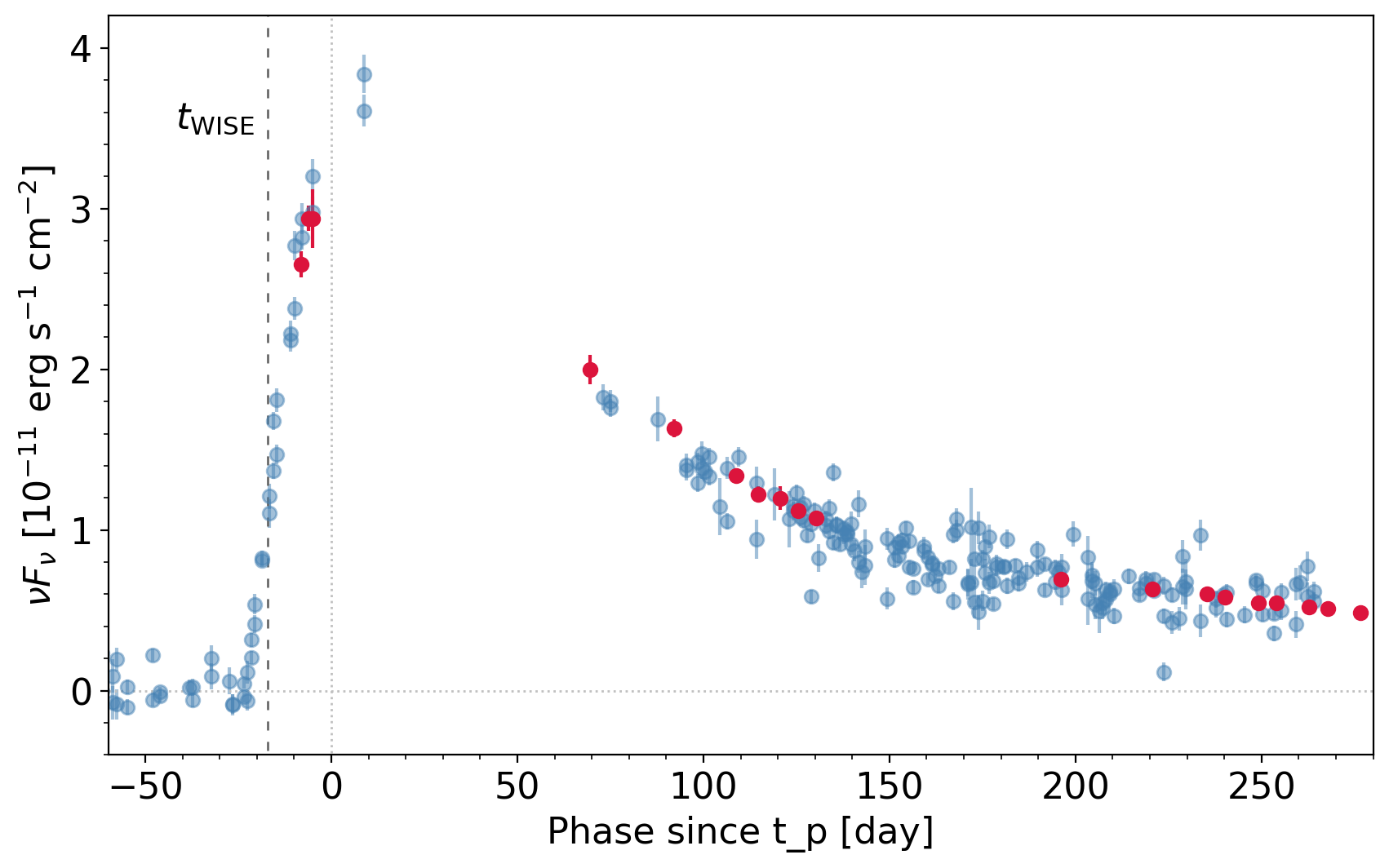}
\caption{ASAS-SN \textit{g}-band flux curve, which includes the early phase of the explosion (blue), scaled to the Konkoly \textit{g} data (red).
\label{fig:asassn}}
\end{figure}

\begin{figure}
\centering
\includegraphics[width=1.0\columnwidth]{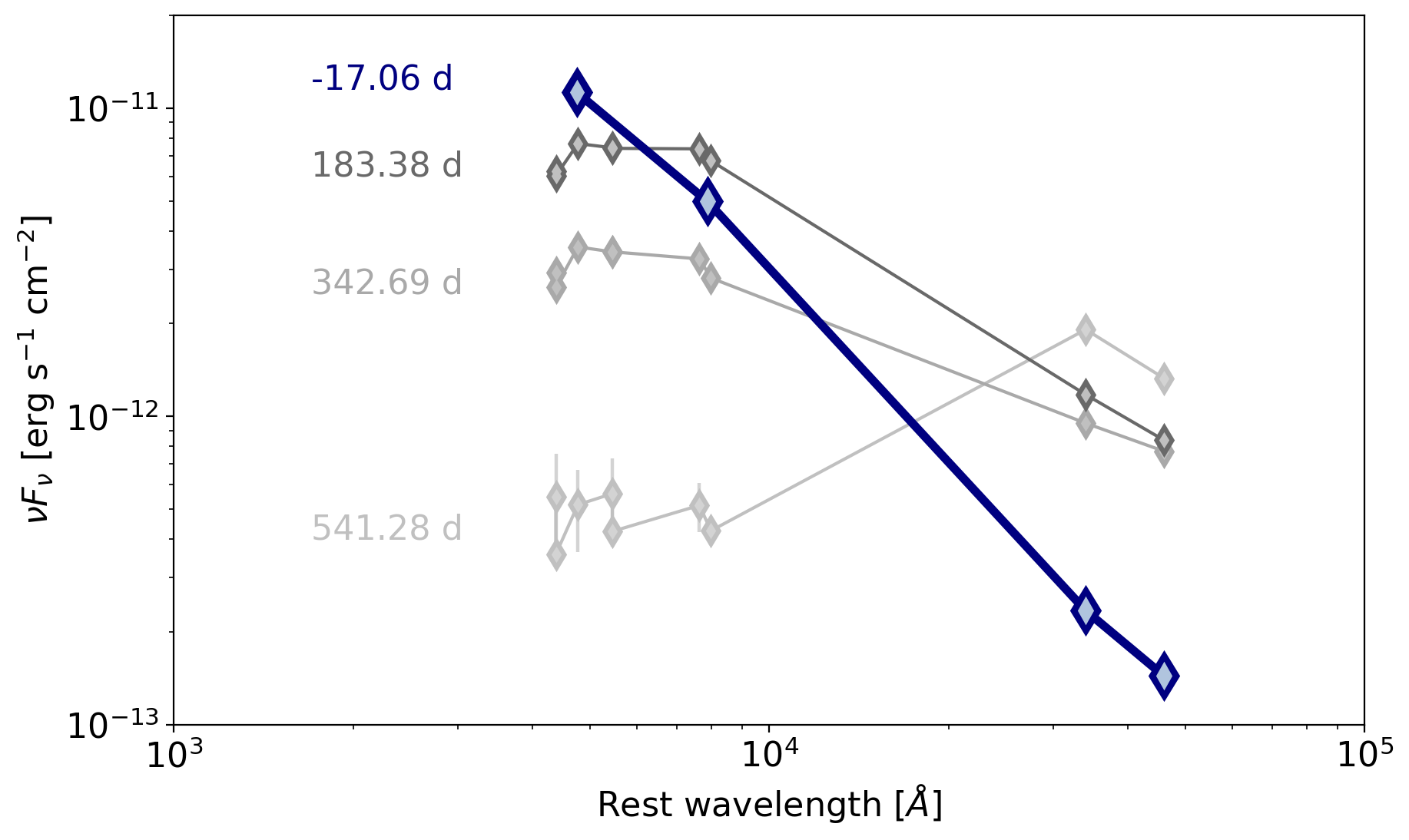}
\caption{The early SED at 58812.25 MJD, -17.06 phase, constructed from the ASAS--SN \textit{g}, TESS \textit{T}, and WISE W1 and W2 photometric data, in blue. We show the later SEDs that include WISE data from Fig.~\ref{fig:double_BB_fit} in grey colors for comparison. 
\label{fig:early_SED}}
\end{figure}

\begin{table*}[]
\centering
\caption{Rectified ASAS--SN relative flux density of SN\,2019vxm, and fluxes scaled to the Konkoly \textit{g} data.}
\label{tab:asassn-data}
\begin{tabular}{c c c c c c c }
\hline\hline
HJD & Rel.\ Flux Density & Flux Density Error & Scaled Flux & Scaled Flux Error & Quality & Camera \\
(day) & (mJy) & (mJy) & ($10^{-11}$ erg\,s$^{-1}$\,cm$^{-2}$) & ($10^{-11}$ erg\,s$^{-1}$\,cm$^{-2}$) & Flag & ~ \\
\hline
2458700.9716678 &   0.03430 & 0.0463 &   0.03038  & 0.04100 & G & bD \\
2458705.7557816 & $-0.0441$ & 0.1147 & $-0.03905$ & 0.10157 & G & bt \\
2458705.7557816 &   0.0770  & 0.0833 &   0.06819  & 0.07377 & G & bs \\
2458706.7881008 &   0.1569  & 0.0744 &   0.13894  & 0.06589 & G & bt \\
2458706.7881008 &   0.1919  & 0.0924 &   0.16994  & 0.08183 & G & bs \\
2458707.7142791 &   0.3372  & 0.0849 &   0.29861  & 0.07518 & G & bt \\
2458709.9021128 & $-0.1426$ & 0.1074 & $-0.12628$ & 0.09511 & B & bD \\
2458709.9021128 &   0.0520  & 0.1346 &   0.04605  & 0.11920 & B & bC \\
2458710.8834419 & $-0.1770$ & 0.2075 & $-0.15674$ & 0.18375 & G & bt \\
2458710.8834419 &   0.5783  & 0.1800 &   0.51212  & 0.15940 & G & bs \\
2458713.9478658 & $-0.0940$ & 0.0789 & $-0.08324$ & 0.06987 & G & bC \\
\dots & & & & & & \\
\hline
\end{tabular}
\end{table*}

For ASAS--SN we used the \textit{g}-band data scaled to the extinction-corrected Konkoly observations, and interpolated the flux density measurements to the WISE measurement time. For TESS, we used the values published by \citet{Lane-2025}, but with an extinction correction consistent with what we applied in other passbands. TESS observes in a wide passband, hence the amount of extinction also becomes a function of the color (or $T_{\rm eff}$) of the source. Since the TESS passband is nearly identical to \textit{Gaia} RP, we used a \textit{Gaia} extinction calculator published by \citet{Anders-2022}\footnote{\url{https://github.com/fjaellet/gaia\_edr3\_photutils}}. Assuming $T_{\rm eff} \sim 15000$\,K at this early phase, and $A_V=0.282$\,mag, we estimate an extinction of $A_T = 0.18\pm0.01$\,mag for this measurement. 

The reconstructed early SED, along with the SEDs from Fig.~\ref{fig:double_BB_fit} for comparison, is shown in Fig.~\ref{fig:early_SED}. The resulting points show the stellar flux steeply rising from the WISE points to the TESS point and then towards the ASAS--SN \textit{g} point, the latter slope reminiscent of the slope of the early SEDs, but at a lower overall flux level, which may suggest that the photosphere was hot already ($>10^4\,K$). The slope changes slightly between the WISE wavelengths, hinting at the possible presence of an IR excess, as well, although at much lower flux levels as later observations. But without detections at more wavelengths, especially towards the blue, reliable BB fits were not possible.

\subsection{WISE cold BB fit}\label{subsec:WISE_cold_db_BBfit}

To estimate the characteristic temperature of the near-IR component, we fitted a single BB function to the \textit{WISE} $W1$ and $W2$ photometric points at each epoch using an MCMC approach. Since only two photometric points are available at each phase, the fits are weakly constrained and should be interpreted with caution. In most epochs, the posterior temperature distribution shows a narrow dominant peak, which we adopt as the characteristic IR color temperaturee. In some cases, however, the MCMC chains also sample secondary solutions, visible as additional families of grey curves in Fig.~\ref{fig:simple_BB_fit_WISE}. 

\begin{figure*}
    \centering
    \includegraphics[width=0.49\textwidth]{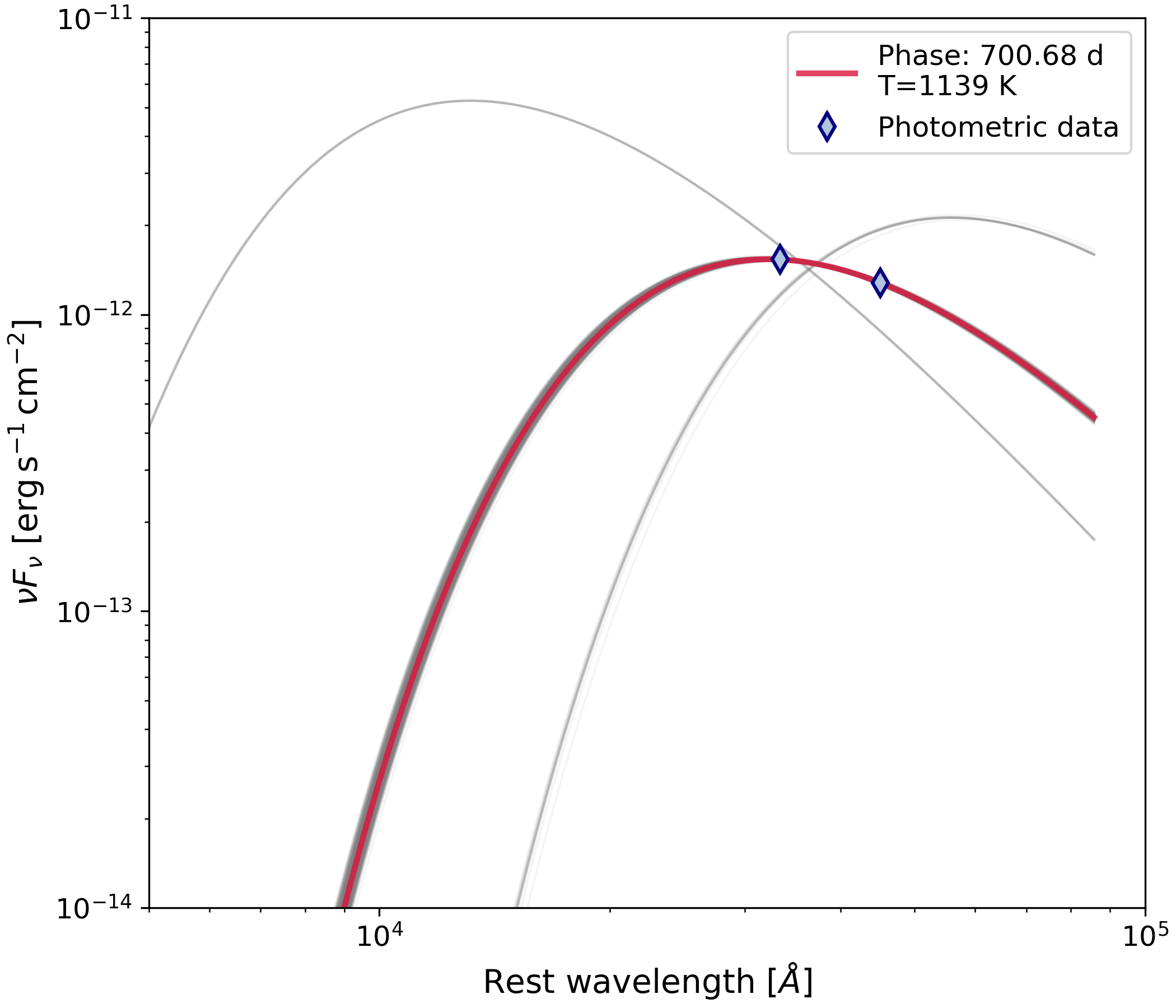}
    \includegraphics[width=0.49\textwidth]{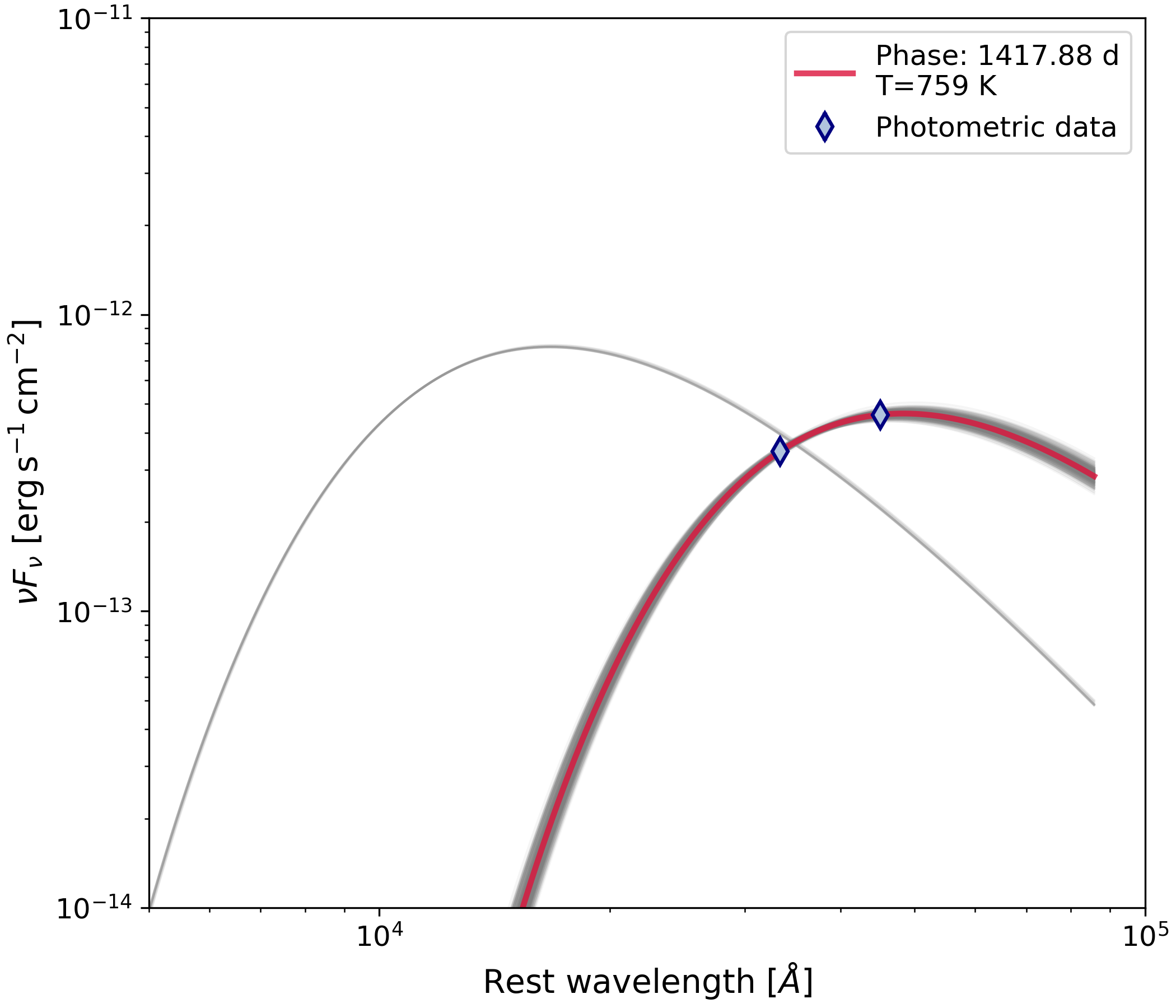}
    \caption{Examples of single BB fits to the IR SED of SN\,2019vxm using the \textit{WISE} $W1$ and $W2$ photometric data. The faint grey curves show 200 fits randomly drawn from the MCMC posterior distribution. The solid red curve represents the best-fit BB model, while blue diamonds indicate the observed photometric data points.}
    \label{fig:simple_BB_fit_WISE}
\end{figure*}

\subsection{Optical SED evolution}\label{subsec:optical_SED_evolution}

The full set of optical SEDs constructed from the Konkoly and Tsvetkov photometry is shown in Fig.~\ref{fig:SED: all Konkoly and Tsvetkov}. This figure complements the representative 12 SEDs shown in Section~\ref{subsec:SED}, in Fig.~\ref{fig:SED_12curves} demonstrating that the selected epochs trace the overall optical evolution well. Because  the two datasets were obtained with different filter sets, they are shown in separate panels, which allows the main evolutionary trends to be compared independently. Around maximum light, the SEDs are dominated by a blue continuum, while at later phases the flux excess in the $r$/$R$ bands becomes increasingly prominent. This excess is consistent with the growing contribution of H$\alpha$ emission to the broad-band optical flux.

\begin{figure*}
\centering
\includegraphics[width=0.99\textwidth]{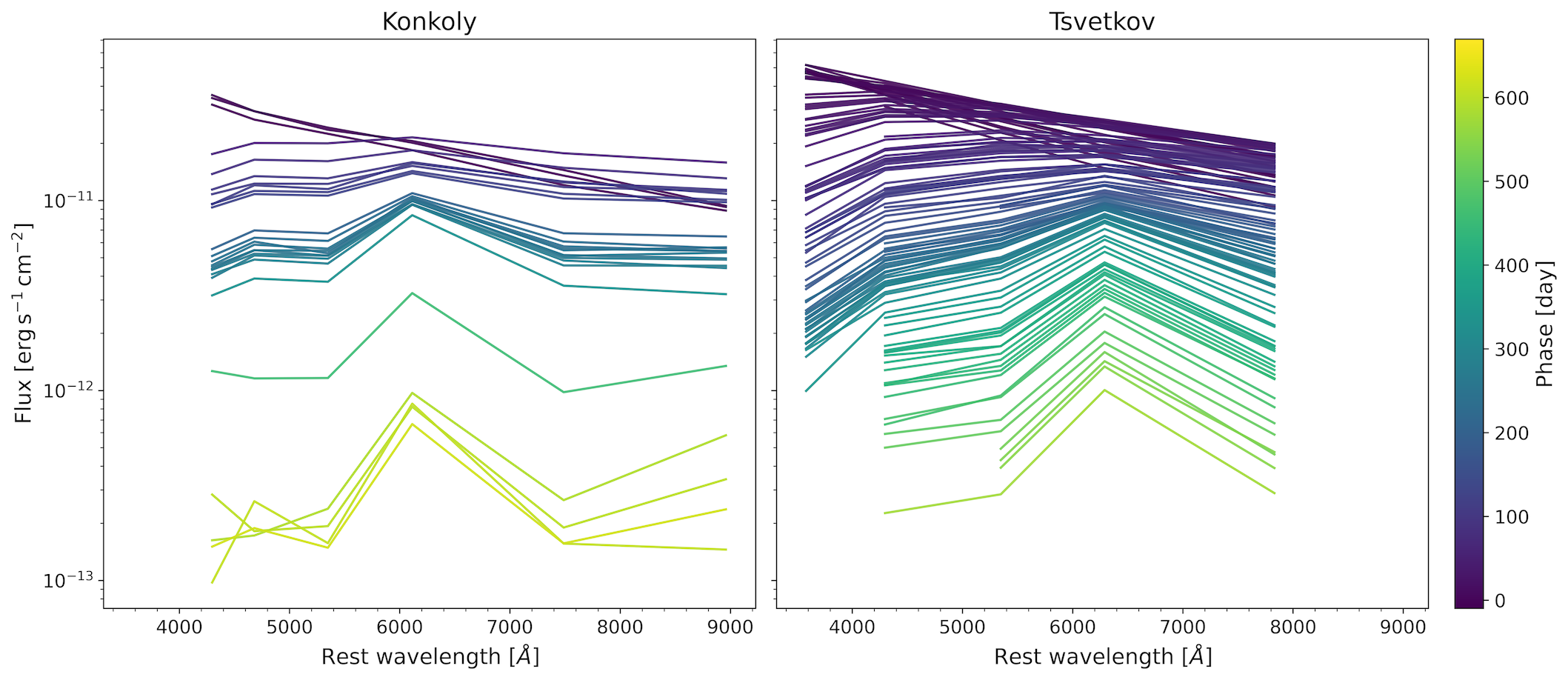}
\caption{Optical spectral energy distributions of SN\,2019vxm constructed from the full photometric datasets. The left panel shows the Konkoly dataset ($BgVriz$ filters), while the right panel shows the Tsvetkov dataset ($UBVRI$ filters). Colors indicate the phase measured relative to the $B$-band maximum ($t_p$).
\label{fig:SED: all Konkoly and Tsvetkov}}
\end{figure*}

\newpage
\section{Tables}\label{sec:Tables}

This section presents the tabulated observational and derived quantities used in the analysis. Table~\ref{tab:Konkoly_photometry} lists the uncorrected optical photometry obtained at Konkoly Observatory, while Table~\ref{tab:LTR} summarizes the bolometric luminosities and the corresponding BB parameters and velocities derived from the combined photometric data. These tables are publicly available on GitHub\footnote{\url{https://github.com/LelkesKlara/SN2019vxm}}.

\begin{deluxetable*}{c c c c c c c c c c c c c}
\centering
\tablecaption{Uncorrected $BVgriz$ photometry of SN\,2019vxm obtained at Piszkéstető Mountain Station of Konkoly Observatory (\textit{Konkoly dataset}). The magnitudes are not corrected for Galactic extinction, and do not include the rescaling to the \textit{Gaia} in $B$-band.}
\label{tab:Konkoly_photometry}
\tablecolumns{13}
\tablehead{
\colhead{MJD} & \colhead{B} & \colhead{$\sigma$B} & \colhead{V} & \colhead{$\sigma$V} & \colhead{g} & \colhead{$\sigma$g} & \colhead{r} & \colhead{$\sigma$r} & \colhead{i} & \colhead{$\sigma$i} & \colhead{z} & \colhead{$\sigma$z} \\
\colhead{(day)} & \colhead{(mag)} & \colhead{(mag)} & \colhead{(mag)} &  \colhead{(mag)} & \colhead{(mag)} & \colhead{(mag)} & \colhead{(mag)} & \colhead{(mag)} & \colhead{(mag)} & \colhead{(mag)} & \colhead{(mag)} & \colhead{(mag)} }
\startdata
58820.86 & 15.39 & 0.06 & 15.14 & 0.04 & 15.17 & 0.03 & 15.18 & 0.03 & 15.36 & 0.03 & 15.46 & 0.07 \\
58822.86 & 15.30 & 0.06 & 15.06 & 0.03 & 15.06 & 0.03 & 15.08 & 0.02 & 15.24 & 0.02 & 15.41 & 0.06 \\
58823.90 & 15.28 & 0.13 & 15.09 & 0.07 & 15.06 & 0.07 & 15.06 & 0.06 & 15.16 & 0.07 & 15.39 & 0.12 \\
58899.88 & 15.92 & 0.07 & 15.27 & 0.04 & 15.48 & 0.05 & 15.01 & 0.03 & 14.94 & 0.04 & 14.82 & 0.12 \\
58922.94 & 16.18 & 0.08 & 15.51 & 0.04 & 15.70 & 0.04 & 15.18 & 0.02 & 15.13 & 0.04 & 15.03 & 0.06 \\
58940.00 & 16.39 & 0.06 & 15.73 & 0.04 & 15.92 & 0.03 & 15.34 & 0.02 & 15.33 & 0.02 & 15.18 & 0.04 \\
58945.97 & 16.45 & 0.07 & 15.80 & 0.03 & 16.02 & 0.04 & 15.39 & 0.02 & 15.39 & 0.02 & 15.20 & 0.03 \\
58951.96 & 16.57 & 0.08 & 15.87 & 0.03 & 16.04 & 0.06 & 15.35 & 0.08 & 15.31 & 0.18 & 15.23 & 0.10 \\
58956.91 & 16.57 & 0.06 & 15.91 & 0.03 & 16.11 & 0.03 & 15.45 & 0.02 & 15.48 & 0.02 & 15.32 & 0.04 \\
58961.96 & 16.62 & 0.05 & 15.96 & 0.03 & 16.15 & 0.03 & 15.48 & 0.02 & 15.54 & 0.02 & 15.35 & 0.04 \\
59028.90 & 17.16 & 0.06 & 16.46 & 0.04 & 16.63 & 0.03 & 15.74 & 0.02 & 15.99 & 0.03 & 15.80 & 0.04 \\
59054.01 & 17.25 & 0.05 & 16.55 & 0.04 & 16.73 & 0.03 & 15.79 & 0.02 & 16.10 & 0.03 & 15.95 & 0.04 \\
59060.96 & --- & --- & --- & --- & --- & --- & 15.82 & 0.03 & 16.17 & 0.04 & 15.99 & 0.04 \\
59068.94 & 17.34 & 0.09 & 16.72 & 0.06 & 16.78 & 0.07 & 15.80 & 0.04 & 16.22 & 0.06 & 15.94 & 0.10 \\
59073.93 & 17.37 & 0.07 & 16.66 & 0.05 & 16.82 & 0.04 & 15.83 & 0.03 & 16.20 & 0.04 & 15.99 & 0.06 \\
59083.00 & 17.41 & 0.06 & 16.68 & 0.04 & 16.89 & 0.03 & 15.85 & 0.02 & 16.28 & 0.03 & 16.09 & 0.04 \\
59088.03 & 17.40 & 0.07 & 16.75 & 0.04 & 16.89 & 0.03 & 15.85 & 0.02 & 16.29 & 0.03 & 16.01 & 0.07 \\
59096.97 & 17.43 & 0.06 & 16.75 & 0.04 & 16.94 & 0.03 & 15.89 & 0.03 & 16.32 & 0.04 & 16.10 & 0.05 \\
59101.97 & 17.52 & 0.06 & 16.79 & 0.04 & 16.96 & 0.03 & 15.89 & 0.02 & 16.35 & 0.03 & 16.21 & 0.05 \\
59110.97 & 17.50 & 0.06 & 16.85 & 0.04 & 17.02 & 0.03 & 15.90 & 0.02 & 16.42 & 0.03 & 16.18 & 0.04 \\
59117.10 & 17.32 & 0.18 & 16.93 & 0.42 & 16.84 & 0.47 & 15.91 & 0.08 & 16.49 & 0.25 & 15.94 & 0.53 \\
59162.01 & 17.77 & 0.07 & 17.09 & 0.04 & 17.26 & 0.04 & 16.03 & 0.02 & 16.68 & 0.03 & 16.56 & 0.05 \\
59297.91 & 18.83 & 0.28 & 18.36 & 0.19 & 18.58 & 0.17 & 17.06 & 0.04 & 18.09 & 0.09 & 17.50 & 0.13 \\
59426.01 & 20.94 & 1.35 & 20.08 & 0.44 & 20.65 & 0.53 & 18.37 & 0.07 & 19.51 & 0.34 & 18.41 & 0.54 \\
59436.97 & 20.53 & 0.20 & 20.31 & 0.17 & 20.59 & 0.18 & 18.56 & 0.03 & 19.87 & 0.15 & 18.99 & 0.33 \\
59447.05 & 21.47 & 3.41 & 20.53 & 0.61 & 20.20 & 0.31 & 18.52 & 0.06 & 20.08 & 0.35 & 19.92 & 0.85 \\
59460.95 & 21.12 & 0.40 & 20.59 & 0.23 & 20.55 & 0.16 & 18.78 & 0.04 & 20.08 & 0.29 & 19.39 & 0.51 \\
59512.03 & --- & --- & --- & --- & --- & --- & 19.05 & 0.15 & 20.26 & 0.74 & --- & --- \\
59563.14 & 17.28 & 0.55 & 18.07 & 1.05 & 19.03 & 0.26 & 18.05 & 0.25 & 17.11 & 1.01 & 15.15 & 0.86 \\
59581.15 & --- & --- & --- & --- & 21.90 & 1.53 & 19.66 & 0.12 & --- & --- & 20.76 & 0.04 \\
59595.10 & --- & --- & 21.63 & 0.04 & 21.34 & 0.03 & 20.11 & 0.54 & 19.98 & 0.69 & 18.80 & 0.99 \\
59608.12 & 23.60 & 0.05 & --- & --- & 21.52 & 0.61 & 19.84 & 0.12 & --- & --- & 22.55 & 0.03 \\
59617.77 & --- & --- & --- & --- & 23.36 & 0.03 & 19.82 & 0.18 & --- & --- & --- & --- \\
59634.03 & --- & --- & 22.99 & 0.02 & --- & --- & 19.86 & 0.16 & 22.20 & 0.02 & --- & --- \\
59673.99 & --- & --- & 22.68 & 0.03 & --- & --- & 20.54 & 0.33 & --- & --- & --- & --- 
\enddata
\end{deluxetable*}

\pagebreak

\startlongtable
\begin{deluxetable}{c c c c c c c c c c c c}
\tablecaption{Bolometric luminosities and BB parameters of SN~2019vxm derived from the Konkoly and Tsvetkov datasets. The table lists the phase relative to $t_p$, the $B$-band maximum, the bolometric luminosity, the BB temperature, the BB radius and the expansion velocity inferred from the radius evolution.}
\label{tab:LTR}
\tablecolumns{12}
\tablehead{
\colhead{MJD} & \colhead{phase} & \colhead{L} & \colhead{$\sigma$L} & \colhead{T} & \colhead{$\sigma$T$_-$} & \colhead{$\sigma$T$_+$} & \colhead{R} & \colhead{$\sigma$R} & \colhead{v} & \colhead{$\sigma$v}& \colhead{source} \\
\colhead{(day)} & \colhead{(day)} & \colhead{($10^{43}$ erg/s)} & \colhead{($10^{43}$ erg/s)} & \colhead{(K)} & \colhead{(K)} & \colhead{(K)} & \colhead{($10^{13}$ cm)} & \colhead{($10^{13}$ cm)} & \colhead{(km/s)} & \colhead{(km/s)} & }
\startdata
58820.86 & -8.62 & 2.502 & 0.376 & 12671 & 833 & 995 & 116.72 & 17.68 & --- & --- & Konkoly \\
58822.65 & -6.86 & 2.696 & 0.207 & 14467 & 707 & 782 & 92.95 & 9.76 & 6800 & 572 & Tsvetkov \\
58822.86 & -6.65 & 2.626 & 0.377 & 12651 & 681 & 783 & 119.96 & 15.51 & --- & --- & Konkoly \\
58823.64 & -5.89 & 2.638 & 0.217 & 14535 & 1013 & 1209 & 91.07 & 13.23 & 6787 & 568 & Tsvetkov \\
58823.90 & -5.63 & 2.657 & 0.446 & 12295 & 1403 & 2088 & 127.74 & 31.05 & --- & --- & Konkoly \\
58824.67 & -4.88 & 2.712 & 0.211 & 13887 & 649 & 619 & 101.17 & 10.25 & 6762 & 560 & Tsvetkov \\
58825.64 & -3.93 & 2.752 & 0.207 & 13421 & 505 & 538 & 109.10 & 9.18 & 6734 & 552 & Tsvetkov \\
58826.64 & -2.94 & 2.801 & 0.211 & 13024 & 488 & 515 & 116.89 & 9.81 & 6707 & 543 & Tsvetkov \\
58827.66 & -1.94 & 2.900 & 0.228 & 13405 & 600 & 653 & 112.26 & 10.98 & 6680 & 535 & Tsvetkov \\
58828.64 & -0.98 & 2.906 & 0.228 & 12139 & 506 & 540 & 137.03 & 12.64 & 6652 & 526 & Tsvetkov \\
58829.64 & 0.00 & 3.110 & 0.252 & 11733 & 657 & 758 & 151.77 & 18.08 & 6594 & 506 & Tsvetkov \\
58832.68 & 2.98 & 2.957 & 0.228 & 11929 & 394 & 407 & 143.16 & 10.95 & 6564 & 495 & Tsvetkov \\
58833.72 & 4.00 & 2.917 & 0.224 & 11218 & 315 & 273 & 160.76 & 10.93 & 6465 & 460 & Tsvetkov \\
58836.65 & 6.88 & 2.902 & 0.228 & 10585 & 366 & 378 & 180.12 & 14.33 & 6431 & 448 & Tsvetkov \\
58837.64 & 7.85 & 2.950 & 0.222 & 10424 & 292 & 306 & 187.26 & 12.64 & 6014 & 348 & Tsvetkov \\
58843.65 & 13.75 & 2.762 & 0.212 & 9663 & 594 & 390 & 210.84 & 27.15 & 5619 & 333 & Tsvetkov \\
58848.66 & 18.67 & 2.720 & 0.210 & 9191 & 215 & 177 & 231.28 & 14.01 & 4919 & 330 & Tsvetkov \\
58851.63 & 21.58 & 2.650 & 0.207 & 8755 & 219 & 209 & 251.59 & 16.00 & 4290 & 411 & Tsvetkov \\
58853.67 & 23.58 & 2.611 & 0.210 & 8631 & 190 & 195 & 256.98 & 15.35 & 4027 & 470 & Tsvetkov \\
58854.63 & 24.52 & 2.572 & 0.206 & 8613 & 227 & 238 & 256.11 & 16.94 & 3751 & 507 & Tsvetkov \\
58858.63 & 28.45 & 2.479 & 0.203 & 8300 & 243 & 207 & 270.79 & 19.33 & 3676 & 510 & Tsvetkov \\
58862.62 & 32.37 & 2.442 & 0.211 & 8113 & 180 & 178 & 281.19 & 17.40 & 3394 & 503 & Tsvetkov \\
58864.64 & 34.35 & 2.371 & 0.199 & 8071 & 193 & 183 & 280.03 & 17.82 & 2842 & 502 & Tsvetkov \\
58865.64 & 35.33 & 2.336 & 0.193 & 7968 & 163 & 167 & 285.17 & 16.57 & 2627 & 500 & Tsvetkov \\
58867.68 & 37.33 & 2.299 & 0.195 & 7901 & 506 & 373 & 287.70 & 38.82 & 2390 & 468 & Tsvetkov \\
58868.64 & 38.27 & 2.284 & 0.198 & 7690 & 167 & 160 & 302.80 & 18.59 & 1819 & 557 & Tsvetkov \\
58869.64 & 39.25 & 2.272 & 0.193 & 7782 & 121 & 115 & 294.80 & 15.52 & 1896 & 475 & Tsvetkov \\
58874.08 & 43.61 & 2.161 & 0.184 & 7626 & 122 & 114 & 299.43 & 15.98 & 1641 & 343 & Tsvetkov \\
58890.08 & 59.31 & 1.909 & 0.185 & 7231 & 267 & 228 & 313.01 & 27.68 & 1431 & 303 & Tsvetkov \\
58896.10 & 65.22 & 1.802 & 0.184 & 6671 & 163 & 176 & 357.41 & 25.31 & 1087 & 309 & Tsvetkov \\
58898.08 & 67.16 & 1.763 & 0.179 & 6912 & 155 & 158 & 329.21 & 22.30 & 975 & 340 & Tsvetkov \\
58899.88 & 68.93 & 1.808 & 0.208 & 6579 & 258 & 293 & 368.00 & 35.82 & --- & --- & Konkoly \\
58905.05 & 74.00 & 1.665 & 0.166 & 6628 & 108 & 118 & 347.88 & 20.77 & 893 & 361 & Tsvetkov \\
58906.05 & 74.99 & 1.688 & 0.171 & 6752 & 136 & 140 & 337.58 & 21.85 & 269 & 477 & Tsvetkov \\
58910.05 & 78.91 & 1.608 & 0.160 & 6778 & 123 & 120 & 326.99 & 20.17 & 192 & 477 & Tsvetkov \\
58911.06 & 79.90 & 1.583 & 0.160 & 6683 & 114 & 117 & 333.74 & 20.34 & -115 & 425 & Tsvetkov \\
58913.03 & 81.84 & 1.564 & 0.156 & 6606 & 107 & 109 & 339.39 & 20.19 & -368 & 371 & Tsvetkov \\
58917.09 & 85.82 & 1.512 & 0.153 & 6481 & 103 & 106 & 346.83 & 20.70 & -402 & 364 & Tsvetkov \\
58918.08 & 86.79 & 1.503 & 0.155 & 6600 & 129 & 136 & 333.43 & 21.63 & -318 & 386 & Tsvetkov \\
58919.00 & 87.69 & 1.436 & 0.153 & 6501 & 107 & 120 & 335.92 & 21.10 & -596 & 331 & Tsvetkov \\
58922.94 & 91.56 & 1.466 & 0.145 & 6363 & 196 & 217 & 354.29 & 28.01 & --- & --- & Konkoly \\
58933.00 & 101.43 & 1.271 & 0.132 & 6368 & 100 & 103 & 329.31 & 20.00 & -664 & 293 & Tsvetkov \\
58940.00 & 108.30 & 1.214 & 0.105 & 6167 & 173 & 266 & 343.14 & 24.29 & --- & --- & Konkoly \\
58942.00 & 110.26 & 1.174 & 0.124 & 6503 & 132 & 141 & 303.62 & 20.21 & -739 & 282 & Tsvetkov \\
58942.97 & 111.22 & 1.155 & 0.123 & 6229 & 116 & 120 & 328.14 & 21.27 & -771 & 276 & Tsvetkov \\
58945.96 & 114.15 & 1.149 & 0.123 & 6424 & 125 & 127 & 307.73 & 20.32 & -803 & 282 & Tsvetkov \\
58945.97 & 114.16 & 1.160 & 0.101 & 5986 & 151 & 271 & 356.08 & 23.76 & --- & --- & Konkoly \\
58947.92 & 116.07 & 1.110 & 0.119 & 6199 & 95 & 97 & 324.80 & 20.07 & -820 & 307 & Tsvetkov \\
58949.93 & 118.05 & 1.171 & 0.128 & 6471 & 183 & 185 & 306.15 & 24.04 & -831 & 297 & Tsvetkov \\
58951.96 & 120.04 & 1.139 & 0.147 & 6063 & 315 & 387 & 343.86 & 42.08 & --- & --- & Konkoly \\
58955.93 & 123.94 & 1.051 & 0.109 & 6218 & 94 & 98 & 314.13 & 18.90 & -835 & 253 & Tsvetkov \\
58956.91 & 124.90 & 1.058 & 0.092 & 6083 & 203 & 417 & 329.21 & 26.22 & --- & --- & Konkoly \\
58961.96 & 129.85 & 1.017 & 0.088 & 6168 & 253 & 467 & 313.96 & 29.18 & --- & --- & Konkoly \\
58965.98 & 133.80 & 0.970 & 0.101 & 6251 & 97 & 97 & 298.55 & 18.11 & -830 & 224 & Tsvetkov \\
58972.93 & 140.62 & 0.915 & 0.096 & 6155 & 96 & 98 & 299.13 & 18.32 & -855 & 184 & Tsvetkov \\
58983.89 & 151.37 & 0.869 & 0.094 & 6135 & 113 & 119 & 293.40 & 19.19 & -886 & 169 & Tsvetkov \\
58990.01 & 157.38 & 0.808 & 0.088 & 6162 & 103 & 104 & 280.39 & 17.96 & -880 & 158 & Tsvetkov \\
59003.99 & 171.10 & 0.747 & 0.081 & 6102 & 91 & 95 & 274.95 & 17.04 & -856 & 136 & Tsvetkov \\
59009.98 & 176.98 & 0.728 & 0.081 & 6140 & 155 & 168 & 268.15 & 20.22 & -804 & 159 & Tsvetkov \\
59028.90 & 195.54 & 0.665 & 0.055 & 6022 & 301 & 542 & 266.41 & 28.88 & --- & --- & Konkoly \\
59030.94 & 197.55 & 0.632 & 0.070 & 6240 & 154 & 160 & 241.77 & 18.01 & -799 & 160 & Tsvetkov \\
59032.93 & 199.50 & 0.623 & 0.070 & 6057 & 131 & 138 & 254.81 & 18.03 & -746 & 204 & Tsvetkov \\
59049.91 & 216.16 & 0.583 & 0.061 & 6093 & 317 & 209 & 243.56 & 28.39 & -716 & 214 & Tsvetkov \\
59054.01 & 220.19 & 0.602 & 0.051 & 6002 & 197 & 256 & 255.16 & 19.92 & --- & --- & Konkoly \\
59056.88 & 223.00 & 0.591 & 0.062 & 6190 & 100 & 104 & 237.68 & 14.59 & -653 & 254 & Tsvetkov \\
59059.94 & 226.01 & 0.576 & 0.061 & 6203 & 204 & 159 & 233.63 & 19.77 & -549 & 260 & Tsvetkov \\
59066.83 & 232.77 & 0.560 & 0.061 & 6227 & 184 & 168 & 228.75 & 18.37 & -504 & 284 & Tsvetkov \\
59068.94 & 234.84 & 0.578 & 0.063 & 6000 & 367 & 620 & 250.17 & 33.44 & --- & --- & Konkoly \\
59069.85 & 235.73 & 0.552 & 0.059 & 6176 & 385 & 227 & 230.76 & 31.27 & -494 & 284 & Tsvetkov \\
59073.93 & 239.74 & 0.566 & 0.051 & 5912 & 220 & 302 & 254.93 & 22.16 & --- & --- & Konkoly \\
59077.01 & 242.76 & 0.545 & 0.058 & 6019 & 304 & 260 & 241.39 & 27.58 & -492 & 290 & Tsvetkov \\
59079.89 & 245.58 & 0.531 & 0.059 & 5998 & 625 & 753 & 240.00 & 51.77 & -482 & 312 & Tsvetkov \\
59083.00 & 248.64 & 0.535 & 0.044 & 6025 & 172 & 216 & 238.75 & 16.81 & --- & --- & Konkoly \\
59083.86 & 249.48 & 0.527 & 0.054 & 5997 & 444 & 505 & 239.22 & 37.53 & -506 & 304 & Tsvetkov \\
59088.03 & 253.57 & 0.543 & 0.048 & 6171 & 421 & 691 & 229.16 & 32.88 & --- & --- & Konkoly \\
59093.94 & 259.37 & 0.513 & 0.059 & 6006 & 590 & 608 & 235.33 & 48.16 & -508 & 303 & Tsvetkov \\
59096.97 & 262.35 & 0.518 & 0.045 & 6007 & 218 & 324 & 236.25 & 19.93 & --- & --- & Konkoly \\
59098.77 & 264.11 & 0.479 & 0.052 & 6082 & 406 & 373 & 221.72 & 31.95 & -582 & 283 & Tsvetkov \\
59101.97 & 267.25 & 0.493 & 0.042 & 6070 & 193 & 261 & 225.87 & 17.25 & --- & --- & Konkoly \\
59102.74 & 268.01 & 0.483 & 0.053 & 6239 & 169 & 167 & 211.55 & 16.35 & -656 & 329 & Tsvetkov \\
59109.75 & 274.89 & 0.466 & 0.048 & 5999 & 489 & 577 & 224.72 & 38.44 & -658 & 325 & Tsvetkov \\
59110.97 & 276.08 & 0.486 & 0.040 & 6269 & 428 & 612 & 210.20 & 29.99 & --- & --- & Konkoly \\
59114.72 & 279.76 & 0.460 & 0.047 & 6012 & 483 & 609 & 222.35 & 37.52 & -684 & 355 & Tsvetkov \\
59119.83 & 284.78 & 0.444 & 0.048 & 6054 & 445 & 439 & 215.34 & 33.72 & -677 & 334 & Tsvetkov \\
59127.79 & 292.59 & 0.440 & 0.048 & 6195 & 362 & 257 & 204.86 & 26.45 & -676 & 319 & Tsvetkov \\
59133.71 & 298.40 & 0.425 & 0.045 & 6573 & 432 & 264 & 178.78 & 25.32 & -665 & 282 & Tsvetkov \\
59139.68 & 304.26 & 0.411 & 0.041 & 6090 & 540 & 667 & 204.88 & 37.78 & -658 & 260 & Tsvetkov \\
59146.71 & 311.16 & 0.394 & 0.040 & 6145 & 462 & 486 & 196.89 & 31.25 & -659 & 222 & Tsvetkov \\
59161.69 & 325.86 & 0.366 & 0.036 & 6046 & 530 & 663 & 196.12 & 35.74 & -667 & 174 & Tsvetkov \\
59162.01 & 326.17 & 0.381 & 0.032 & 6196 & 186 & 224 & 190.45 & 13.91 & --- & --- & Konkoly \\
59180.69 & 344.50 & 0.318 & 0.035 & 6156 & 510 & 524 & 176.33 & 30.77 & -679 & 140 & Tsvetkov \\
59193.65 & 357.22 & 0.310 & 0.032 & 6281 & 401 & 286 & 167.23 & 23.04 & -673 & 165 & Tsvetkov \\
59202.63 & 366.04 & 0.285 & 0.030 & 6358 & 515 & 462 & 156.39 & 26.61 & -689 & 186 & Tsvetkov \\
59215.64 & 378.80 & 0.264 & 0.027 & 6083 & 457 & 502 & 164.64 & 26.10 & -694 & 180 & Tsvetkov \\
59234.10 & 396.92 & 0.224 & 0.024 & 6348 & 430 & 270 & 139.22 & 20.27 & -680 & 189 & Tsvetkov \\
59236.08 & 398.86 & 0.230 & 0.023 & 6332 & 241 & 150 & 141.83 & 12.97 & -644 & 190 & Tsvetkov \\
59238.05 & 400.79 & 0.222 & 0.023 & 6273 & 284 & 195 & 141.73 & 14.75 & -560 & 152 & Tsvetkov \\
59246.08 & 408.68 & 0.213 & 0.023 & 6455 & 150 & 146 & 131.17 & 9.28 & -541 & 145 & Tsvetkov \\
59248.08 & 410.64 & 0.205 & 0.023 & 6238 & 218 & 254 & 137.71 & 12.29 & -571 & 152 & Tsvetkov \\
59255.03 & 417.46 & 0.194 & 0.020 & 6507 & 182 & 154 & 123.33 & 9.48 & -531 & 131 & Tsvetkov \\
59265.03 & 427.27 & 0.180 & 0.019 & 6378 & 129 & 134 & 123.68 & 8.33 & -485 & 124 & Tsvetkov \\
59273.02 & 435.11 & 0.166 & 0.021 & 6152 & 278 & 323 & 127.38 & 13.96 & -474 & 103 & Tsvetkov \\
59278.94 & 440.92 & 0.159 & 0.017 & 6191 & 120 & 123 & 123.44 & 8.10 & -474 & 92 & Tsvetkov \\
59280.99 & 442.93 & 0.154 & 0.017 & 6239 & 153 & 170 & 119.37 & 8.87 & -480 & 87 & Tsvetkov \\
59287.99 & 449.80 & 0.144 & 0.016 & 5866 & 122 & 131 & 130.77 & 8.91 & -471 & 86 & Tsvetkov \\
59297.91 & 459.54 & 0.145 & 0.022 & 6325 & 713 & 894 & 112.58 & 26.83 & --- & --- & Konkoly \\
59304.95 & 466.45 & 0.120 & 0.015 & 5773 & 202 & 230 & 122.94 & 11.56 & -463 & 85 & Tsvetkov \\
59311.90 & 473.27 & 0.115 & 0.014 & 6125 & 190 & 204 & 107.12 & 9.15 & -435 & 66 & Tsvetkov \\
59335.00 & 495.94 & 0.096 & 0.011 & 6028 & 174 & 198 & 100.82 & 8.40 & -431 & 77 & Tsvetkov \\
59347.97 & 508.67 & 0.084 & 0.011 & 6014 & 181 & 198 & 95.20 & 8.25 & -439 & 70 & Tsvetkov \\
59411.98 & 571.48 & 0.045 & 0.006 & 5784 & 237 & 261 & 75.25 & 7.99 & -440 & 69 & Tsvetkov \\
59426.01 & 585.25 & 0.041 & 0.026 & 5503 & 1245 & 1569 & 78.96 & 43.45 & --- & --- & Konkoly \\
59436.97 & 596.01 & 0.033 & 0.007 & 6462 & 920 & 951 & 51.55 & 15.59 & --- & --- & Konkoly \\
59447.05 & 605.90 & 0.023 & 0.065 & 6094 & 1416 & 1281 & 48.53 & 71.57 & --- & --- & Konkoly \\
59460.95 & 619.54 & 0.024 & 0.008 & 6408 & 1185 & 1082 & 44.76 & 18.01 & --- & --- & Konkoly 
\enddata
\end{deluxetable}

\end{document}